%% file: main.tex
\documentclass[sigconf]{acmart}
\AtBeginDocument{%
  }

\copyrightyear{2026}
\acmYear{2026}
\setcopyright{cc}
\setcctype{by}
\acmConference[CHI '26]{Proceedings of the 2026 CHI Conference on Human Factors in Computing Systems}{April 13--17, 2026}{Barcelona, Spain}
\acmBooktitle{Proceedings of the 2026 CHI Conference on Human Factors in Computing Systems (CHI '26), April 13--17, 2026, Barcelona, Spain}
\acmPrice{}
\acmDOI{10.1145/3772318.3791291}
\acmISBN{979-8-4007-2278-3/2026/04}

\usepackage{xspace}
\usepackage{xcolor}
\usepackage{soul}
\usepackage{graphicx}
\usepackage{multirow}
\usepackage{bbding}
\usepackage{hyperref}
\usepackage{colortbl}
\usepackage{duckuments}
\usepackage{framed}
\usepackage{cleveref}
\usepackage{placeins}

\makeatletter
\@ifpackageloaded{arydshln}{
}{
  
}

\begin{document}

%%
%% The "title" command has an optional parameter,
%% allowing the author to define a "short title" to be used in page headers.
\title{A Design Space for Live Music Agents}

%%
%% The "author" command and its associated commands are used to define
%% the authors and their affiliations.
%% Of note is the shared affiliation of the first two authors, and the
%% "authornote" and "authornotemark" commands
%% used to denote shared contribution to the research.

\settopmatter{authorsperrow=4}

\author{Yewon Kim}
\affiliation{
    \institution{CMU}
    \city{Pittsburgh}
    \state{PA}
    \country{USA}
}

\author{Stephen Brade}
\affiliation{
    \institution{MIT}
    \city{Cambridge}
    \state{MA}
    \country{USA}
}

\author{Alexander Wang}
\affiliation{
    \institution{CMU}
    \city{Pittsburgh}
    \state{PA}
    \country{USA}
}

\author{David Zhou}
\affiliation{
    \institution{UIUC}
    \city{Urbana}
    \state{IL}
    \country{USA}
}

\author{Haven Kim}
\affiliation{
    \institution{UCSD}
    \city{La Jolla}
    \state{CA}
    \country{USA}
}

\author{Bill Wang}
\affiliation{
    \institution{UCSD}
    \city{La Jolla}
    \state{CA}
    \country{USA}
}

\author{Sung-Ju Lee}
\affiliation{
    \institution{KAIST}
    \city{Daejeon}
    \country{Republic of Korea}
}

\author{Hugo F Flores Garcia}
\affiliation{
    \institution{Northwestern University}
    \city{Evanston}
    \state{IL}
    \country{USA}
}

\author{Cheng-Zhi Anna Huang}
\affiliation{
    \institution{MIT}
    \city{Cambridge}
    \state{MA}
    \country{USA}
}

\author{Chris Donahue}
\affiliation{
    \institution{CMU}
    \city{Pittsburgh}
    \state{PA}
    \country{USA}
}

\renewcommand{\shortauthors}{Kim et al.}

%%
%% The code below is generated by the tool at http://dl.acm.org/ccs.cfm.
%% Please copy and paste the code instead of the example below.
%%
\input{macro}

\input{sections/00_abstract}



%%
%% Keywords. The author(s) should pick words that accurately describe
%% the work being presented. Separate the keywords with commas.
%% A "teaser" image appears between the author and affiliation
%% information and the body of the document, and typically spans the
%% page.
% \begin{teaserfigure}
%   \includegraphics[width=\textwidth]{sampleteaser}
%   \caption{Seattle Mariners at Spring Training, 2010.}
%   \Description{Enjoying the baseball game from the third-base
%   seats. Ichiro Suzuki preparing to bat.}
%   \label{fig:teaser}
% \end{teaserfigure}

%%
%% This command processes the author and affiliation and title
%% information and builds the first part of the formatted document.
\maketitle

\input{sections/01_introduction}
\input{sections/02_background}
\input{sections/03_definition}

\input{sections/04_methodology}
\input{sections/05_00_designspace}
\input{sections/06_casestudies}
\input{sections/07_discussion}
\input{sections/08_conclusion}

\begin{acks}
    We thank the CHI 2026 ACs and anonymous reviewers for their constructive feedback on the manuscript, and Mina Lee and Sihyun Yu for insightful discussions on the design space creation.
    The visual framing of the design space draws inspiration from prior work by Mina Lee and her coauthors.
    We disclose the use of generative AI tools in the writing process; these tools were used solely to edit the authors' own text, and the authors ensured that the final content is free from plagiarism, misrepresentation, fabrication, and falsification. 
    This work was supported by the Humanities and AI Virtual Institute (HAVI) at Schmidt Sciences and by the National Research Foundation of Korea (NRF) grant funded by the Korea government (MSIT) (RS-2024-00337007). 
\end{acks}

\bibliographystyle{ACM-Reference-Format}
\bibliography{_references,otherrefs}

\newpage
\input{sections/_appendix}

\end{document}

%% file: macro.tex
\definecolor{aliceblue}{rgb}{0.91, 0.94, 0.97}

\definecolor{pinegreen}{HTML}{01796F}
\newcommand{\greenurl}[1]{\href{#1}{\textcolor{pinegreen}{#1}}}

\newcommand{\note}[1]{{\color{red}\noindent\textbf{NOTE:} #1}\xspace}
\newcommand{\todo}[1]{{\color{blue}\emph{TODO: #1}}\xspace}
\newcommand{\todocite}[0]{{\color{blue}\textbf{[CITE]}}\xspace}
\newcommand{\placeholder}[1]{{\color{red}#1}\xspace}

\definecolor{revc}{RGB}{0, 0, 0} %{108, 0, 255} %revision color
\newcommand{\rev}[1]{{\color{revc}#1}}

\newcommand{\chris}[1]{{\color{magenta}\textbf{Chris}: #1}\xspace}
\newcommand{\yewon}[1]{{\color{blue}\textbf{Yewon}: #1}\xspace}
\newcommand{\stephen}[1]{{\color{pinegreen}\textbf{Stephen}: #1}\xspace}

\newcommand\eg{e.g.,~}
\newcommand\ie{i.e.,~}
\newcommand\dash{---}

% stats
\newenvironment{change}{}{\xspace}
\newcommand{\numpaper}{{\change{151}}\xspace}
\newcommand{\numvideo}{{\change{31}}\xspace}
\newcommand{\numsystem}{{\change{153}}\xspace}
\newcommand{\numkeypaper}{{\change{34}}\xspace}
\newcommand{\numkeysystem}{{\change{36}}\xspace}
\newcommand{\numretrievedpaper}{{\change{731}}\xspace}
\newcommand{\numfilteredpaper}{{\change{117}}\xspace}
\newcommand{\numtotalsystem}{{\change{184}}\xspace}

\newcommand{\agreementrate}{{\change{89.77}}\xspace}

\newcommand{\numdim}{{\change{31}}\xspace}
\newcommand{\numcode}{{\change{165}}\xspace}
\newcommand{\numplaceholder}{{\color{red}n}\xspace}

% % Custom fonts
\definecolor{mina-skyblue}{rgb}{0.1098, 0.5686, 0.7411}
% original color
% \definecolor{mina-yellow}{rgb}{0.9803, 0.7372, 0.0352}
% slightly darker color for readability
\definecolor{mina-yellow}{rgb}{0.9450, 0.6117, 0}
\definecolor{mina-blue}{rgb}{0.3333, 0.3764, 0.6627}
% original color
% \definecolor{mina-orange}{rgb}{0.9450, 0.5647, 0.1254}
% slightly darker color for readability
\definecolor{mina-orange}{rgb}{0.8392, 0.3921, 0.2666}
% original color
% \definecolor{mina-green}{rgb}{0.4078, 0.6392, 0.4274}
% slightly darker color for readability
\definecolor{mina-green}{rgb}{0.0274, 0.4274, 0.2352}

\definecolor{url-blue}{rgb}{0, 0, 0.5429}
\newcommand{\website}{\textcolor{url-blue}{\url{https://writing-assistant.github.io}}\xspace}

\newcommand{\textft}[1]{#1}
\newcommand{\customul}[2][black]{\setulcolor{#1}\ul{#2}\setulcolor{black}}

% Yewon's addition
% define your four colors
\definecolor{mutedviolet}{HTML}{EEDAFF} 
\definecolor{mutedpink}{HTML}{FFD9D9} 
\definecolor{mutedblue}{HTML}{CCE5FF} 
\definecolor{mutedgreen}{HTML}{CEF8C2} 
\definecolor{forestgreen}{rgb}{0.13, 0.55, 0.13}
\definecolor{hookergreen}{rgb}{0.0, 0.44, 0.0}
\definecolor{cozygreen}{RGB}{170, 210, 160} 

\definecolor{vviolet}{HTML}{676aa4} 
\definecolor{vpink}{HTML}{bb7699} 
\definecolor{vblue}{HTML}{618eb4} 
\definecolor{vgreen}{HTML}{367660} 

% now define four commands

\newcommand{\contextaspect}[1]{\textcolor{vviolet}{\textbf{#1}}}
\newcommand{\interactionaspect}[1]{\textcolor{vpink}{\textbf{#1}}}
\newcommand{\technologyaspect}[1]{\textcolor{vblue}{\textbf{#1}}}
\newcommand{\ecosystemaspect}[1]{\textcolor{vgreen}{\textbf{#1}}}

\newcommand{\contextdim}[1]{{\sethlcolor{mutedviolet}\hl{\textbf{#1}}}}
\newcommand{\interactiondim}[1]{{\sethlcolor{mutedpink}\hl{\textbf{#1}}}}
\newcommand{\technologydim}[1]{{\sethlcolor{mutedblue}\hl{\textbf{#1}}}}
\newcommand{\ecosystemdim}[1]
{{\sethlcolor{cozygreen!50}\hl{\textbf{#1}}}}

\newcommand{\contextcode}[1]{{\textft{\customul[vviolet]{#1}}}}
\newcommand{\interactioncode}[1]{{\textft{\customul[vpink]{#1}}}}
\newcommand{\technologycode}[1]{{\textft{\customul[vblue]{#1}}}}
\newcommand{\ecosystemcode}[1]{{\textft{\customul[vgreen]{#1}}}}

\newcommand{\question}[1]{\textbf{#1}}

\newcommand{\systemname}[1]{\textsc{#1}}

\newcommand{\autocitep}[1]{\citep{#1}}

\input{tables/macros}

%% file: tables/macros.tex
% Source material:
    % Spreadsheet: https://docs.google.com/spreadsheets/d/1FQWXAhaoJLxRXp3NMVx1oVLSiLKFzIlWCnAN7IdyyIU/edit?gid=0#gid=0
    % Processing Colab: https://colab.research.google.com/drive/1ABOSW5qM8_K4y0Qh0VOjLXYcj7nqJSJR?usp=sharing

\input{tables/data}
\newcolumntype{C}[1]{>{\centering\arraybackslash}p{#1}}
\newcommand{\aspectwithcaption}[7]{%
    % #1: Name
    % #2: Label for this aspect table
    % #3: Other labeled tables in this aspect
    % #4: Number relevant
    % #5: Percentage relevant
    % #6: Dimensions (content)
    % #7: Extra caption info
    \begin{table*}
    \caption[
          #1 dimensions, codes, and definitions. %
          #1 was relevant for #4 papers (#5). %
        ]{%
          \if\relax\detokenize{#3}\relax
            #1 dimensions, codes, and definitions. %
            #1 was relevant for #4 papers (#5). %
            #7%
          \else
            #1 dimensions, codes, and definitions (continues in \protect\Cref{#3}). %
            #1 was relevant for #4 papers (#5). %
            #7%
          \fi
        }
        \resizebox{1.\linewidth}{!}{%
            \renewcommand{\arraystretch}{1.45}
            \setlength{\tabcolsep}{4pt}
            \begin{tabular}{p{0.01\textwidth}p{0.22\textwidth}p{0.73\textwidth}C{0.03\textwidth}C{0.03\textwidth}C{0.03\textwidth}}
                \toprule
                & \textbf{Code} & \textbf{Definition} & \multicolumn{1}{c}{e.g.} & \multicolumn{1}{c}{\%} & \multicolumn{1}{c}{\#}\\
                #6
                \arrayrulecolor{black}\bottomrule
            \end{tabular}
        }
        % \caption[
        %   #1 dimensions, codes, and definitions. %
        %   #1 was relevant for #4 papers (#5). %
        % ]{%
        %   \if\relax\detokenize{#3}\relax
        %     #1 dimensions, codes, and definitions. %
        %     #1 was relevant for #4 papers (#5). %
        %     #7%
        %   \else
        %     #1 dimensions, codes, and definitions (continues in \protect\Cref{#3}). %
        %     #1 was relevant for #4 papers (#5). %
        %     #7%
        %   \fi
        % }        
    \label{#2}
    \end{table*}
}

\newcommand{\aspect}[6]{\aspectwithcaption{#1}{#2}{#3}{#4}{#5}{#6}{}}

\newcommand{\dimension}[7]{%
    % #1: Spreadsheet Name (DON'T CHANGE THIS, USED TO UPDATE NUMBERS FROM SPREADSHEET)
    % #2: Paper Name (e.g., Agent Role)
    % #3: Color (e.g., red!15, blue!15, yellow!15, green!15)
    % #4: Defining question (e.g., What musical function does the agent play in real time?)
    % #5: Number relevant
    % #6: Percentage relevant
    % #7: Codes (content)
    \arrayrulecolor{black}\midrule
    \multicolumn{4}{l}{{\sethlcolor{#3}\hl{\textbf{#2}}}: \textit{#4}} & \multicolumn{1}{c}{#6} & \multicolumn{1}{c}{#5} \\
    \arrayrulecolor{black!30}\midrule
    #7
}

\newcommand{\code}[6]{%
    % #1: Spreadsheet Name (DON'T CHANGE THIS, USED TO UPDATE NUMBERS FROM SPREADSHEET)
    % #2: Paper Name
    % #3: Definition
    % #4: Example citation
    % #5: Number
    % #6: Percentage
    & {#2} & #3 & \multicolumn{1}{c}{#4} & \multicolumn{1}{c}{#6} & \multicolumn{1}{c}{#5} \\
}

\newcommand{\codeio}[9]{
    & {#2} & {#3} & {#4 #7} & {#6 #9} & {#5 #8} \\
    \noalign{\setlength\aboverulesep{0pt}\setlength\belowrulesep{0pt}}%
    \cline{4-6}%
 }

\newcommand{\codeiolast}[9]{
    & {#2} & {#3} & {#4 #7} & {#6 #9} & {#5 #8} \\
    \noalign{\setlength\aboverulesep{0pt}\setlength\belowrulesep{0pt}}%
 }

\newcommand{\codeiopast}[9]{
\code{#1}{#2}{#3}{#4 #7}{#5 #8}{#6 #9}
}

%% file: tables/data.tex
% Progress: 184 / 184 (100.0%)
% Removed papers: 0
\newcommand{\aspUseContextNum}{182}
\newcommand{\aspUseContextPct}{99\%}
\newcommand{\aspInteractionNum}{184}
\newcommand{\aspInteractionPct}{100\%}
\newcommand{\aspTechnologyNum}{178}
\newcommand{\aspTechnologyPct}{97\%}
\newcommand{\aspEcosystemNum}{147}
\newcommand{\aspEcosystemPct}{80\%}
\newcommand{\dimUseContextUsePurposeNum}{162}
\newcommand{\dimUseContextUsePurposePct}{88\%}
\newcommand{\codUseContextUsePurposeLivePerformanceNum}{130}
\newcommand{\codUseContextUsePurposeLivePerformancePct}{80\%}
\newcommand{\codUseContextUsePurposeCompositionNum}{42}
\newcommand{\codUseContextUsePurposeCompositionPct}{26\%}
\newcommand{\codUseContextUsePurposeSkillAcquisitionNum}{6}
\newcommand{\codUseContextUsePurposeSkillAcquisitionPct}{4\%}
\newcommand{\codUseContextUsePurposeEntertainmentNum}{11}
\newcommand{\codUseContextUsePurposeEntertainmentPct}{7\%}
\newcommand{\codUseContextUsePurposeArtInstallationNum}{5}
\newcommand{\codUseContextUsePurposeArtInstallationPct}{3\%}
\newcommand{\codUseContextUsePurposeOtherNum}{2}
\newcommand{\codUseContextUsePurposeOtherPct}{1\%}
\newcommand{\dimUseContextMusicalContextNum}{177}
\newcommand{\dimUseContextMusicalContextPct}{96\%}
\newcommand{\codUseContextMusicalContextJazzImprovisationNum}{17}
\newcommand{\codUseContextMusicalContextJazzImprovisationPct}{10\%}
\newcommand{\codUseContextMusicalContextElectronicMusicNum}{40}
\newcommand{\codUseContextMusicalContextElectronicMusicPct}{23\%}
\newcommand{\codUseContextMusicalContextNewMusicNum}{31}
\newcommand{\codUseContextMusicalContextNewMusicPct}{18\%}
\newcommand{\codUseContextMusicalContextWesternClassicalNum}{17}
\newcommand{\codUseContextMusicalContextWesternClassicalPct}{10\%}
\newcommand{\codUseContextMusicalContextPopularMusicNum}{11}
\newcommand{\codUseContextMusicalContextPopularMusicPct}{6\%}
\newcommand{\codUseContextMusicalContextTraditionalMusicNum}{8}
\newcommand{\codUseContextMusicalContextTraditionalMusicPct}{5\%}
\newcommand{\codUseContextMusicalContextLiveCodingNum}{6}
\newcommand{\codUseContextMusicalContextLiveCodingPct}{3\%}
\newcommand{\codUseContextMusicalContextVirtuosicPracticeNum}{4}
\newcommand{\codUseContextMusicalContextVirtuosicPracticePct}{2\%}
\newcommand{\codUseContextMusicalContextNonSpecificNum}{68}
\newcommand{\codUseContextMusicalContextNonSpecificPct}{38\%}
\newcommand{\codUseContextMusicalContextOtherNum}{0}
\newcommand{\codUseContextMusicalContextOtherPct}{0\%}
\newcommand{\codUseContextMusicalContextScoreBasedNum}{7}
\newcommand{\codUseContextMusicalContextScoreBasedPct}{4\%}
\newcommand{\dimUseContextTargetUserNum}{177}
\newcommand{\dimUseContextTargetUserPct}{96\%}
\newcommand{\codUseContextTargetUserMusiciansNum}{150}
\newcommand{\codUseContextTargetUserMusiciansPct}{85\%}
\newcommand{\codUseContextTargetUserNoviceUsersNum}{30}
\newcommand{\codUseContextTargetUserNoviceUsersPct}{17\%}
\newcommand{\codUseContextTargetUserAudienceNum}{13}
\newcommand{\codUseContextTargetUserAudiencePct}{7\%}
\newcommand{\codUseContextTargetUserOtherNum}{2}
\newcommand{\codUseContextTargetUserOtherPct}{1\%}
\newcommand{\dimUseContextCollaborationStructureNum}{182}
\newcommand{\dimUseContextCollaborationStructurePct}{99\%}
\newcommand{\codUseContextCollaborationStructureOneOneCollaborationNum}{137}
\newcommand{\codUseContextCollaborationStructureOneOneCollaborationPct}{75\%}
\newcommand{\codUseContextCollaborationStructureOneNCollaborationNum}{7}
\newcommand{\codUseContextCollaborationStructureOneNCollaborationPct}{4\%}
\newcommand{\codUseContextCollaborationStructureNOneCollaborationNum}{32}
\newcommand{\codUseContextCollaborationStructureNOneCollaborationPct}{18\%}
\newcommand{\codUseContextCollaborationStructureNNCollaborationNum}{18}
\newcommand{\codUseContextCollaborationStructureNNCollaborationPct}{10\%}
\newcommand{\codUseContextCollaborationStructureAgentOnlyNum}{5}
\newcommand{\codUseContextCollaborationStructureAgentOnlyPct}{3\%}
\newcommand{\codUseContextCollaborationStructureOtherNum}{0}
\newcommand{\codUseContextCollaborationStructureOtherPct}{0\%}
\newcommand{\dimUseContextAgentRoleNum}{182}
\newcommand{\dimUseContextAgentRolePct}{99\%}
\newcommand{\codUseContextAgentRoleLeadNum}{47}
\newcommand{\codUseContextAgentRoleLeadPct}{26\%}
\newcommand{\codUseContextAgentRoleAccompanistNum}{59}
\newcommand{\codUseContextAgentRoleAccompanistPct}{32\%}
\newcommand{\codUseContextAgentRoleControllerNum}{20}
\newcommand{\codUseContextAgentRoleControllerPct}{11\%}
\newcommand{\codUseContextAgentRoleMapperNum}{49}
\newcommand{\codUseContextAgentRoleMapperPct}{27\%}
\newcommand{\codUseContextAgentRoleRemixerNum}{44}
\newcommand{\codUseContextAgentRoleRemixerPct}{24\%}
\newcommand{\codUseContextAgentRoleInterpreterNum}{9}
\newcommand{\codUseContextAgentRoleInterpreterPct}{5\%}
\newcommand{\codUseContextAgentRoleConductorNum}{5}
\newcommand{\codUseContextAgentRoleConductorPct}{3\%}
\newcommand{\codUseContextAgentRoleEvaluatorNum}{2}
\newcommand{\codUseContextAgentRoleEvaluatorPct}{1\%}
\newcommand{\codUseContextAgentRoleOtherNum}{1}
\newcommand{\codUseContextAgentRoleOtherPct}{1\%}
\newcommand{\dimUseContextUserRoleNum}{177}
\newcommand{\dimUseContextUserRolePct}{96\%}
\newcommand{\codUseContextUserRoleLeadNum}{118}
\newcommand{\codUseContextUserRoleLeadPct}{67\%}
\newcommand{\codUseContextUserRoleAccompanistNum}{8}
\newcommand{\codUseContextUserRoleAccompanistPct}{5\%}
\newcommand{\codUseContextUserRoleNonMusicalPerformerNum}{26}
\newcommand{\codUseContextUserRoleNonMusicalPerformerPct}{15\%}
\newcommand{\codUseContextUserRoleCoderNum}{9}
\newcommand{\codUseContextUserRoleCoderPct}{5\%}
\newcommand{\codUseContextUserRoleMixerNum}{4}
\newcommand{\codUseContextUserRoleMixerPct}{2\%}
\newcommand{\codUseContextUserRoleConductorNum}{15}
\newcommand{\codUseContextUserRoleConductorPct}{8\%}
\newcommand{\codUseContextUserRoleManipulatorNum}{37}
\newcommand{\codUseContextUserRoleManipulatorPct}{21\%}
\newcommand{\codUseContextUserRoleOtherNum}{0}
\newcommand{\codUseContextUserRoleOtherPct}{0\%}
\newcommand{\dimUseContextUserPreferenceNum}{153}
\newcommand{\dimUseContextUserPreferencePct}{83\%}
\newcommand{\codUseContextUserPreferenceControlNum}{59}
\newcommand{\codUseContextUserPreferenceControlPct}{39\%}
\newcommand{\codUseContextUserPreferenceDiversityNum}{43}
\newcommand{\codUseContextUserPreferenceDiversityPct}{28\%}
\newcommand{\codUseContextUserPreferenceCoherenceNum}{58}
\newcommand{\codUseContextUserPreferenceCoherencePct}{38\%}
\newcommand{\codUseContextUserPreferenceNoveltyNum}{56}
\newcommand{\codUseContextUserPreferenceNoveltyPct}{37\%}
\newcommand{\codUseContextUserPreferencePersonalizationNum}{36}
\newcommand{\codUseContextUserPreferencePersonalizationPct}{24\%}
\newcommand{\codUseContextUserPreferenceOtherNum}{4}
\newcommand{\codUseContextUserPreferenceOtherPct}{3\%}
\newcommand{\dimUseContextExpectedImpactNum}{150}
\newcommand{\dimUseContextExpectedImpactPct}{82\%}
\newcommand{\codUseContextExpectedImpactExpressionNum}{40}
\newcommand{\codUseContextExpectedImpactExpressionPct}{27\%}
\newcommand{\codUseContextExpectedImpactImmersionNum}{15}
\newcommand{\codUseContextExpectedImpactImmersionPct}{10\%}
\newcommand{\codUseContextExpectedImpactExplorationNum}{78}
\newcommand{\codUseContextExpectedImpactExplorationPct}{52\%}
\newcommand{\codUseContextExpectedImpactReflectionNum}{15}
\newcommand{\codUseContextExpectedImpactReflectionPct}{10\%}
\newcommand{\codUseContextExpectedImpactEngagementNum}{41}
\newcommand{\codUseContextExpectedImpactEngagementPct}{27\%}
\newcommand{\codUseContextExpectedImpactDelegationNum}{14}
\newcommand{\codUseContextExpectedImpactDelegationPct}{9\%}
\newcommand{\codUseContextExpectedImpactEmpowermentNum}{53}
\newcommand{\codUseContextExpectedImpactEmpowermentPct}{35\%}
\newcommand{\codUseContextExpectedImpactOtherNum}{0}
\newcommand{\codUseContextExpectedImpactOtherPct}{0\%}
\newcommand{\dimInteractionInterfaceNum}{176}
\newcommand{\dimInteractionInterfacePct}{96\%}
\newcommand{\codInteractionInterfaceGraphicalUserInterfaceNum}{84}
\newcommand{\codInteractionInterfaceGraphicalUserInterfacePct}{48\%}
\newcommand{\codInteractionInterfaceProgrammingInterfaceNum}{9}
\newcommand{\codInteractionInterfaceProgrammingInterfacePct}{5\%}
\newcommand{\codInteractionInterfaceConventionalInstrumentNum}{85}
\newcommand{\codInteractionInterfaceConventionalInstrumentPct}{48\%}
\newcommand{\codInteractionInterfaceCustomInstrumentNum}{27}
\newcommand{\codInteractionInterfaceCustomInstrumentPct}{15\%}
\newcommand{\codInteractionInterfaceDJGearNum}{2}
\newcommand{\codInteractionInterfaceDJGearPct}{1\%}
\newcommand{\codInteractionInterfaceEmbodiedAgentNum}{7}
\newcommand{\codInteractionInterfaceEmbodiedAgentPct}{4\%}
\newcommand{\codInteractionInterfaceStageVisualNum}{16}
\newcommand{\codInteractionInterfaceStageVisualPct}{9\%}
\newcommand{\codInteractionInterfaceSensorDeviceNum}{39}
\newcommand{\codInteractionInterfaceSensorDevicePct}{22\%}
\newcommand{\codInteractionInterfaceXRInterfaceNum}{2}
\newcommand{\codInteractionInterfaceXRInterfacePct}{1\%}
\newcommand{\codInteractionInterfaceOtherNum}{2}
\newcommand{\codInteractionInterfaceOtherPct}{1\%}
\newcommand{\dimInteractionInputModalityNum}{180}
\newcommand{\dimInteractionInputModalityPct}{98\%}
\newcommand{\codInteractionInputModalitySymbolicMusicNum}{49}
\newcommand{\codInteractionInputModalitySymbolicMusicPct}{27\%}
\newcommand{\codInteractionInputModalityRawAudioWaveformNum}{70}
\newcommand{\codInteractionInputModalityRawAudioWaveformPct}{39\%}
\newcommand{\codInteractionInputModalityGestureNum}{39}
\newcommand{\codInteractionInputModalityGesturePct}{22\%}
\newcommand{\codInteractionInputModalityVisualNum}{7}
\newcommand{\codInteractionInputModalityVisualPct}{4\%}
\newcommand{\codInteractionInputModalityControlSignalsNum}{47}
\newcommand{\codInteractionInputModalityControlSignalsPct}{26\%}
\newcommand{\codInteractionInputModalityNaturalLanguageNum}{13}
\newcommand{\codInteractionInputModalityNaturalLanguagePct}{7\%}
\newcommand{\codInteractionInputModalityProgrammingLanguageNum}{1}
\newcommand{\codInteractionInputModalityProgrammingLanguagePct}{1\%}
\newcommand{\codInteractionInputModalityOnBodySensorDataNum}{12}
\newcommand{\codInteractionInputModalityOnBodySensorDataPct}{7\%}
\newcommand{\codInteractionInputModalityExogenousSensorDataNum}{4}
\newcommand{\codInteractionInputModalityExogenousSensorDataPct}{2\%}
\newcommand{\codInteractionInputModalityOtherNum}{2}
\newcommand{\codInteractionInputModalityOtherPct}{1\%}
\newcommand{\dimInteractionOutputModalityNum}{184}
\newcommand{\dimInteractionOutputModalityPct}{100\%}
\newcommand{\codInteractionOutputModalitySymbolicMusicNum}{68}
\newcommand{\codInteractionOutputModalitySymbolicMusicPct}{37\%}
\newcommand{\codInteractionOutputModalityRawAudioWaveformNum}{107}
\newcommand{\codInteractionOutputModalityRawAudioWaveformPct}{58\%}
\newcommand{\codInteractionOutputModalityGestureNum}{4}
\newcommand{\codInteractionOutputModalityGesturePct}{2\%}
\newcommand{\codInteractionOutputModalityVisualNum}{33}
\newcommand{\codInteractionOutputModalityVisualPct}{18\%}
\newcommand{\codInteractionOutputModalityControlSignalsNum}{16}
\newcommand{\codInteractionOutputModalityControlSignalsPct}{9\%}
\newcommand{\codInteractionOutputModalityNaturalLanguageNum}{3}
\newcommand{\codInteractionOutputModalityNaturalLanguagePct}{2\%}
\newcommand{\codInteractionOutputModalityProgrammingLanguageNum}{1}
\newcommand{\codInteractionOutputModalityProgrammingLanguagePct}{1\%}
\newcommand{\codInteractionOutputModalityOnBodySensorDataNum}{0}
\newcommand{\codInteractionOutputModalityOnBodySensorDataPct}{0\%}
\newcommand{\codInteractionOutputModalityExogenousSensorDataNum}{0}
\newcommand{\codInteractionOutputModalityExogenousSensorDataPct}{0\%}
\newcommand{\codInteractionOutputModalityOtherNum}{2}
\newcommand{\codInteractionOutputModalityOtherPct}{1\%}
\newcommand{\dimInteractionInputMusicalElementNum}{146}
\newcommand{\dimInteractionInputMusicalElementPct}{79\%}
\newcommand{\codInteractionInputMusicalElementMelodyNum}{75}
\newcommand{\codInteractionInputMusicalElementMelodyPct}{51\%}
\newcommand{\codInteractionInputMusicalElementHarmonyNum}{25}
\newcommand{\codInteractionInputMusicalElementHarmonyPct}{17\%}
\newcommand{\codInteractionInputMusicalElementRhythmNum}{32}
\newcommand{\codInteractionInputMusicalElementRhythmPct}{22\%}
\newcommand{\codInteractionInputMusicalElementTimbreNum}{17}
\newcommand{\codInteractionInputMusicalElementTimbrePct}{12\%}
\newcommand{\codInteractionInputMusicalElementControlParametersNum}{39}
\newcommand{\codInteractionInputMusicalElementControlParametersPct}{27\%}
\newcommand{\codInteractionInputMusicalElementLyricsNum}{3}
\newcommand{\codInteractionInputMusicalElementLyricsPct}{2\%}
\newcommand{\codInteractionInputMusicalElementSoundTextureNum}{26}
\newcommand{\codInteractionInputMusicalElementSoundTexturePct}{18\%}
\newcommand{\codInteractionInputMusicalElementRenderedTrackNum}{5}
\newcommand{\codInteractionInputMusicalElementRenderedTrackPct}{3\%}
\newcommand{\codInteractionInputMusicalElementOtherNum}{9}
\newcommand{\codInteractionInputMusicalElementOtherPct}{6\%}
\newcommand{\dimInteractionOutputMusicalElementNum}{179}
\newcommand{\dimInteractionOutputMusicalElementPct}{97\%}
\newcommand{\codInteractionOutputMusicalElementMelodyNum}{65}
\newcommand{\codInteractionOutputMusicalElementMelodyPct}{36\%}
\newcommand{\codInteractionOutputMusicalElementHarmonyNum}{52}
\newcommand{\codInteractionOutputMusicalElementHarmonyPct}{29\%}
\newcommand{\codInteractionOutputMusicalElementRhythmNum}{51}
\newcommand{\codInteractionOutputMusicalElementRhythmPct}{28\%}
\newcommand{\codInteractionOutputMusicalElementTimbreNum}{51}
\newcommand{\codInteractionOutputMusicalElementTimbrePct}{28\%}
\newcommand{\codInteractionOutputMusicalElementControlParametersNum}{20}
\newcommand{\codInteractionOutputMusicalElementControlParametersPct}{11\%}
\newcommand{\codInteractionOutputMusicalElementLyricsNum}{3}
\newcommand{\codInteractionOutputMusicalElementLyricsPct}{2\%}
\newcommand{\codInteractionOutputMusicalElementSoundTextureNum}{53}
\newcommand{\codInteractionOutputMusicalElementSoundTexturePct}{30\%}
\newcommand{\codInteractionOutputMusicalElementRenderedTrackNum}{18}
\newcommand{\codInteractionOutputMusicalElementRenderedTrackPct}{10\%}
\newcommand{\codInteractionOutputMusicalElementOtherNum}{0}
\newcommand{\codInteractionOutputMusicalElementOtherPct}{0\%}
\newcommand{\dimInteractionMusicalOutcomeNum}{178}
\newcommand{\dimInteractionMusicalOutcomePct}{97\%}
\newcommand{\codInteractionMusicalOutcomeMonophonyNum}{24}
\newcommand{\codInteractionMusicalOutcomeMonophonyPct}{13\%}
\newcommand{\codInteractionMusicalOutcomeHomophonyNum}{46}
\newcommand{\codInteractionMusicalOutcomeHomophonyPct}{26\%}
\newcommand{\codInteractionMusicalOutcomePolyphonyNum}{42}
\newcommand{\codInteractionMusicalOutcomePolyphonyPct}{24\%}
\newcommand{\codInteractionMusicalOutcomeHeterophonyNum}{3}
\newcommand{\codInteractionMusicalOutcomeHeterophonyPct}{2\%}
\newcommand{\codInteractionMusicalOutcomeTexturalNum}{50}
\newcommand{\codInteractionMusicalOutcomeTexturalPct}{28\%}
\newcommand{\codInteractionMusicalOutcomeNonSpecificNum}{27}
\newcommand{\codInteractionMusicalOutcomeNonSpecificPct}{15\%}
\newcommand{\codInteractionMusicalOutcomeOtherNum}{1}
\newcommand{\codInteractionMusicalOutcomeOtherPct}{1\%}
\newcommand{\dimInteractionPlanningNum}{165}
\newcommand{\dimInteractionPlanningPct}{90\%}
\newcommand{\codInteractionPlanningScorePlanningNum}{33}
\newcommand{\codInteractionPlanningScorePlanningPct}{20\%}
\newcommand{\codInteractionPlanningTimelineNum}{3}
\newcommand{\codInteractionPlanningTimelinePct}{2\%}
\newcommand{\codInteractionPlanningTailoringNum}{35}
\newcommand{\codInteractionPlanningTailoringPct}{21\%}
\newcommand{\codInteractionPlanningUserConfigurationNum}{40}
\newcommand{\codInteractionPlanningUserConfigurationPct}{24\%}
\newcommand{\codInteractionPlanningPredefinedCuesNum}{18}
\newcommand{\codInteractionPlanningPredefinedCuesPct}{11\%}
\newcommand{\codInteractionPlanningMaterialPrepNum}{26}
\newcommand{\codInteractionPlanningMaterialPrepPct}{16\%}
\newcommand{\codInteractionPlanningNoPlanningNum}{59}
\newcommand{\codInteractionPlanningNoPlanningPct}{36\%}
\newcommand{\codInteractionPlanningOtherNum}{0}
\newcommand{\codInteractionPlanningOtherPct}{0\%}
\newcommand{\dimInteractionTemporalStructureNum}{175}
\newcommand{\dimInteractionTemporalStructurePct}{95\%}
\newcommand{\codInteractionTemporalStructureDenseParallelNum}{112}
\newcommand{\codInteractionTemporalStructureDenseParallelPct}{64\%}
\newcommand{\codInteractionTemporalStructureSparseParallelNum}{21}
\newcommand{\codInteractionTemporalStructureSparseParallelPct}{12\%}
\newcommand{\codInteractionTemporalStructureTurnTakingNum}{8}
\newcommand{\codInteractionTemporalStructureTurnTakingPct}{5\%}
\newcommand{\codInteractionTemporalStructureHybridNum}{21}
\newcommand{\codInteractionTemporalStructureHybridPct}{12\%}
\newcommand{\codInteractionTemporalStructureUnstructuredNum}{18}
\newcommand{\codInteractionTemporalStructureUnstructuredPct}{10\%}
\newcommand{\codInteractionTemporalStructureOtherNum}{0}
\newcommand{\codInteractionTemporalStructureOtherPct}{0\%}
\newcommand{\dimInteractionDataAlignmentNum}{166}
\newcommand{\dimInteractionDataAlignmentPct}{90\%}
\newcommand{\codInteractionDataAlignmentContinuousStreamNum}{117}
\newcommand{\codInteractionDataAlignmentContinuousStreamPct}{70\%}
\newcommand{\codInteractionDataAlignmentPeriodicNum}{12}
\newcommand{\codInteractionDataAlignmentPeriodicPct}{7\%}
\newcommand{\codInteractionDataAlignmentBackgroundTriggerNum}{40}
\newcommand{\codInteractionDataAlignmentBackgroundTriggerPct}{24\%}
\newcommand{\codInteractionDataAlignmentOtherNum}{1}
\newcommand{\codInteractionDataAlignmentOtherPct}{1\%}
\newcommand{\dimInteractionControlModeNum}{181}
\newcommand{\dimInteractionControlModePct}{98\%}
\newcommand{\codInteractionControlModeExplicitNum}{87}
\newcommand{\codInteractionControlModeExplicitPct}{48\%}
\newcommand{\codInteractionControlModeImplicitNum}{117}
\newcommand{\codInteractionControlModeImplicitPct}{65\%}
\newcommand{\codInteractionControlModeNoControlNum}{7}
\newcommand{\codInteractionControlModeNoControlPct}{4\%}
\newcommand{\codInteractionControlModeOtherNum}{0}
\newcommand{\codInteractionControlModeOtherPct}{0\%}
\newcommand{\dimInteractionControlScopeNum}{175}
\newcommand{\dimInteractionControlScopePct}{95\%}
\newcommand{\codInteractionControlScopeBehavioralDirectionNum}{36}
\newcommand{\codInteractionControlScopeBehavioralDirectionPct}{21\%}
\newcommand{\codInteractionControlScopeStyleDirectionNum}{67}
\newcommand{\codInteractionControlScopeStyleDirectionPct}{38\%}
\newcommand{\codInteractionControlScopeMusicalStructureNum}{29}
\newcommand{\codInteractionControlScopeMusicalStructurePct}{17\%}
\newcommand{\codInteractionControlScopeMusicalMaterialNum}{128}
\newcommand{\codInteractionControlScopeMusicalMaterialPct}{73\%}
\newcommand{\codInteractionControlScopeOtherNum}{5}
\newcommand{\codInteractionControlScopeOtherPct}{3\%}
\newcommand{\dimInteractionSystemInitiativeNum}{180}
\newcommand{\dimInteractionSystemInitiativePct}{98\%}
\newcommand{\codInteractionSystemInitiativeReactiveNum}{108}
\newcommand{\codInteractionSystemInitiativeReactivePct}{60\%}
\newcommand{\codInteractionSystemInitiativeProactiveNum}{6}
\newcommand{\codInteractionSystemInitiativeProactivePct}{3\%}
\newcommand{\codInteractionSystemInitiativeMixedInitiativeNum}{66}
\newcommand{\codInteractionSystemInitiativeMixedInitiativePct}{37\%}
\newcommand{\codInteractionSystemInitiativeOtherNum}{0}
\newcommand{\codInteractionSystemInitiativeOtherPct}{0\%}
\newcommand{\dimInteractionSystemGenerationStrategyNum}{162}
\newcommand{\dimInteractionSystemGenerationStrategyPct}{88\%}
\newcommand{\codInteractionSystemGenerationStrategyConvergingNum}{84}
\newcommand{\codInteractionSystemGenerationStrategyConvergingPct}{52\%}
\newcommand{\codInteractionSystemGenerationStrategyDivergingNum}{4}
\newcommand{\codInteractionSystemGenerationStrategyDivergingPct}{2\%}
\newcommand{\codInteractionSystemGenerationStrategyFlexibleNum}{75}
\newcommand{\codInteractionSystemGenerationStrategyFlexiblePct}{46\%}
\newcommand{\codInteractionSystemGenerationStrategyOtherNum}{0}
\newcommand{\codInteractionSystemGenerationStrategyOtherPct}{0\%}
\newcommand{\dimInteractionAgencyFramingNum}{177}
\newcommand{\dimInteractionAgencyFramingPct}{96\%}
\newcommand{\codInteractionAgencyFramingPartnerNum}{43}
\newcommand{\codInteractionAgencyFramingPartnerPct}{24\%}
\newcommand{\codInteractionAgencyFramingToolNum}{71}
\newcommand{\codInteractionAgencyFramingToolPct}{40\%}
\newcommand{\codInteractionAgencyFramingHybridNum}{63}
\newcommand{\codInteractionAgencyFramingHybridPct}{36\%}
\newcommand{\codInteractionAgencyFramingOtherNum}{1}
\newcommand{\codInteractionAgencyFramingOtherPct}{1\%}
\newcommand{\dimTechnologyModelNum}{162}
\newcommand{\dimTechnologyModelPct}{88\%}
\newcommand{\codTechnologyModelRuleBasedNum}{37}
\newcommand{\codTechnologyModelRuleBasedPct}{23\%}
\newcommand{\codTechnologyModelStochasticProcessNum}{54}
\newcommand{\codTechnologyModelStochasticProcessPct}{33\%}
\newcommand{\codTechnologyModelClassicalMLNum}{40}
\newcommand{\codTechnologyModelClassicalMLPct}{25\%}
\newcommand{\codTechnologyModelShallowNeuralNetworkNum}{23}
\newcommand{\codTechnologyModelShallowNeuralNetworkPct}{14\%}
\newcommand{\codTechnologyModelTaskSpecificDNNNum}{45}
\newcommand{\codTechnologyModelTaskSpecificDNNPct}{28\%}
\newcommand{\codTechnologyModelTransformerNum}{10}
\newcommand{\codTechnologyModelTransformerPct}{6\%}
\newcommand{\codTechnologyModelOtherNum}{0}
\newcommand{\codTechnologyModelOtherPct}{0\%}
\newcommand{\dimTechnologyLearningAlgorithmNum}{113}
\newcommand{\dimTechnologyLearningAlgorithmPct}{61\%}
\newcommand{\codTechnologyLearningAlgorithmSupervisedLearningNum}{60}
\newcommand{\codTechnologyLearningAlgorithmSupervisedLearningPct}{53\%}
\newcommand{\codTechnologyLearningAlgorithmUnsupervisedLearningNum}{28}
\newcommand{\codTechnologyLearningAlgorithmUnsupervisedLearningPct}{25\%}
\newcommand{\codTechnologyLearningAlgorithmSelfSupervisedLearningNum}{19}
\newcommand{\codTechnologyLearningAlgorithmSelfSupervisedLearningPct}{17\%}
\newcommand{\codTechnologyLearningAlgorithmReinforcementLearningNum}{6}
\newcommand{\codTechnologyLearningAlgorithmReinforcementLearningPct}{5\%}
\newcommand{\codTechnologyLearningAlgorithmOtherNum}{8}
\newcommand{\codTechnologyLearningAlgorithmOtherPct}{7\%}
\newcommand{\dimTechnologyInferenceObjectiveNum}{168}
\newcommand{\dimTechnologyInferenceObjectivePct}{91\%}
\newcommand{\codTechnologyInferenceObjectiveRegressionNum}{24}
\newcommand{\codTechnologyInferenceObjectiveRegressionPct}{14\%}
\newcommand{\codTechnologyInferenceObjectiveClassificationNum}{41}
\newcommand{\codTechnologyInferenceObjectiveClassificationPct}{24\%}
\newcommand{\codTechnologyInferenceObjectiveRetrievalNum}{13}
\newcommand{\codTechnologyInferenceObjectiveRetrievalPct}{8\%}
\newcommand{\codTechnologyInferenceObjectiveUnimodalGenerationNum}{88}
\newcommand{\codTechnologyInferenceObjectiveUnimodalGenerationPct}{52\%}
\newcommand{\codTechnologyInferenceObjectiveCrossModalGenerationNum}{33}
\newcommand{\codTechnologyInferenceObjectiveCrossModalGenerationPct}{20\%}
\newcommand{\codTechnologyInferenceObjectiveOtherNum}{1}
\newcommand{\codTechnologyInferenceObjectiveOtherPct}{1\%}
\newcommand{\dimTechnologyAdaptationNum}{165}
\newcommand{\dimTechnologyAdaptationPct}{90\%}
\newcommand{\codTechnologyAdaptationNoAdaptationNum}{105}
\newcommand{\codTechnologyAdaptationNoAdaptationPct}{64\%}
\newcommand{\codTechnologyAdaptationOnlineAdaptationNum}{18}
\newcommand{\codTechnologyAdaptationOnlineAdaptationPct}{11\%}
\newcommand{\codTechnologyAdaptationOfflineAdaptationNum}{42}
\newcommand{\codTechnologyAdaptationOfflineAdaptationPct}{25\%}
\newcommand{\codTechnologyAdaptationContinualAdaptationNum}{5}
\newcommand{\codTechnologyAdaptationContinualAdaptationPct}{3\%}
\newcommand{\codTechnologyAdaptationOtherNum}{0}
\newcommand{\codTechnologyAdaptationOtherPct}{0\%}
\newcommand{\dimTechnologyTechnologyInfrastructureNum}{135}
\newcommand{\dimTechnologyTechnologyInfrastructurePct}{73\%}
\newcommand{\codTechnologyTechnologyInfrastructureMusicProgrammingEnvironmentNum}{84}
\newcommand{\codTechnologyTechnologyInfrastructureMusicProgrammingEnvironmentPct}{62\%}
\newcommand{\codTechnologyTechnologyInfrastructureGeneralProgrammingEnvironmentNum}{54}
\newcommand{\codTechnologyTechnologyInfrastructureGeneralProgrammingEnvironmentPct}{40\%}
\newcommand{\codTechnologyTechnologyInfrastructureProtocolNum}{42}
\newcommand{\codTechnologyTechnologyInfrastructureProtocolPct}{31\%}
\newcommand{\codTechnologyTechnologyInfrastructureSoftwareToolkitNum}{22}
\newcommand{\codTechnologyTechnologyInfrastructureSoftwareToolkitPct}{16\%}
\newcommand{\codTechnologyTechnologyInfrastructureHardwareToolkitNum}{13}
\newcommand{\codTechnologyTechnologyInfrastructureHardwareToolkitPct}{10\%}
\newcommand{\codTechnologyTechnologyInfrastructureAIMLFrameworkNum}{44}
\newcommand{\codTechnologyTechnologyInfrastructureAIMLFrameworkPct}{33\%}
\newcommand{\codTechnologyTechnologyInfrastructureOtherNum}{1}
\newcommand{\codTechnologyTechnologyInfrastructureOtherPct}{1\%}
\newcommand{\dimTechnologyTechnologyDesiderataNum}{124}
\newcommand{\dimTechnologyTechnologyDesiderataPct}{67\%}
\newcommand{\codTechnologyTechnologyDesiderataLatencyNum}{106}
\newcommand{\codTechnologyTechnologyDesiderataLatencyPct}{85\%}
\newcommand{\codTechnologyTechnologyDesiderataEfficiencyNum}{33}
\newcommand{\codTechnologyTechnologyDesiderataEfficiencyPct}{27\%}
\newcommand{\codTechnologyTechnologyDesiderataErrorHandlingNum}{15}
\newcommand{\codTechnologyTechnologyDesiderataErrorHandlingPct}{12\%}
\newcommand{\codTechnologyTechnologyDesiderataTempoAdaptabilityNum}{30}
\newcommand{\codTechnologyTechnologyDesiderataTempoAdaptabilityPct}{24\%}
\newcommand{\codTechnologyTechnologyDesiderataOtherNum}{0}
\newcommand{\codTechnologyTechnologyDesiderataOtherPct}{0\%}
\newcommand{\dimTechnologyRuntimeRequirementsNum}{151}
\newcommand{\dimTechnologyRuntimeRequirementsPct}{82\%}
\newcommand{\codTechnologyRuntimeRequirementsCommodityMachineNum}{116}
\newcommand{\codTechnologyRuntimeRequirementsCommodityMachinePct}{77\%}
\newcommand{\codTechnologyRuntimeRequirementsHighPerformanceComputeNum}{1}
\newcommand{\codTechnologyRuntimeRequirementsHighPerformanceComputePct}{1\%}
\newcommand{\codTechnologyRuntimeRequirementsDedicatedCommodityHardwareNum}{39}
\newcommand{\codTechnologyRuntimeRequirementsDedicatedCommodityHardwarePct}{26\%}
\newcommand{\codTechnologyRuntimeRequirementsCustomHardwareNum}{30}
\newcommand{\codTechnologyRuntimeRequirementsCustomHardwarePct}{20\%}
\newcommand{\codTechnologyRuntimeRequirementsCloudAPINum}{6}
\newcommand{\codTechnologyRuntimeRequirementsCloudAPIPct}{4\%}
\newcommand{\codTechnologyRuntimeRequirementsOtherNum}{0}
\newcommand{\codTechnologyRuntimeRequirementsOtherPct}{0\%}
\newcommand{\dimTechnologyWorkflowIntegrationNum}{147}
\newcommand{\dimTechnologyWorkflowIntegrationPct}{80\%}
\newcommand{\codTechnologyWorkflowIntegrationBespokeSetupNum}{64}
\newcommand{\codTechnologyWorkflowIntegrationBespokeSetupPct}{44\%}
\newcommand{\codTechnologyWorkflowIntegrationSourceOnlyPrototypeNum}{27}
\newcommand{\codTechnologyWorkflowIntegrationSourceOnlyPrototypePct}{18\%}
\newcommand{\codTechnologyWorkflowIntegrationDeveloperToolkitNum}{7}
\newcommand{\codTechnologyWorkflowIntegrationDeveloperToolkitPct}{5\%}
\newcommand{\codTechnologyWorkflowIntegrationPackagedStandaloneNum}{37}
\newcommand{\codTechnologyWorkflowIntegrationPackagedStandalonePct}{25\%}
\newcommand{\codTechnologyWorkflowIntegrationToolIntegratedNum}{25}
\newcommand{\codTechnologyWorkflowIntegrationToolIntegratedPct}{17\%}
\newcommand{\codTechnologyWorkflowIntegrationOtherNum}{0}
\newcommand{\codTechnologyWorkflowIntegrationOtherPct}{0\%}
\newcommand{\dimEcosystemSocioculturalFactorsNum}{91}
\newcommand{\dimEcosystemSocioculturalFactorsPct}{49\%}
\newcommand{\codEcosystemSocioculturalFactorsMusicalGenreNum}{44}
\newcommand{\codEcosystemSocioculturalFactorsMusicalGenrePct}{48\%}
\newcommand{\codEcosystemSocioculturalFactorsMusicalPracticeNum}{57}
\newcommand{\codEcosystemSocioculturalFactorsMusicalPracticePct}{63\%}
\newcommand{\codEcosystemSocioculturalFactorsAIPerceptionNum}{19}
\newcommand{\codEcosystemSocioculturalFactorsAIPerceptionPct}{21\%}
\newcommand{\codEcosystemSocioculturalFactorsCulturalConservatismNum}{2}
\newcommand{\codEcosystemSocioculturalFactorsCulturalConservatismPct}{2\%}
\newcommand{\codEcosystemSocioculturalFactorsOtherNum}{1}
\newcommand{\codEcosystemSocioculturalFactorsOtherPct}{1\%}
\newcommand{\dimEcosystemPolicyFactorsNum}{14}
\newcommand{\dimEcosystemPolicyFactorsPct}{8\%}
\newcommand{\codEcosystemPolicyFactorsCopyrightConcernsNum}{9}
\newcommand{\codEcosystemPolicyFactorsCopyrightConcernsPct}{64\%}
\newcommand{\codEcosystemPolicyFactorsPlagiarismNum}{2}
\newcommand{\codEcosystemPolicyFactorsPlagiarismPct}{14\%}
\newcommand{\codEcosystemPolicyFactorsDataPrivacyNum}{5}
\newcommand{\codEcosystemPolicyFactorsDataPrivacyPct}{36\%}
\newcommand{\codEcosystemPolicyFactorsPersonalityRightsNum}{1}
\newcommand{\codEcosystemPolicyFactorsPersonalityRightsPct}{7\%}
\newcommand{\codEcosystemPolicyFactorsOtherNum}{0}
\newcommand{\codEcosystemPolicyFactorsOtherPct}{0\%}
\newcommand{\dimEcosystemEconomicConsequencesNum}{10}
\newcommand{\dimEcosystemEconomicConsequencesPct}{5\%}
\newcommand{\codEcosystemEconomicConsequencesJobReplacementNum}{5}
\newcommand{\codEcosystemEconomicConsequencesJobReplacementPct}{50\%}
\newcommand{\codEcosystemEconomicConsequencesDevaluingNum}{7}
\newcommand{\codEcosystemEconomicConsequencesDevaluingPct}{70\%}
\newcommand{\codEcosystemEconomicConsequencesOtherNum}{0}
\newcommand{\codEcosystemEconomicConsequencesOtherPct}{0\%}
\newcommand{\dimEcosystemMusicalSocietalConsequencesNum}{91}
\newcommand{\dimEcosystemMusicalSocietalConsequencesPct}{49\%}
\newcommand{\codEcosystemMusicalSocietalConsequencesReshapingIdiomsNum}{65}
\newcommand{\codEcosystemMusicalSocietalConsequencesReshapingIdiomsPct}{71\%}
\newcommand{\codEcosystemMusicalSocietalConsequencesCulturalExchangeNum}{4}
\newcommand{\codEcosystemMusicalSocietalConsequencesCulturalExchangePct}{4\%}
\newcommand{\codEcosystemMusicalSocietalConsequencesCrossDomainCollaborationNum}{12}
\newcommand{\codEcosystemMusicalSocietalConsequencesCrossDomainCollaborationPct}{13\%}
\newcommand{\codEcosystemMusicalSocietalConsequencesRevitalizationNum}{2}
\newcommand{\codEcosystemMusicalSocietalConsequencesRevitalizationPct}{2\%}
\newcommand{\codEcosystemMusicalSocietalConsequencesSkillDilutionNum}{4}
\newcommand{\codEcosystemMusicalSocietalConsequencesSkillDilutionPct}{4\%}
\newcommand{\codEcosystemMusicalSocietalConsequencesDemocratizationNum}{23}
\newcommand{\codEcosystemMusicalSocietalConsequencesDemocratizationPct}{25\%}
\newcommand{\codEcosystemMusicalSocietalConsequencesHomogenizationNum}{0}
\newcommand{\codEcosystemMusicalSocietalConsequencesHomogenizationPct}{0\%}
\newcommand{\codEcosystemMusicalSocietalConsequencesMisrepresentationNum}{2}
\newcommand{\codEcosystemMusicalSocietalConsequencesMisrepresentationPct}{2\%}
\newcommand{\codEcosystemMusicalSocietalConsequencesOverRelianceNum}{1}
\newcommand{\codEcosystemMusicalSocietalConsequencesOverReliancePct}{1\%}
\newcommand{\codEcosystemMusicalSocietalConsequencesOtherNum}{0}
\newcommand{\codEcosystemMusicalSocietalConsequencesOtherPct}{0\%}

%% file: sections/00_abstract.tex
\begin{abstract}
Live music provides a uniquely rich setting for studying 
creativity and interaction due to its spontaneous nature. 
The pursuit of \emph{live music agents}---intelligent systems supporting real-time music performance and interaction---has captivated researchers across HCI, AI, and computer music for decades, and recent advancements in AI suggest unprecedented opportunities to evolve their design. 
However, the interdisciplinary nature of music 
has led to fragmented development across 
research communities, hindering 
effective communication and collaborative progress. 
In this work, we bring together perspectives from these diverse fields to 
map the current landscape of live music agents. 
Based on our analysis of \numtotalsystem systems across both academic literature and video, we develop a comprehensive design space that categorizes dimensions spanning usage contexts, interactions, technologies, and ecosystems. 
By highlighting trends and gaps in live music agents, our design space offers researchers, designers, and musicians a structured lens to 
understand existing systems and shape future directions in real-time human-AI music co-creation. We release our annotated systems as a living artifact at \greenurl{https://live-music-agents.github.io}.
\end{abstract}

%% file: sections/01_introduction.tex
\section{Introduction}
\label{sec:intro}

\input{figures/visual_abstract}

% paragraph 1: importance of live music and current research landscape
Live music stands as one of humanity's oldest and most fundamental forms of creative expression and social connection, weaving together spontaneous composition, collaborative improvisation, and dynamic performance in ways that recorded media cannot replicate~\citep{swarbrick2019live,trehub2015cross}. 
From intimate jam sessions in basement studios to large-scale concerts, live music creates unique temporal experiences that emerge from the intersection of human creativity, technical skill, and social interaction. 
As digital technologies increasingly permeate musical practices, there has been growing interest in developing \emph{live music agents}: 
intelligent and interactive systems that listen and respond to a human musician in real time~\citep{Collins08a, oscar, collins2006towards}.
This emerging research landscape encompasses diverse approaches, 
from AI-driven accompaniment systems~\citep{dannenberg1984line, raphael2001bayesian, xia2017improvised, realchords} 
and intelligent instruments~\citep{aiterity, privato2024stacco} 
to collaborative composition tools~\citep{marley2015gestroviser, spiremuse}, 
all aimed at enhancing and extending the possibilities of 
real-time musical experience through live music agents.
\rev{
The capabilities of these systems are even more rapidly evolving as advances in AI music generation~\citep{huang2018music,dhariwal2020jukebox,musicfxdj, agostinelli2023musiclm, midillm, wu2025generative}
and multimodal understanding~\citep{ghosh2025musicflamingoscalingmusic, rubenstein2023audiopalm, radford2022whisper}
overcome longstanding technical barriers in live settings, including realistic music audio generation, diversity of inputs and outputs, and reaction latency~\citep{musicfxdj, zhou2024local, novack2025fast, wu2025streaming}. 
As these barriers fall, we approach the possibility of highly general agents that can perform alongside musicians across diverse styles, opening new frontiers for both musical practice and research~\citep{Blanchard2024Developing}.
}

% paragraph 2: fragmentation across HCI, AI, and computer music communities, and also, musicians in practice
\rev{
Despite this momentum, 
realizing the potential of live music agents
requires navigating significant challenges rooted in the field's fragmented structure. 
Live music agents sit at the intersection of 
Human-Computer Interaction (HCI),
Artificial Intelligence (AI), 
and Computer Music: 
fields that have developed largely in parallel, with limited integration of insights, methods, and tooling. 
Each community studies live music agents with distinct priorities: 
HCI focuses on user experience and interaction design~\citep{inasilentway, reflexivelooper, khallaghi2025squishysonics},
AI on algorithmic performance and efficiency~\citep{realchords, jiang2020when, rave},
and Computer Music on musical theory and audio processing techniques~\citep{turczan2019scale, Young07-2, privato2024stacco}.
Without holistic frameworks to bridge these perspectives, 
researchers and designers risk developing systems that excel in one dimension while neglecting essential considerations from others~\citep{leavitt1965applied, dsiiwa}.
The consequences of such fragmentation are evident in adjacent creative domains: isolated technical progress in visual arts, voice acting, and motion capture has generated economic disruption and ethical controversies among creative professionals~\citep{mocap, theft, labor_speech}. 
Music domain shows similar warning signs, with fewer than 10\% of generative audio papers discussing negative ethical impacts of proposed systems~\citep{barnett2023ethical} 
and AI music systems lacking interaction components despite musicians' needs of controllability, transparency, and personalization in co-creation process~\citep{huang2020aisongcontesthumanai}. 
These patterns highlight the need for holistic frameworks that integrate interdisciplinary knowledge and guide responsible development.
}

In this paper, we contribute a \textit{design space}
that brings together contributions from HCI, AI, and computer music in a structured framework. 
Rather than simply cataloging prior systems, 
our design space identifies the design decisions, recurring patterns, and points of tension that shape how live music agents are built and used. 
By providing a shared vocabulary and structured perspective, it enables researchers and practitioners across communities to imagine, compare, and explore new types of live music agents while supporting both responsible development and creative innovation.
Our design space centers 
on four key \emph{aspects} of live music systems: Usage Context, Interaction, Technology, and Ecosystem. 
Within each aspect, we identify \emph{dimensions} (i.e.,~fundamental components of an aspect) and \emph{codes} 
(i.e.,~design decisions for each dimension)~\citep{dsiiwa}
by systematically 
analyzing \numtotalsystem{} systems from HCI, AI, and Computer Music papers as well as online videos. 
%As a result, our design space contains 
In total, the design space comprises \numdim{} dimensions and \numcode{} codes, providing a structured view of the field that surfaces trends and future research opportunities.

Our main contributions are threefold.
First, we develop a design space for live music agents through systematic literature review and video analysis of systems from HCI, AI, and Computer Music communities.
Second, we demonstrate the design space's utility through multiple use cases, providing guidelines for generating new insights and supporting practitioners in ideating on or critiquing their designs.
Third, we document concrete insights and design alternatives derived from our design space.
In addition, we publicly release our annotated systems as a living artifact at \greenurl{https://live-music-agents.github.io} 
to promote exploration and community involvement in refining and extending the design space.

%% file: figures/visual_abstract.tex
\begin{figure*}[t!]
  \includegraphics[width=0.95\textwidth]{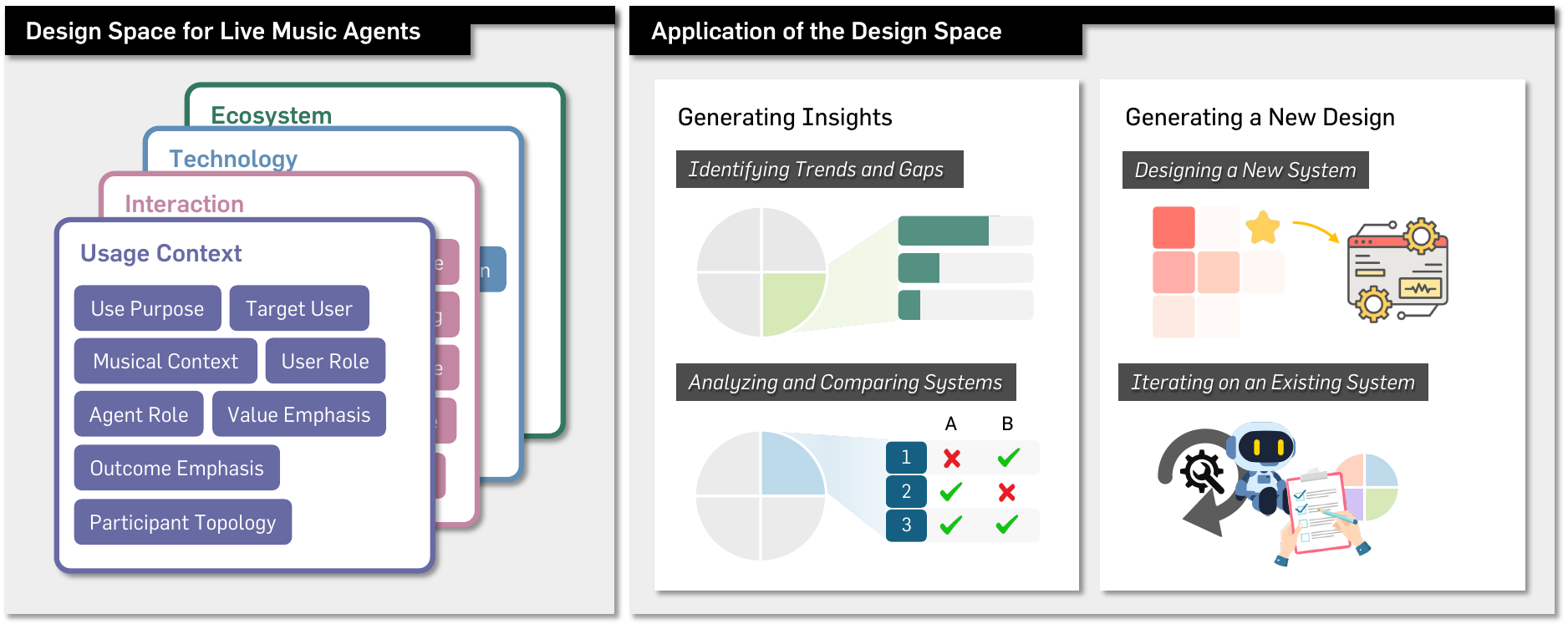}
  % \vspace{.3em}
  \caption{
  \rev{
  Visual abstract summarizing our work on constructing a Design Space for Live Music Agents.
Drawing on an analysis of 184 systems across HCI, AI, and computer music research, we introduce a design space organized around four aspects: \contextaspect{Usage Context}, \interactionaspect{Interaction}, \technologyaspect{Technology}, and \ecosystemaspect{Ecosystem}.
We also illustrate how this design space can be applied through concrete use cases, and we summarize key insights and design opportunities surfaced through our analysis.
  }
  }
  % \vspace{.5em}
  \Description{
  This figure shows the structure of our design space for live music agents and examples of how it can be applied. On the left, four stacked layers—Usage Context, Interaction, Technology, and Ecosystem—represent the aspects and their dimensions. On the right, two panels illustrate how the design space supports generating insights (such as identifying trends and comparing systems) and generating new designs (creating new systems or iterating on existing ones). The figure highlights both the components of the design space and its practical uses.
  }
  \label{fig:abstract}
\end{figure*}

%% file: sections/02_background.tex
\section{Background and Related Work}
\label{sec:background}

\subsection{Research Landscape of Live Music Agents}

Here we outline related work on live music agents, as distinct from research on real-time computer music synthesis (\textit{not agents}) and offline music AI generation (\textit{not live}). 
% paragraph 1: introduce foundational works 
Live music agents have evolved over 
%three decades \cd{closer to four (Voyager) or five (League of Automatic Music Composers) than three} 
at least four decades 
of interdisciplinary research, encompassing diverse computational approaches to real-time musical interaction. 
%\cd{suddenly recalling that the League of Automatic Music Composers (1977)~\citep{perkis2007league} may predate other systems by a solid decade: \url{https://perkis.com/wpc/leagueCDnotes.pdf}} 
Some of the earliest experiments include performances by the \emph{League of Automatic Music Composers}~\citep{perkis2007league} where rule-based algorithms were networked together to compose music in a live setting. 
% \cd{Need to mention score following here IMO, and clarify the distinction from systems like Voyager, i.e., that score following expands the potential of live music agents beyond systems that generate notes}
Early research centered around \emph{score following}~\citep{dannenberg1984line,vercoe1984synthetic}: computing systems capable of reconciling live human music performances against a known score, enabling use cases like automatic accompaniment. 
% The earliest explicit \cd{wdym by explicit? there's probably a good bit of subjectivity to what can be considered the earliest LMA.} live music agent, Oscar~\citep{oscar}, 
Shortly after, pioneering work such as George Lewis's \emph{Voyager}~\citep{voyager} and \textit{Oscar}~\citep{oscar} offered a vision of live music agents that could not just follow along with but \emph{contribute directly} to a musical performance alongside human musicians. 
This vision was broadened via new methods for machine listening, improvisation, and stylistic adaptation~\citep{bob,omax}. 
% Perhaps the earliest musical agent is the foundational work on
% established the foundational paradigm of a virtual musician designed to listen and react to human performers through real-time accompaniment. 
% This vision expanded through notable systems like Voyager~\citep{voyager}, Band-out-of-the-Box~\citep{bob}, and Omax~\citep{omax}, each contributing distinct approaches to listening, improvisation, and stylistic adaptation as autonomous performance partners. 
In parallel, Collins~\citep{Collins08a} %formally codified the field by defining 
offered a definition for 
live music agents as: 
\begin{quote}
    ``\textit{\ldots autonomous agents for interactive music, which can at a minimum operate independently of composer intervention during performance \ldots}''
\end{quote} 
Building on these foundations, today’s landscape encompasses a rich taxonomy of live music agents including intelligent score followers~\citep{jiang2025improvised}, gesture-to-sound mappers~\citep{borovik2023realtime, dooley2021mytrombone}, adaptive accompanists~\citep{rlduet, realchords}, intelligent instruments~\citep{aiterity, khallaghi2025squishysonics}, synthesizers~\citep{rave, shier2024real}, and collaborative improvisational partners~\citep{jambot, wang2025ai}, each addressing different facets of the 
%complex challenge of creating real-time musical systems.
associated technical and creative challenges.

This increasingly broad space 
now intersects with a transformative moment in computing history, where rapid advances in 
AI create \textbf{unprecedented opportunities for sophisticated live music agents}. 
Prior systems are constrained in their capabilities and generality, often limited to specific use cases or musical contexts, e.g.,~generating MIDI improvisations for a fixed chord progression~\citep{bob}.  
Now, 
radical improvements in offline music AI generation capabilities~\citep{huang2018music,dhariwal2020jukebox,agostinelli2023musiclm,evans2024fast} 
and multimodal understanding~\citep{gardner2021mt3,radford2021learning,alayrac2022flamingo,radford2022whisper,rubenstein2023audiopalm} 
are nearing a confluence with the technical constraints of the live music setting~\citep{Blanchard2024Developing,jambot,zhou2024local,novack2025fast,musicfxdj}. 
This suggests the potential for extremely general live music agents; for example, systems that can perform alongside (or in lieu of) musicians in broad range of music styles with human-like musical consistency, diverse inputs and outputs, and superhuman reaction times. 
Beyond live performance, such agents may assume increasingly broad roles, 
such as acting as a conversational studio collaborator or music teacher. 
New roles may involve higher levels of initiative and creative agency than the roles of human participants, e.g.,~composing sheet music for a human ensemble to perform in real time. 
Collectively, these expanded capabilities point to a fundamental shift from agents as reactive accompanists to proactive creative partners and first class participants in musical creativity.

While we may be on the cusp of the next era of live music agents, 
the field's interdisciplinary nature---spanning computer music, HCI, and AI---has resulted in fragmented knowledge communities with limited integration of 
insights, methods, and tooling.
As technological barriers dissolve and interest intensifies, 
the absence of a unified theoretical framework and shared understanding of design principles 
becomes a critical bottleneck at odds with the anticipated impact of modern AI methods to the development of live music agents. 
Motivated by this need, 
we aim to develop a comprehensive design space that integrates existing cross-disciplinary works.

%%%
\subsection{Design Spaces in HCI}

% Paragraph 1: Definition and purpose
Design spaces are taxonomies that organize and characterize the design choices and properties of artifacts created for similar purposes~\citep{maclean1991questions, 10.1145/97243.97263, card1991morphological, romer2004thedesignspace}.
Design spaces serve three key functionalities: (i) establishing a \textit{shared vocabulary} among interdisciplinary stakeholders, (ii) synthesizing \textit{existing} systems to 
examine past 
design decisions, and (iii) revealing gaps and opportunities for \textit{future} exploration~\citep{dsiiwa}.
Design spaces thus serve both retrospective and prospective roles: as analytical tools for understanding what has been created and as generative frameworks for envisioning what could be created within their domains~\citep{MacLean01091991}---the latter may be of particular relevance at this critical moment in the evolution of live music agents.

% paragraph 2: introduce existing design spaces and how they are created
Building on this foundation, HCI researchers have developed design spaces across diverse domains. 
\rev{
One of the primary methods for constructing design spaces is through systematic literature reviews, which help effectively map the breadth of academic work within a subfield~\citep{dsiiwa, dsida, chung2021creativity, frich2019mapping, gero-etal-2022-design, 10.1145/3491102.3501939, 10.1145/3290605.3300792, 10.1145/3449194}.
For instance, Frich et al.~\citep{frich2019mapping} systematically reviewed 143 papers from the ACM Digital Library to create a design space of creativity support tools, 
while Lee et al.~\citep{dsiiwa} analyzed 115 papers from natural language processing and human-computer interaction venues 
to construct a design space for writing assistants.
By extracting dimensions and organizing existing systems within this framework, design spaces constructed through systematic literature review move beyond descriptive cataloging to provide analytical tools for understanding design trade-offs and identifying unexplored opportunities.
}
Complementing these academic reviews, another thread of work captures design practices beyond research, drawing from practitioner-oriented sources such as commercial software, blogs, product websites, and videos~\citep{10.1145/3532106.3533489, Lau2020TheDS, 5613452, 10.1145/3411764.3445511, 10.1007/s00779-009-0244-5}. For example, Zhao et al.~\citep{10.1145/3532106.3533489} analyzed 40 streamer-produced videos to map livestreaming equipment setups, while Vogel and Balakrishnan~\citep{10.1007/s00779-009-0244-5} combined academic papers and commercial products to characterize interactive display technologies. These approaches demonstrate how design spaces can capture both scholarly and real-world practices, offering a comprehensive view of existing systems. Following this practice, we combine literature review with video analysis to characterize the state of practice in our domain.

% paragraph 3: design space in live music agent. 
Within the music AI domain, 
existing taxonomies 
have primarily focused on generation capabilities and creative outcomes, 
rather than the interactive dynamics that characterize live musical collaboration. 
Although comprehensive surveys have mapped the technical landscape of music foundation models~\citep{ma2024foundationmodelsmusicsurvey}
and broader taxonomies have addressed human-AI co-creation across diverse musical contexts~\citep{musictaxonomy, Tatar01012019}, 
these frameworks reflect predominant paradigms of \textit{offline} content generation. 
Additional work has explored AI's role in artistic contexts~\citep{huang2020aisongcontesthumanai}, including performance environments~\citep{pons2025musicartificialintelligenceartistic}. 
These scholarly efforts provide essential groundwork while leaving room for more granular investigation of specific interaction paradigms. 
Closer to our scope, Gifford et al.~\citep{Gifford02012018} mapped the design of fundamental computational improvisation systems. Similarly, Jung~\citep{doi:10.18261/smn.49.1.3} proposed a taxonomy of intelligent performance systems, identifying three design aspects---embodiment, participation, and autonomy---that influence the dynamics of human-AI interaction. 
While these works mark important milestones, their focus on particular exemplars and dimensions means they capture only fragments of a rapidly diversifying landscape of live music agents.
Our work complements this depth with breadth. 
By systematically analyzing \numtotalsystem{} systems drawn from both research and practice, 
we extend current taxonomic understanding into a comprehensive design space. 
This framework integrates perspectives across HCI, AI, and computer music, while also accounting for practitioner practices, offering a foundation for interdisciplinary dialogue and future innovation in live music agents.

%% file: sections/03_definition.tex
\newpage
\section{Scope of Live Music Agents}
% here to define live music agents 
\label{sec:scope}

We define the scope of our work by specifying what we consider as \textit{live music agents} through three key dimensions. 
% This scoping framework serves to establish clear boundaries for our study and ensures alignment with our study objectives, while acknowledging that alternative conceptualizations may exist within the broader literature. 

\begin{itemize}
    \item \textbf{Live: } We consider a system to be ``live'' if it operates in \textit{real time} with human or non-human collaborators. This real-time constraint requires the system to process inputs, make decisions, and generate outputs %within the temporal constraints of real-time musical interaction, typically allowing for 
    with minimal latency that does not disrupt the natural flow of live music creation.

    \item \textbf{Music}: We define a system to be related to ``music'' if it \textit{responds to} and/or \textit{generates} music through symbolic (e.g., MIDI) or audio representations, with the explicit goal of contributing to or shaping a \textit{shared musical outcome}.

    \item \textbf{Agents}: We define ``agents'' as \textit{intelligent} and \textit{interactive} systems capable of autonomous decision-making and/or music generation, which can interact with human users and/or other agents during the creative process. 
    
\end{itemize}

Amid competing definitions of ``agents'' in literature~\citep{aiterity, Tatar01012019, Wooldridge_Jennings_1995, 10.1007/BFb0013570, bent2025termagentdilutedutility, weiss1999multiagent, russell1995modern}, 
a widely used definition~\citep{Wooldridge_Jennings_1995} 
%of agent 
lists
autonomy, social ability, reactivity, and proactivity as core characteristics of agents. 
We adopt this definition and restrict 
%attention % CHRIS: too overloaded w/ attention in transformers
our focus
to interactive systems that are \emph{non-deterministic} and/or \emph{learning-based}.
This choice aligns even with the recent resurgence of the word ``agent'', in reference to AI systems powered by large language models that stress adaptation and initiative in complex environments~\citep{guo2024large}. 

%% file: sections/04_methodology.tex
\section{Methodology}
\label{sec:method}

We construct our design space 
combining a systematic literature review with an analysis of practice in online videos.
We first describe our sampling of papers and videos (Section~\ref{sec:method:sampling}) and then our procedure for creating the design space (Section~\ref{sec:method:analysis}).

\subsection{Sampling}
\label{sec:method:sampling}

We review \numtotalsystem systems from two sources to create a comprehensive picture of practice: (i) papers published across HCI, AI, and Computer Music venues ($N=\numsystem$), and (ii) online videos ($N=\numvideo$).
We include both sources because musical activity today is widely shared through audiovisual media~\citep{cross_2023, cayari2011youtube} and creative-practice trends complement insights from research~\citep{pons2025musicartificialintelligenceartistic, huang2020aisongcontesthumanai}.

For literature, we use a two-step strategy: (i) curate a foundational set of key papers and (ii) systematically retrieve papers from peer-reviewed conferences, following mixed-method approaches in design space research~\citep{chung2021creativity, frich2019mapping}.
For videos, we sample demonstrations, performances, and personal recordings featuring human–AI live music systems.
Following prior video-based design space work~\citep{xambo2023video, heath2010video}, we use YouTube as the primary source due to its central role in music sharing~\citep{cayari2011youtube}.
Figure~\ref{fig:method} illustrates the pipeline, and Appendix~\ref{appendix:data:resources} lists all resources.

\paragraph{Key Paper Curation}
We first curated a foundational set of papers closely aligned with our definition of live music agents.
This process established a shared understanding among authors and captured contributions across diverse disciplinary traditions.
Four researchers in computer music, HCI, and AI worked in two institutional pairs to independently curate candidate papers.
The lists were merged through two collaborative sessions to mitigate community bias.
The final set included \numkeypaper key papers.

\input{figures/method_visualization}

\paragraph{Systematic Paper Retrieval and Filtering}
We identified target peer-reviewed conferences across HCI, AI, and Computer Music (Appendix~\ref{appendix:data:venues}).
We retrieved \numretrievedpaper candidate papers (as of June 2025) using Boolean queries combining three concept clusters: \textit{live} (\eg `real-time,' `jamming'), \textit{music}, and \textit{agents} (\eg `AI,' `autonomous'); full queries appear in Appendix~\ref{appendix:data:queries:papers}.
Six researchers filtered papers by project scope (Section~\ref{sec:scope}) and excluded works already in the key paper list. 
When papers appeared across venues, we grouped them for joint analysis, as different versions often emphasized distinct aspects. 
This process yielded \numfilteredpaper filtered papers; with the key papers ($N=\numkeysystem$ systems), the final dataset comprised \numsystem systems.
Appendix~\ref{appendix:data:resources:crawledpapers} provides the full list of retrieved papers, including grouped duplicates.

\input{figures/design_space_overview}

\paragraph{Video Sampling}
In July 2025, we executed keyword searches (e.g., `AI jam session,' `improvisation with AI'; full list in Appendix~\ref{appendix:data:queries:videos}) on YouTube and reviewed the first 20 results per query.
We included videos depicting or describing a music-making scenario with an explicit live music agent.
We excluded advertisements and reviews, favoring demonstrations, performances, and personal recordings as more naturalistic data.
For each included video, we analyzed transcripts, descriptions, on-screen interactions, and linked materials.
This process yielded \numvideo systems from videos.

\subsection{Analysis and design space creation}
\label{sec:method:analysis}

Six authors conducted an iterative, inductive analysis of the combined corpus.
Guided by prior sociotechnical framings of AI-assisted systems~\citep{dsiiwa}, we began with five aspects: Task, User, Interaction, Technology, and Ecosystem.
Authors independently open coded key papers and then collaboratively performed axial coding to develop an initial design space.
During this process, we merged \emph{Task} and \emph{User} aspects into a single \emph{Usage Context} aspect, as many codes spanned both.
From the resulting codes, we derived higher-level themes that became the initial dimensions of the design space.

We then divided the full set of papers and videos among the authors and conducted coding using the evolving aspects and dimensions.
We applied constant comparative analysis: 
when new items did not fit existing codes, we created new ones and iteratively refined, merged, or reorganized dimensions and codes through weekly discussions. 
Once the design space was finalized, the authors re-annotated the entire corpus.
\rev{
If a paper did not address a dimension or aspect, it was marked as irrelevant.\footnote{
\rev{
For example, despite RAVE's~\citep{rave} widespread adoption among musicians, the original paper is a primarily technical contribution that does not engage with the \contextaspect{Usage Context} or \ecosystemaspect{Ecosystem} aspects (see Section~\ref{sec:designspace} for definitions). Therefore, all dimensions within these two aspects were marked as N/A for this paper.
}}
Codes were generally mutually exclusive within a category, but we allowed multiple codes per dimension to reflect the multifunctionality of live agents.
To ensure reliability, following prior work~\citep{dsida}, about 10\% ($N=19$ among 184 papers) of the corpus was double-coded, yielding an inter-annotator agreement of \agreementrate\%.}

% What exactly was coded and how? words? passages?
% What was the process of open codes getting into groups? collaboratively I assume but still. I buy that you doublecoded 10% of the papers but how about the other papers? were they just solo coded and did you do anything to combat any sort of drift there? If so what?

% I would also, now that the veil of anonymity can be lifted, encourage the authors to offer a positionality statement of sorts to make it even clearer how their competence and positions plays into the coding.

%% file: figures/method_visualization.tex
\begin{figure}[t!]
  \includegraphics[width=.95\linewidth]{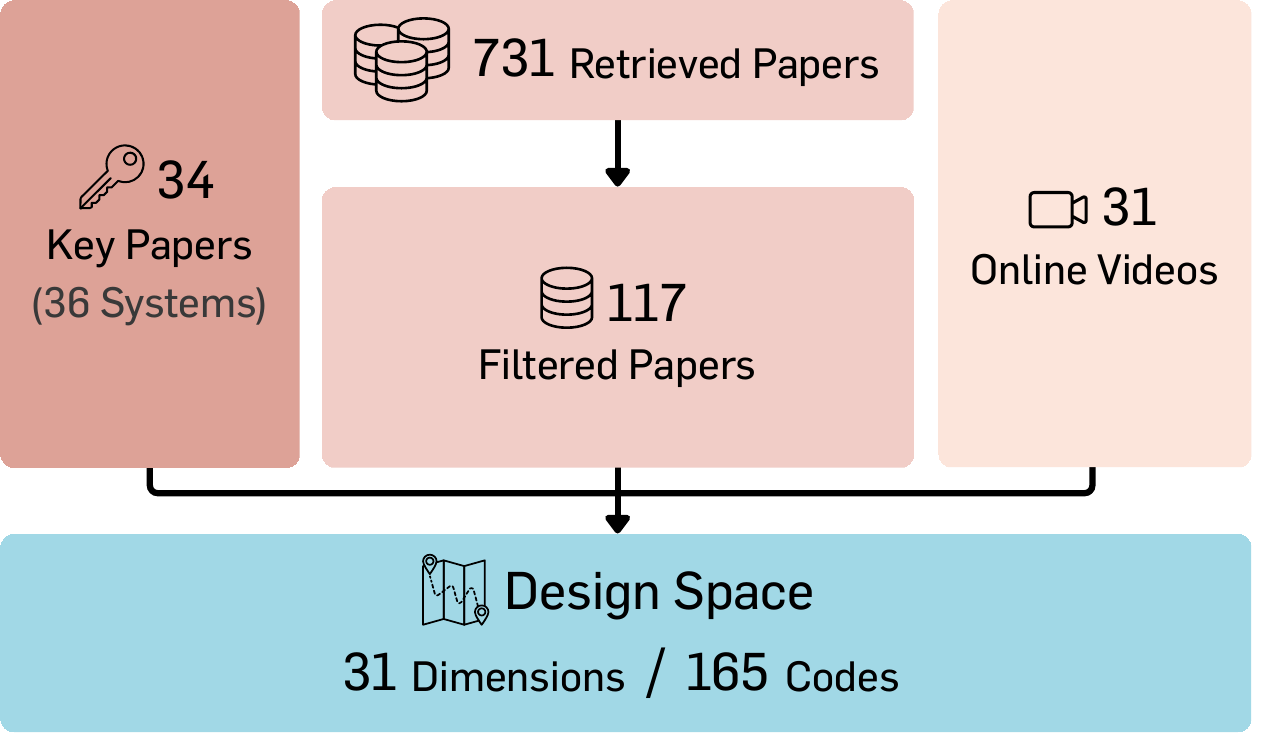}
  % \vspace{1em}
  \caption{We construct a comprehensive design space for live music agents by analyzing both academic literature and online videos. Our dataset includes 34 key papers (36 systems) that closely align with the scope of live music agents, 117 filtered papers retrieved from peer-reviewed HCI, AI, and computer music venues, and 31 online videos of live performances, demonstrations, or personal recordings. In total, we analyze 184 systems to derive a design space comprising 31 dimensions and 165 codes.
  % CHRIS: feels like this is missing the steps beyond the design space. can we dazzle by highlighting both actual (in this paper) and potential (in other papers) downstream outcomes? e.g. (1, actual) uncover research opportunities by identifying sparse regions of design space, (2, potential) provide common vocabulary among interdisciplinary stakeholders, (3, potential) inform the development of new LMAs. there might be more, these are just front of mind for me
  }
  \vspace{-.5em}
  \Description{Figure shows a flow diagram of sources used to build the design space. On the left, 34 key papers covering 36 systems. In the middle, 731 retrieved papers narrowed to 117 filtered papers. On the right, 31 online videos. All sources contribute to the design space at the bottom, consisting of 31 dimensions and 165 codes.}
  \label{fig:method}
\end{figure}

%% file: figures/design_space_overview.tex
\begin{figure*}[t!]
  \includegraphics[width=\textwidth]{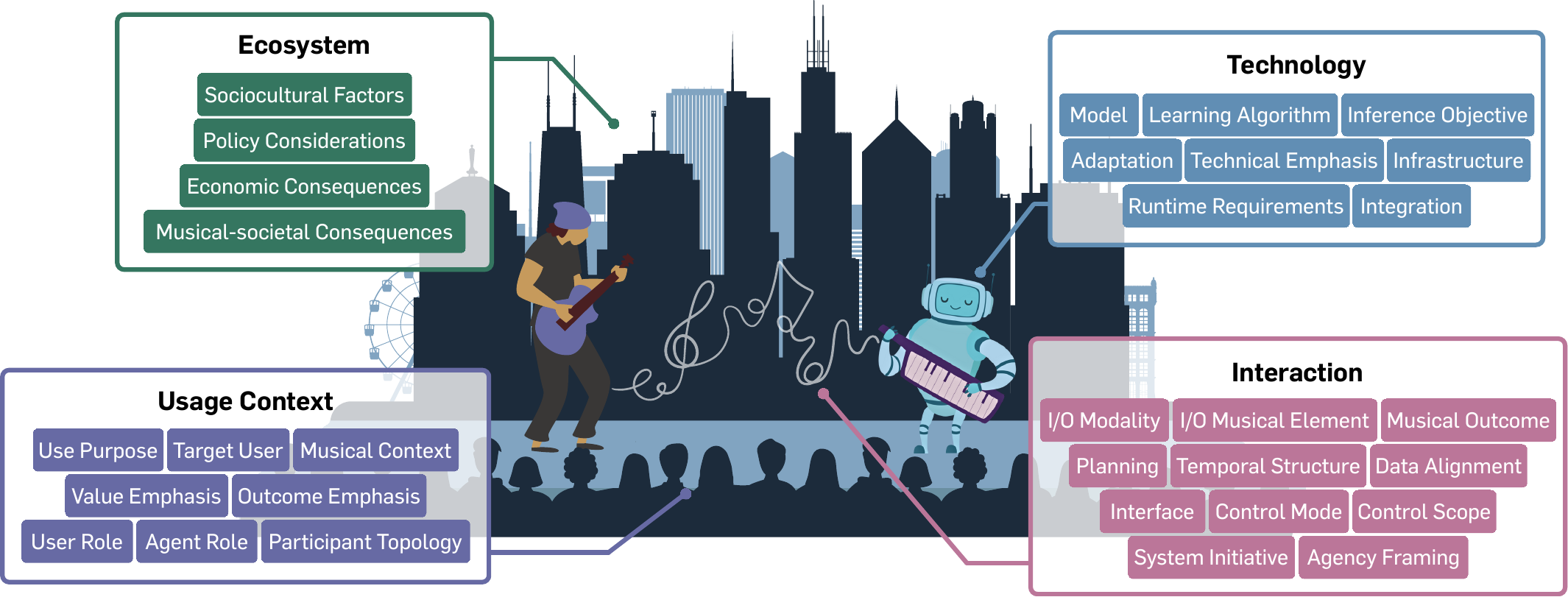}
  % \vspace{.1em}
  \caption{Overview of our proposed design space for \emph{live music agents}, 
  % CHRIS: I think it's important to have a summary definition of LMAs here, as readers may glance at figure 1 first
  intelligent and interactive music systems that execute musical tasks in real time. 
  %The framework 
  Our design space 
  is organized into four aspects---\contextaspect{Usage Context}, \interactionaspect{Interaction}, \technologyaspect{Technology}, and \ecosystemaspect{Ecosystem}---each composed of dimensions (shown in blocks) that capture fundamental components of system design.
  }
  % \vspace{-.5em}
  \Description{An illustration shows a human guitarist and a robot keyboardist performing on stage in front of a city skyline and an audience. Around the performers, four colored boxes represent the design space aspects for live music agents. The bottom-left box (Usage Context) lists Use Purpose, Target User, Musical Context, Value Emphasis, Outcome Emphasis, User Role, Agent Role, and Participant Topology. The bottom-right box (Interaction) lists I/O Modality, I/O Musical Element, Musical Outcome, Planning, Temporal Structure, Data Alignment, Interface, Control Mode, Control Scope, System Initiative, and Agency Framing.The top-right box (Technology) lists Model, Learning Algorithm, Inference Objective, Adaptation, Technical Emphasis, Infrastructure, Runtime Requirements, and Integration. The top-left box (Ecosystem) lists Sociocultural Factors, Policy Considerations, Economic Consequences, and Musical-societal Consequences. Connecting lines link these categories to the performers on stage, illustrating how the design space dimensions relate to human–AI musical interaction.}
  \label{fig:main}
\end{figure*}

%% file: sections/05_00_designspace.tex
\section{The Design Space}
\label{sec:designspace}

% figure

Our analysis yields four high-level aspects of live music agents:

\begin{itemize}
    \item \contextaspect{Usage Context}: Musical purposes and settings that motivate using live music agents, including who is involved and the roles assumed by users and 
    %systems 
    agents.

    \item \interactionaspect{Interaction}: Functional specifications describing how live music agents work from the user's perspective, including interfaces and control mechanisms.
    
    \item \technologyaspect{Technology}: Technical specifications detailing how live music agents are implemented and enabled, including underlying models and technology infrastructures.

    \item \ecosystemaspect{Ecosystem}: The broader environment surrounding live music agents, including social, cultural, economic, and policy factors that influence their adoption and evolution, as well as the impacts these systems produce.
\end{itemize}

\rev{
In this section, we present a design space as a structured framework for analyzing the multidimensional landscape of live music agents (Figure~\ref{fig:main}).
For each aspect, we describe its dimensions (\textit{fundamental components of an aspect}) and codes (\textit{possible options for each dimension}), together with their proportions and representative examples. 
Percentages are calculated relative to the total number of coded systems ($N=\numtotalsystem$). 
Paper counts per aspect and dimension exclude those marked as N/A. 
Code percentages within a dimension may sum to $\geq 100$\%, as papers can be assigned multiple codes per dimension. 
We additionally provide case studies of representative live music agents to support the interpretation of the design space and illustrate its practical use.\footnote{The papers used for the case studies were selected through discussions among the authors, with consensus on works that best represent each aspect.}
A summarized version of the full design space is included in Appendix~\ref{appendix:tables} (Tables~\ref{tab:task-designspace-1}–\ref{tab:ecosystem-designspace}).
}

%%% Context
\subsection{\contextaspect{Usage Context}}
\label{sec:designspace:context}

\input{sections/05_01_designspace_context}

%%% Interaction
\subsection{\interactionaspect{Interaction}}
\label{sec:designspace:interaction}

\input{sections/05_02_designspace_interaction}

%%% Technology
\subsection{\technologyaspect{Technology}}
\label{sec:designspace:technology}

\input{sections/05_03_designspace_technology}

%%% Environment
\subsection{\ecosystemaspect{Ecosystem}}
\label{sec:designspace:environment}
\input{sections/05_04_designspace_environment}

%% file: sections/05_01_designspace_context.tex
This aspect defines the situational framework of human-AI musical collaboration, including musical purposes and settings. 
We outline the dimensions and codes that characterize usage context, 
with a full summary provided in Tables~\ref{tab:task-designspace-1}–\ref{tab:task-designspace-2} in Appendix~\ref{appendix:tables}.

\subsubsection{Dimensions and Codes}

\paragraph{\contextdim{Use Purpose}}
A design of live music agents can be motivated by a goal of the musical activity.
A prominent goal is supporting \contextcode{live performance} (\codUseContextUsePurposeLivePerformancePct), often with an emphasis on responsiveness and avoiding errors~\citep{marchini2017rethinking, jambot, tsiros2020towards}. 
Another common goal is \contextcode{composition} (\codUseContextUsePurposeCompositionPct), where systems help users explore ideas and expand sketches into larger structures~\citep{spiremuse, gioti2019imitation}. 
Some systems emphasize \contextcode{recreation} (\codUseContextUsePurposeEntertainmentPct), enabling playful music-making via games~\citep{harmonix} or simplified interfaces~\citep{pianogenie, marley2015gestroviser}.
A few systems are designed for \contextcode{skill acquisition} (\codUseContextUsePurposeSkillAcquisitionPct), offering real-time feedback on performance~\citep{almeida2019amigo} or play with users to build improvisation skills~\citep{kondak2016active, bachduet}. 
 Finally, \contextcode{art installations} (\codUseContextUsePurposeArtInstallationPct) place agents in spatial sound environments where visitor actions (\eg movement) are turned into music~\citep{nash2020crowd, Schedel2021RhumbLine}.

% One prominent goal is \contextcode{live performance} (\codUseContextUsePurposeLivePerformancePct), where the primary aim is to perform in front of an audience~\citep{marchini2017rethinking, evans2025repurposing}, often emphasizing responsiveness and error avoidance~\citep{jambot, tsiros2020towards}. Another common purpose is \contextcode{composition} (\codUseContextUsePurposeCompositionPct), where musicians interact with systems in jam sessions to explore ideas and expand them into larger formal structures~\citep{spiremuse, gioti2019imitation}.
% Less attention is given to \contextcode{skill acquisition}, where users engage in real time with systems to receive feedback~\citep{almeida2019amigo} or to practice, such as improving improvisation skills~\citep{kondak2016active, bachduet}. \contextcode{Recreation} systems focus on playful music-making, whether through games~\citep{harmonix} or simplified interactions~\citep{pianogenie, marley2015gestroviser}. 
% Finally, \contextcode{art installations} embed agents into spatial sound environments, translating visitors’ collective actions into musical experiences~\citep{nash2020crowd, Schedel2021RhumbLine}.

\paragraph{\contextdim{Target User}}
%revised
This dimension describes the intended users of a system, whose characteristics shape its design. 
Most systems target \contextcode{musicians} (\codUseContextTargetUserMusiciansPct), assuming sufficient musical background to create music or play instruments~\citep{aidj, kitani2010improvgenerator, jambot}. 
Others focus on \contextcode{novice} users (\codUseContextTargetUserNoviceUsersPct) by providing scaffolds that let users with little musical background create music through simplified controls, 
such as triggering short piano phrases with a single key~\citep{tsuruoka2019soundwriter} or mapping an eight-button controller to a full keyboard~\citep{pianogenie}. 
A smaller group involves the \contextcode{audience} (\codUseContextTargetUserAudiencePct), translating collective actions (\eg, chats in streaming services) into sound~\citep{fruitgenie, dahl2011tweetdreams, commentsai} or giving audience members direct control of the system~\citep{turczan2019scale}.

\paragraph{\contextdim{Musical Context}}
A specific musical style or practices can shape the design of the systems.
For example, \contextcode{jazz} systems (\codUseContextMusicalContextJazzImprovisationPct) often feature bar trading practices~\citep{kondak2016active, bob} or shared lead sheets~\citep{reflexivelooper}.
In electronic music (\codUseContextMusicalContextElectronicMusicPct), electroacoustic practices inspire augmentation of acoustic instruments with digital sound~\citep{Lepri16, van2012mapping, brown2010network, sophtar}, while others follow DJing practices~\citep{vogl2017an, davies2014improvasher, dadabots, aidj}.
Systems for \contextcode{new music} (\codUseContextMusicalContextNewMusicPct) center on novel sound design~\citep{aiterity, fiebrink2020reflections, privato2024stacco} and experimental approaches such as laptop orchestras~\citep{proctor2020a}.
Unlike improvisation-focused systems, \contextcode{repertoire}-based systems (\codUseContextMusicalContextScoreBasedPct) aid performance of pre-composed works, often through score following~\citep{raphael2001bayesian, dannenberg1984line, TerasakiTH17}.
\contextcode{Western classical} systems (\codUseContextMusicalContextWesternClassicalPct) similarly assume fixed scores~\citep{EvansFGMO14} and specific techniques such as counterpoint~\citep{bachduet}. 
\contextcode{Popular music} systems (\codUseContextMusicalContextPopularMusicPct) draw on forms such as pop~\citep{marchini2017rethinking, knotts2021algorithmic} and rock~\citep{fruitgenie}, while \contextcode{traditional music} systems (\codUseContextMusicalContextTraditionalMusicPct) reference folk practices~\citep{jiang2020when, xia2017improvised} and regional instruments~\citep{wu2025gesturedriven, 10.1145/1254960.1254990}.
There are emerging practices as well. \contextcode{Live coding} (\codUseContextMusicalContextLiveCodingPct)
situates performance in programming environments such as Strudel~\citep{strudel2022}, with systems supporting coding activity~\citep{Xambo2021Live, livecoding}.
A few recent systems explore \contextcode{virtuosic practice} (\codUseContextMusicalContextVirtuosicPracticePct), highlighting free improvisation with highly skilled performers~\citep{jambot}.
Lastly, there are general, \contextcode{non-specific} systems (\codUseContextMusicalContextNonSpecificPct) that do not assume a particular musical context~\citep{musicfxdj, BakhtB09, borovik2023realtime}.\footnote{Although some contexts are implied, we only code those explicitly stated.}

\paragraph{\contextdim{Value Emphasis}}
% \rev{
This dimension captures values prioritized in design. 
\contextcode{Control} (\codUseContextUserPreferenceControlPct) lets users steer agent output through interfaces such as knobs~\citep{naess2019physical} or gestures~\citep{dooley2021mytrombone, juregui2019automatic}. 
\contextcode{Diversity} (\codUseContextUserPreferenceDiversityPct) promotes varied output to support creativity~\citep{shepardson2025evolving, sioros2011automatic, Fay15-1}, while \contextcode{coherence} (\codUseContextUserPreferenceCoherencePct) ensures musical consistency and quality~\citep{realchords, nime2025_54}. 
\contextcode{Novelty} (\codUseContextUserPreferenceNoveltyPct) favors surprising or serendipitous results~\citep{kobayashi2020exsampling, van2012mapping}, sometimes embracing unpredictability~\citep{privato2024stacco, Gresham-Lancaster15, marley2015gestroviser}. 
\contextcode{Personalization} (\codUseContextUserPreferencePersonalizationPct) tailors system behavior to individual styles, often through fine-tuning on user data~\citep{jambot, fiebrink2010wekinator}.
% }

\paragraph{\contextdim{Outcome Emphasis}}
This dimension describes the intended impacts of human–agent collaboration. 
Most systems highlight \contextcode{exploration} (\codUseContextExpectedImpactExplorationPct) which encourages users to pursue new musical ideas or directions~\citep{kitani2010improvgenerator, khallaghi2025squishysonics}. 
\contextcode{Empowerment} (\codUseContextExpectedImpactEmpowermentPct) emphasizes extending users’ musical capabilities or expanding the roles of instruments~\citep{pianogenie, stefani2024esteso, memachine}. 
\contextcode{Engagement} (\codUseContextExpectedImpactEngagementPct) values strengthening users’ sense of involvement in performance~\citep{donnarumma2012music, borovik2023realtime}, while \contextcode{expression} (\codUseContextExpectedImpactExpressionPct) supports freedom in conveying emotions, ideas, and identities~\citep{godbehere2008wearable, lucas2023a}. 
\contextcode{Immersion} (\codUseContextExpectedImpactImmersionPct) seeks to facilitate flow states during interaction~\citep{Kalonaris16, filandrianos2020brainwaves, spiremuse}. 
Fewer systems prioritize \contextcode{reflection} (\codUseContextExpectedImpactReflectionPct), prompting critical examination of one’s musical practice~\citep{jambot}, or \contextcode{delegation} (\codUseContextExpectedImpactDelegationPct), which reduces cognitive load so that users can focus on core musical work~\citep{johnson2023musical, byuksel2015braahms}.

\paragraph{\contextdim{User Role}}
This dimension describes the roles users take in musical interaction. 
As \contextcode{lead} (\codUseContextUserRoleLeadPct), users create and guide primary musical material, often through instrumental improvisation~\citep{reflexivelooper, rlduet}. 
As \contextcode{manipulators} (\codUseContextUserRoleManipulatorPct), they shape music at a higher level by adjusting parameters or model controls~\citep{mcauliffe2023stochgran, somax2, bing2017a, naess2019physical}. 
\contextcode{Non-music performers} (\codUseContextUserRoleNonMusicalPerformerPct) contribute through other modalities such as dance~\citep{calmuswaves} or writing~\citep{dahl2011tweetdreams, tsuruoka2019soundwriter}. 
Users sometimes act as \contextcode{conductors} (\codUseContextUserRoleConductorPct), coordinating participants or directing structure~\citep{mimi4x, proctor2020a}. 
The \contextcode{accompanist} (\codUseContextUserRoleAccompanistPct) role appears rarely and typically alternates with lead, with no cases of users serving solely as accompanists~\citep{voyager, somax2, jambot}. 
Other roles include \contextcode{coders} (\codUseContextUserRoleCoderPct), who generate music through live coding or prompting~\citep{johnson2023musical, musicfxdj}, and \contextcode{mixers} (\codUseContextUserRoleMixerPct), who balance and combine musical elements~\citep{aidj, kobayashi2020exsampling, tsiros2020towards}.

\paragraph{\contextdim{Agent Role}}
This dimension describes the musical roles taken by agents. 
Their primary role is \contextcode{accompanist} (\codUseContextAgentRoleAccompanistPct), supporting participants through harmonization, rhythmic contribution, or backing material~\citep{reflexivelooper, realchords, trumpetai, evans2025repurposing}. 
Agents also \contextcode{lead} music making (\codUseContextAgentRoleLeadPct), for example by trading solos~\citep{kondak2016active, bob} or generating independent melodic lines~\citep{bachduet}. 
They rarely \contextcode{conduct} (\codUseContextAgentRoleConductorPct), directing structure or coordination~\citep{uncannylove, turczan2019scale}. 
Other roles include \contextcode{mapper} (\codUseContextAgentRoleMapperPct), translating non-musical actions into sound~\citep{shepardson2024tungnaa, easthope2023snakesynth, calmuswaves}; \contextcode{remixer} (\codUseContextAgentRoleRemixerPct), modifying existing material through parameter or playback control~\citep{mimi4x, LeffueK16, dannenberg1984line}; and \contextcode{controller} (\codUseContextAgentRoleControllerPct), where intentional signals such as button presses or knob turns shape output~\citep{pianogenie, martin2016intelligent}. 
Some agents act as \contextcode{interpreter} (\codUseContextAgentRoleInterpreterPct), translating music into other modalities like visualization~\citep{shepardson2025evolving, kobayashi2023improvise+}, while a few serve as \contextcode{evaluators} (\codUseContextAgentRoleEvaluatorPct), giving feedback on generated music~\citep{tsiros2020towards, jamai}.

% figure
\input{figures/design_space_usagecontext}

\paragraph{\contextdim{Participant Topology}}
This dimension describes the intended ratios between humans and agents. 
While most systems use a \contextcode{\texttt{1:1}} setup (\codUseContextCollaborationStructureOneOneCollaborationPct)~\citep{stefani2024esteso, jambot}, 
others adopt \contextcode{\texttt{N:1}} (\codUseContextCollaborationStructureNOneCollaborationPct) structures, where multiple humans interact with one agent, common in band contexts~\citep{shimon, fruitgenie}. 
Some support \contextcode{\texttt{N:N}} (\codUseContextCollaborationStructureNNCollaborationPct) interactions involving multiple humans and agents (e.g., several human performers each playing an AI-augmented instrument)~\citep{NikaDCVA17, brown2018interacting}, while fewer use \contextcode{\texttt{1:N}} (\codUseContextCollaborationStructureOneNCollaborationPct) structures with one human controlling multiple agents~\citep{lucas2023a, derbinsky2011cognitive}. 
A small number feature \contextcode{agent-only} settings (\codUseContextCollaborationStructureAgentOnlyPct), where agents perform autonomously and humans only configure the system before performance~\citep{genjam}.

\subsubsection{Case Study: \texttt{jam\_bot}~\citep{jambot}}
\rev{
The \texttt{jam\_bot} is a real-time performance system designed to support human–AI free improvisation on stage (Figure~\ref{fig:designspace:usagecontext}).
The system is built with a close collaboration with Jordan Rudess, a Grammy-winning professional keyboard \contextcode{musician}~(\contextdim{Target User}). Its primary goal is to facilitate \contextcode{live performance}~(\contextdim{Use Purpose}) in a \contextcode{\texttt{1:1}} collaboration setting~(\contextdim{Participant Topology}), where Rudess and the system jointly improvise in real time.
The system is situated within a \contextcode{virtuosic practice}~(\contextdim{Musical Context}), supporting performances without predetermined musical structure or score.
In performance, the musician and agent dynamically shift between \contextcode{lead} and \contextcode{accompanist} roles~(\contextdim{User/Agent Role}).
These shifts include trading complete melodic–harmonic phrases, letting Rudess lead with melody while \texttt{jam\_bot} harmonizes, or flipping the roles so the agent initiates melodic ideas.
% To support this, the designers implemented explicit interaction cues; for example, pressing a root note in the lowest register signaled the model to rephrase his recent melodic gestures.

The system’s design foregrounds several \contextdim{Value Emphasis} considerations.
A key requirement was \contextcode{control} over the performance: 
Rudess needed reliable ways to shape the agent’s generative behavior and coordinate role transitions.
To support this, the designers implemented explicit interaction cues; for example, pressing a root note in the lowest register signaled the model to rephrase his recent melodic gestures.
\contextcode{Personalization} was also central: Rudess wanted the \texttt{jam\_bot} to ``\textit{feel like a version of myself,}'' which the designers realized by fine-tuning the underlying model on MIDI recordings from his practice sessions.
Finally, 
the collaboration yielded meaningful effects (\contextdim{Outcome Emphasis}): 
interacting with \texttt{jam\_bot} 
fostered self \contextcode{reflection}, offering Rudess an ``\textit{analytical look at how I actually think and play}.''\footnote{Demo videos of \texttt{jam\_bot}: \greenurl{https://jam-bot-ismir-2025.media.mit.edu/}}
} % end of rev

%% file: figures/design_space_usagecontext.tex
\begin{figure*}[t!]
  \includegraphics[width=0.9\textwidth]{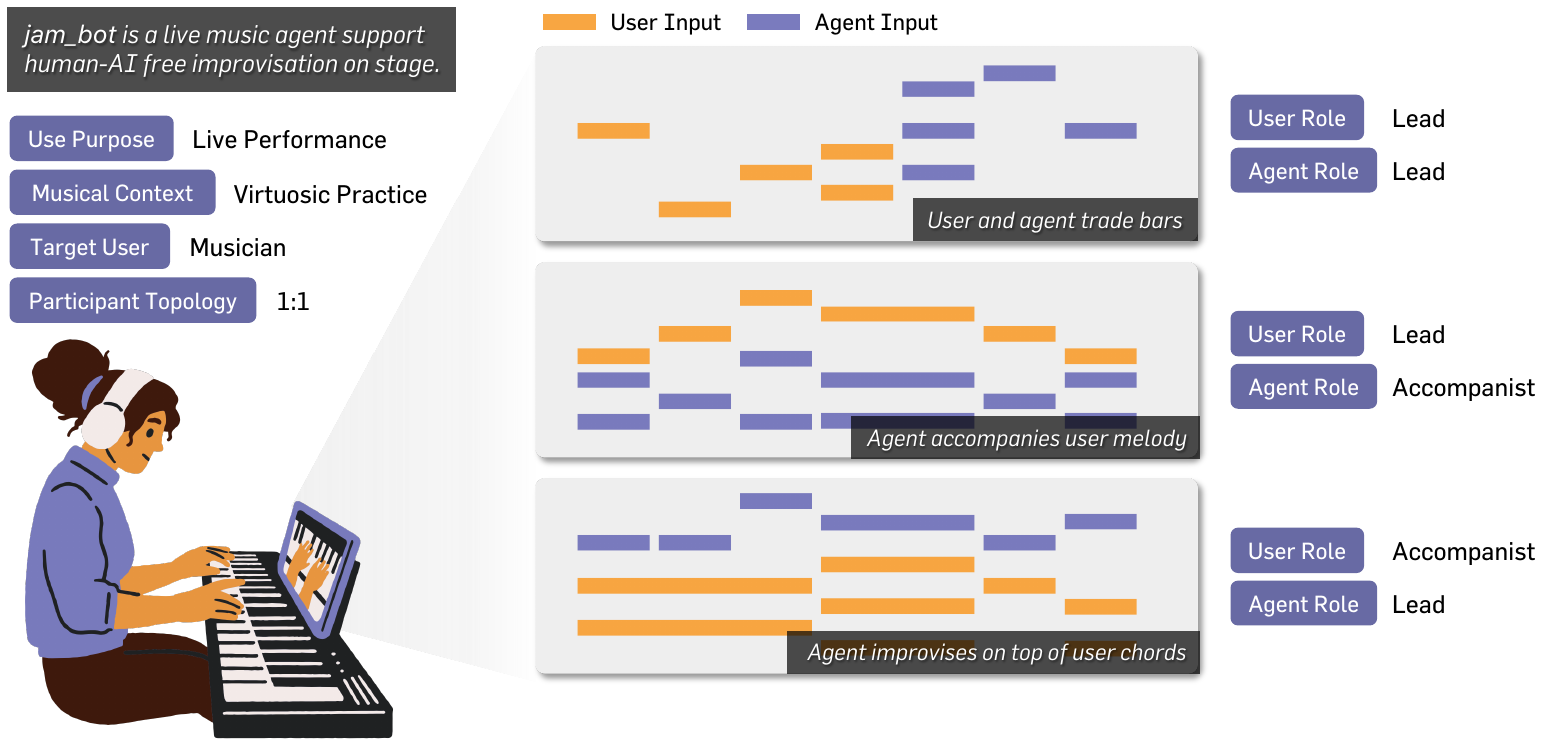}
  % \vspace{-.7em}
  \caption{
  \rev{
    Overview of \texttt{jam\_bot}~\citep{jambot} within the \contextaspect{Usage Context} dimension.
    \texttt{jam\_bot} supports \contextcode{1:1} free improvisation between Jordan Rudess---a professional \contextcode{musician}---and the system in a \contextcode{live performance} setting oriented toward \contextcode{virtuosic practice}. During performance, user and agent alternate between musical roles. Musician and agent can trade bars as shared \contextcode{lead} voices; the agent can serve as \contextcode{accompanist} to the musician’s \contextcode{lead} melody; or the agent can take the \contextcode{lead} by improvising on top of the musician’s accompaniments.
  }
  }
  % \vspace{-.5em}
  \Description{
The figure illustrates three examples of human–AI improvisation supported by \texttt{jam\_bot}. On the left, an illustration shows a musician wearing headphones and playing a keyboard. Above the musician, text states that jam\_bot is a live music agent supporting human–AI free improvisation on stage. Four labels describe the usage context: Use Purpose is Live Performance, Musical Context is Virtuosic Practice, Target User is Musician, and Participant Topology is 1:1.
To the right, three horizontal diagrams show sequences of musical events, where orange bars represent User Input and purple bars represent Agent Input.
In the first example, orange and purple phrases alternate, labeled “User and agent trade bars,” and both user and agent are marked as Lead. In the second example, the agent’s purple notes form an accompaniment under the user’s orange melody, labeled “Agent accompanies user melody,” with the user labeled Lead and the agent labeled Accompanist. In the third example, the musician provides orange chords while the agent plays purple melodic lines above them, labeled “Agent improvises on top of user chords,” with the user labeled Accompanist and the agent labeled Lead.
Overall, the figure shows how musical roles and interaction patterns shift between user and agent during real-time improvisation.
}
  \label{fig:designspace:usagecontext}
\end{figure*}

%% file: sections/05_02_designspace_interaction.tex
In this aspect, we 
focus on functional specifications of live music agents
modeling how 
they interact with users.
\rev{
The summary is provided in Tables~\ref{tab:interaction-designspace-1}-\ref{tab:interaction-designspace-3} in Appendix~\ref{appendix:tables}.
}

\subsubsection{Dimensions and Codes}

\paragraph{\interactiondim{I/O Modality}}
Here we examine the types of raw data modalities that systems take as input or produce as output.
The majority of systems rely on \interactioncode{audio waveforms}~\citep{musicfxdj} (\codInteractionInputModalityRawAudioWaveformPct/\codInteractionOutputModalityRawAudioWaveformPct) and \interactioncode{symbolic music} notation such as MIDI~\citep{jambot} (\codInteractionInputModalitySymbolicMusicPct/\codInteractionOutputModalitySymbolicMusicPct). 
Systems also take \interactioncode{control signals} (\codInteractionInputModalityControlSignalsPct/\codInteractionOutputModalityControlSignalsPct) or \interactioncode{gestures} (\codInteractionInputModalityGesturePct/\codInteractionOutputModalityGesturePct) as input, though these appear infrequently as output. 
\interactioncode{Visuals} are often outputs but rarely inputs, while some systems accept \interactioncode{natural language}. 
A long tail includes modalities such as \interactioncode{physiological data}, \interactioncode{exogenous sensors}, and \interactioncode{programming languages} for live coding.

% SHOULD BE ADDRESSED IN THE REVISION PROCESS
% \todo{\% papers have symmetric input and output modalities, while others are asymmetric.}
% \todo{\% papers that have multiple different types of inputs or outputs}
% \todo{
% I/O modality description paragraph
%   - gesture: Combining audio and gestures for a real-time improviser (improvising gesture; communication practice in improvisation); dance/motion
% }

%
\paragraph{\interactiondim{I/O Musical Element}} Here we discuss the musical elements that systems receive and generate. A majority of systems either receive or respond with 
\interactioncode{melodies} (\codInteractionInputMusicalElementMelodyPct/\codInteractionOutputMusicalElementMelodyPct)~\citep{bachduet, gimenes2007musicianship}. 
Many input or output \interactioncode{rhythms} (\codInteractionInputMusicalElementRhythmPct/\codInteractionOutputMusicalElementRhythmPct)~\cite{franklin2024robocajon, vogl2017an}, 
\interactioncode{harmonies} (\codInteractionInputMusicalElementHarmonyPct/\codInteractionOutputMusicalElementHarmonyPct)~\citep{realchords}, 
\interactioncode{timbres} (\codInteractionInputMusicalElementTimbrePct/\codInteractionOutputMusicalElementTimbrePct)~\citep{rave}, non-melodic 
\interactioncode{sound textures} (\codInteractionInputMusicalElementSoundTexturePct/\codInteractionOutputMusicalElementSoundTexturePct) like vocalizations and soundscapes, or 
\interactioncode{control parameters} that tune explicit musical qualities (\codInteractionInputMusicalElementControlParametersPct/\codInteractionOutputMusicalElementControlParametersPct)~\cite{shepardson2024tungnaa, fiebrink2010wekinator}. 
%\citep{shier2025designing}. 
Some systems leverage \interactioncode{multitrack music} for output (\codInteractionOutputMusicalElementRenderedTrackPct) but few for input (\codInteractionInputMusicalElementRenderedTrackPct)~\citep{davies2014improvasher}. Additionally, very few papers focus on linguistic content like \interactioncode{lyrics} for input or output (\codInteractionInputMusicalElementLyricsPct/\codInteractionOutputMusicalElementLyricsPct)~\citep{shepardson2024tungnaa}.

\paragraph{\interactiondim{Musical Outcome}}
Music often consists of multiple voices, instruments, or tracks; here we capture the texture resulting from the interaction between participants. 
A majority of systems yield outcomes of a tonal nature, i.e.,~one where one or more voices conform to norms found in Western harmony. 
This often manifests as \interactioncode{homophony}~(\codInteractionMusicalOutcomeHomophonyPct). 
In particular, many interactions involve a user providing the melody and the agent generating a harmonic accompaniment~\citep{realchords}.
% \interactioncode{Polyphony}
%  - bach
Alternatively, the user and agent may each provide one or more harmonic voices to create \interactioncode{polyphony} (\codInteractionMusicalOutcomePolyphonyPct)~\citep{bachduet}. 
% \interactioncode{Monophony}
%  - turn-taking 
Less commonly, the user and agent work together to create a \interactioncode{monophonic} (\codInteractionMusicalOutcomeMonophonyPct) outcome, often through a ``trading bars'' (turn taking) setup~\citep{kondak2016active}. 
% CHRIS: Is it worth spending words on something that 2% of papers do?
% Few (x\%) \interactioncode{heterophony}
%  - Few papers target composing \interactioncode{heterophony} music
% \interactioncode{Textural}
%  - instruments~\citep{}?????
%  - synthesis~\citep{shepardson2024tungnaa}
%  - what else
Outcomes from several systems focus less on tonality and more on 
\interactioncode{texture}.
% S are \interactioncode{non-specific}
%  - mentioned but no dtail
Some systems are \interactioncode{non-specific} (\codInteractionMusicalOutcomeNonSpecificPct) and may cover multiple cases of the above. 

\paragraph{\interactiondim{Planning}}

This dimension describes how much information participants share before performing. 
Most systems use \interactioncode{no planning} (\codInteractionPlanningNoPlanningPct), making all decisions live~\citep{pianogenie, HothkerH00, dahl2011tweetdreams}. 
Others allow \interactioncode{user configuration} (\codInteractionPlanningUserConfigurationPct), letting performers set parameters, modes, sounds, or even model personality~\citep{realjam, collins2010musical, spiremuse, roberts2013enabling, Fay15-1, tsiros2020towards, eigenfeldt2008an, strudel}. 
\interactioncode{Tailoring} adapts the system to user data through tuning, fine-tuning, or sensor calibration~\citep{raphael2001bayesian, byuksel2015braahms, jambot, inasilentway, dooley2021mytrombone}. 
With \interactioncode{score} preparation, performers agree on chord grids, structures, or full notations~\citep{marchini2017rethinking, reflexivelooper, borovik2023realtime, jiang2025improvised}. 
\interactioncode{Material prep} involves loading samples or recordings for reuse~\citep{mimi4x, spiremuse, shier2025designing, reflexivelooper}. 
Some systems define \interactioncode{predefined cues} that trigger model behaviors through notes, speech, or gestures~\citep{jambot, roberts2013enabling, godbehere2008wearable, mitchell2011soundgrasp}. 
A few use an \interactioncode{initiative timeline} to schedule turns or leadership roles~\citep{bob, kondak2016active}.

\paragraph{\interactiondim{Temporal Structure}}
Systems vary in how they organize contributions over time.
\interactioncode{Dense parallel} systems allow continuous overlapping activity, typical in accompaniment and intelligent-instrument settings~\citep{realchords, trumpetai, pianogenie}. 
\interactioncode{Sparse parallel} structures also support simultaneous play but with asymmetry, such as agents generating continuously while users issue high-level controls, or the reverse~\citep{naess2019physical, musicfxdj, turczan2019scale}. 
In \interactioncode{turn-taking}, users and agents alternate contributions without overlap, as in call-and-response models~\citep{continuator, spiremuse, bob, DealS13}, though this is least common (\codInteractionTemporalStructureTurnTakingPct). 
\interactioncode{Hybrid} systems flexibly shift between parallel and turn-based interaction~\citep{shimon, magenta, LeffueK16}. 
\interactioncode{Unstructured} systems impose no temporal rules, yielding loosely coupled or intermittent contributions~\citep{Gresham-Lancaster15, jamai, uncannylove, nime2025_54, commentsai}.

\paragraph{\interactiondim{Data Alignment}} 
% This dimension concerns how data flows between participants, which shapes the timing and feel of interaction. 
% Most systems adopt a \interactioncode{continuous stream} (\codInteractionDataAlignmentContinuousStreamPct), transmitting data at fine temporal resolutions to support highly responsive exchanges~\citep{carey2013derivations, gimenes2007musicianship, jamfactory, takase2020support}.
% Others rely on \interactioncode{background triggers}, where the system listens passively and transmits or processes data only when specific events occur, such as when a new note is played~\citep{almeida2019amigo}, a sensor value exceeds a threshold~\citep{van2012mapping}, or an object is detected in vision-based tracking~\citep{roberts2013enabling}.
% Few systems adopt \interactioncode{periodic} alignment, in which data is exchanged at fixed intervals such as every eight bars~\citep{kobayashi2023improvise+} or every 10 seconds~\citep{knotts2021algorithmic}.

This dimension concerns how data flows between participants, shaping the timing and feel of interaction. 
Most systems use a \interactioncode{continuous stream} (\codInteractionDataAlignmentContinuousStreamPct), sending data at fine temporal resolutions for highly responsive behavior~\citep{carey2013derivations, gimenes2007musicianship, jamfactory, takase2020support}. 
Others rely on \interactioncode{background triggers}, processing or transmitting data only when specific events occur—such as note onsets, threshold crossings, or detected objects~\citep{almeida2019amigo, van2012mapping, roberts2013enabling}. 
A smaller group uses \interactioncode{periodic} alignment, exchanging data at fixed intervals, for example every few bars or seconds~\citep{kobayashi2023improvise+, knotts2021algorithmic}.

\paragraph{\interactiondim{Interface}}
Interfaces span from familiar tools to experimental designs. 
\interactioncode{Graphical user interfaces} (GUIs) allow performers to enter inputs~\citep{BakhtB09, kitahara2017jamsketch}
or track agent contributions in real time~\citep{pianogenie, realjam}. 
Users often play through \interactioncode{conventional instruments}, both acoustic and digital~\citep{stefani2024esteso, knotts2021algorithmic, carey2013derivations, xia2017improvised, mimi4x}. 
Some systems rely on \interactioncode{sensor devices} worn by performers, mounted on instruments, or embedded in spaces~\citep{filandrianos2020brainwaves, NIME22_25, savery2024collaborationrobotsinterfaceshumans, juregui2019automatic, tahiroglu2016nonintrusive}. 
Others use \interactioncode{custom instruments} such as deformable surfaces, spring-based controllers, or magnetic boards~\citep{aiterity, fiebrink2020reflections, privato2024stacco}. 
\interactioncode{Stage visuals} support audio-visual work by projecting audience contributions or real-time musical visualizations~\citep{dahl2011tweetdreams, shepardson2025evolving}. 
\interactioncode{Programming interfaces} include live-coding environments and notebooks~\citep{strudel2022, livecoding, magenta}. 
\interactioncode{Embodied agents} use robotic forms as interactive instruments~\citep{shimon, kapur2007integrating, gioti2019imitation, eigenfeldt2008an}. 
Emerging systems explore \interactioncode{XR interfaces} for extended or 3D interaction~\citep{lucas2023a, wang2025ai}. 
Some connect to professional \interactioncode{DJ gear} for performance~\citep{aidj, tsiros2020towards}.

\paragraph{\interactiondim{Control Mode}} Systems differ in how users steer generation.
Many systems rely on \interactioncode{implicit} control (\codInteractionControlModeImplicitPct), where the user’s main musical activity is interpreted by the system and influences generation. For example, a user changing key may cause the system to modulate accordingly~\citep{realchords}, or repeated user gestures may prompt the AI to generate new material~\citep{spiremuse}.
In contrast, \interactioncode{explicit} control enables users to directly guide outcomes, such as sketching melodic contours~\citep{pianogenie, kitahara2017jamsketch} or adjusting generation diversity~\citep{proctor2020a, naess2019physical}.
Finally, some systems offer \interactioncode{no control} (\codInteractionControlModeNoControlPct). Examples include agents generating music fully autonomously~\citep{kobayashi2023improvise+, genjam} or reacting to exogenous signals rather than user input~\citep{Gresham-Lancaster15}.

\paragraph{\interactiondim{Control Scope}}
Among relevant systems, \codInteractionControlScopeMusicalMaterialPct~of them adapt its \interactioncode{musical material} such as melodic phrasing~\citep{continuator} or accompaniment~\citep{kapur2007integrating, martin2019interactive} to user input.
Less commonly, systems offer control over \interactioncode{global style} (\codInteractionControlScopeStyleDirectionPct), such as timbre/FX~\citep{rave, filandrianos2020brainwaves, semilla}, instrumentation~\citep{songdriver}, and stylistic conditioning~\citep{musicfxdj}.
Some also support steering agent's \interactioncode{high-level behavior} (\codInteractionControlScopeBehavioralDirectionPct); for example, by adjusting agent autonomy~\citep{spiremuse} or ensemble role~\citep{voyager}.
The least common scope is \interactioncode{musical layout} (\codInteractionControlScopeMusicalStructurePct), covering structural features like tempo, meter, or key~\citep{proctor2020a, martin2019interactive, BarateHL14}.

\paragraph{\interactiondim{System Initiative}}
In \interactioncode{reactive} systems (\codInteractionSystemInitiativeReactivePct),
users initiate interaction and the system responds by generating material that reinforces or extends user input~\citep{stefani2024esteso, pianogenie}. 
\interactioncode{Proactive} systems (\codInteractionSystemInitiativeProactivePct) instead take the lead, generating material that may diverge from the current context~\citep{Gresham-Lancaster15}, though such systems are rare (\codInteractionSystemInitiativeProactivePct).
\interactioncode{Mixed-initiative} systems (\codInteractionSystemInitiativeMixedInitiativePct) combine both strategies, alternating between responding and initiating~\citep{aiterity, spiremuse, musicfxdj, aidj, gioti2019imitation}.
For example, some intelligent instruments produce sound upon user input (reactive) but generates novel, unpredictable material (proactive)~\citep{aiterity, naess2019physical}.

\paragraph{\interactiondim{Agency Framing}}
The way systems are framed shapes how users relate to them.
As \interactioncode{tools}, they are portrayed as instruments or controllers with limited autonomy~\citep{easthope2023snakesynth, mitchell2011soundgrasp, mcauliffe2023stochgran}.
By contrast, the \interactioncode{partner} framing emphasizes collaboration, either by simulating the role of a co-performer~\citep{jambot, Collins08a}, adding expressive cues such as facial gestures~\citep{inasilentway, shimon}, or embodying a virtual persona~\citep{commentsai, strudel}.
In some cases, both framings coexist (\interactioncode{hybrid}), with systems described interchangeably as instruments and collaborators depending on the context~\citep{gioti2019imitation, Lepri16}.

% figure
\input{figures/design_space_interaction}

\subsubsection{Case Study: Shimon~\citep{shimon}}
\rev{
Shimon is an embodied robotic musician designed to perform alongside human musicians (Figure~\ref{fig:designspace:interaction}).
The system takes the form of an \interactioncode{embodied agent} with four mallet-playing arms and an expressive head, performing on a marimba as its \interactioncode{(conventional) instrument} (\interactiondim{Interface}).
Shimon collaborates with human musicians in jazz settings, producing musical texture of \interactioncode{homophony}
which consists of tightly coupled melodic and harmonic lines (\interactiondim{Musical Outcome}).

Shimon’s behavior is organized into three interaction modules, which determine its \interactioncode{high-level behavior} (e.g., leading or accompanying role) and \interactioncode{musical material} (e.g., rhythmic patterns) (\interactiondim{Control Scope}).
These behaviors are \interactioncode{implicitly} shaped by human musical inputs rather than by explicit commands (\interactiondim{Control Mode}).
In the first module, Shimon detects a human \interactioncode{melody} phrase and responds with accompanying \interactioncode{harmony}. 
The second module uses the collaborator's \interactioncode{rhythm} (typically a bassline pattern) and \interactioncode{harmony} to play a fitting \interactioncode{melody}. 
The third module rephrases the collaborator’s \interactioncode{melody}, preserving rhythmic style while introducing new pitch content (\interactiondim{I/O Musical Element}).
These behaviors are realized by a MIDI listener
that lets Shimon track human \interactioncode{symbolic} input in real time (\interactiondim{Input Modality}).
Shimon's responses are multimodal: 
its arms generating \interactioncode{audio waveform} by striking the marimba, and its head delivering choreographed \interactioncode{gesture} (\interactiondim{Output Modality}). 
Together, these dynamics yield a \interactioncode{hybrid} \interactiondim{Temporal Structure} in which Shimon alternates between turn-taking and parallel accompaniment.

Throughout the paper, 
Shimon is framed as a musical \interactioncode{partner} (\interactiondim{Agency Framing}).
Unlike \interactioncode{tools} typically portrayed as instruments or controllers, 
Shimon has a degree of autonomy that supports a \interactioncode{mixed initiative} interaction (\interactiondim{System Initiative}).
In performance, both human and Shimon drive the musical flow: beyond reacting to human phrases, Shimon communicates proactively using \interactioncode{predefined cues} (\interactiondim{Planning}).
Specifically, its head bobs provide a beat for the human to follow, and its eye contact patterns signal change of initiative: 
turning its head toward the marimba signals that it will lead the next phrase, and turning back toward the musician indicates that it expects the human to take the initiative.\footnote{Demo video of Shimon play: \greenurl{https://www.youtube.com/watch?v=l9OUbqWHOSk}}

} % end of rev

%% file: figures/design_space_interaction.tex
\begin{figure*}[t!]
  \includegraphics[width=\textwidth]{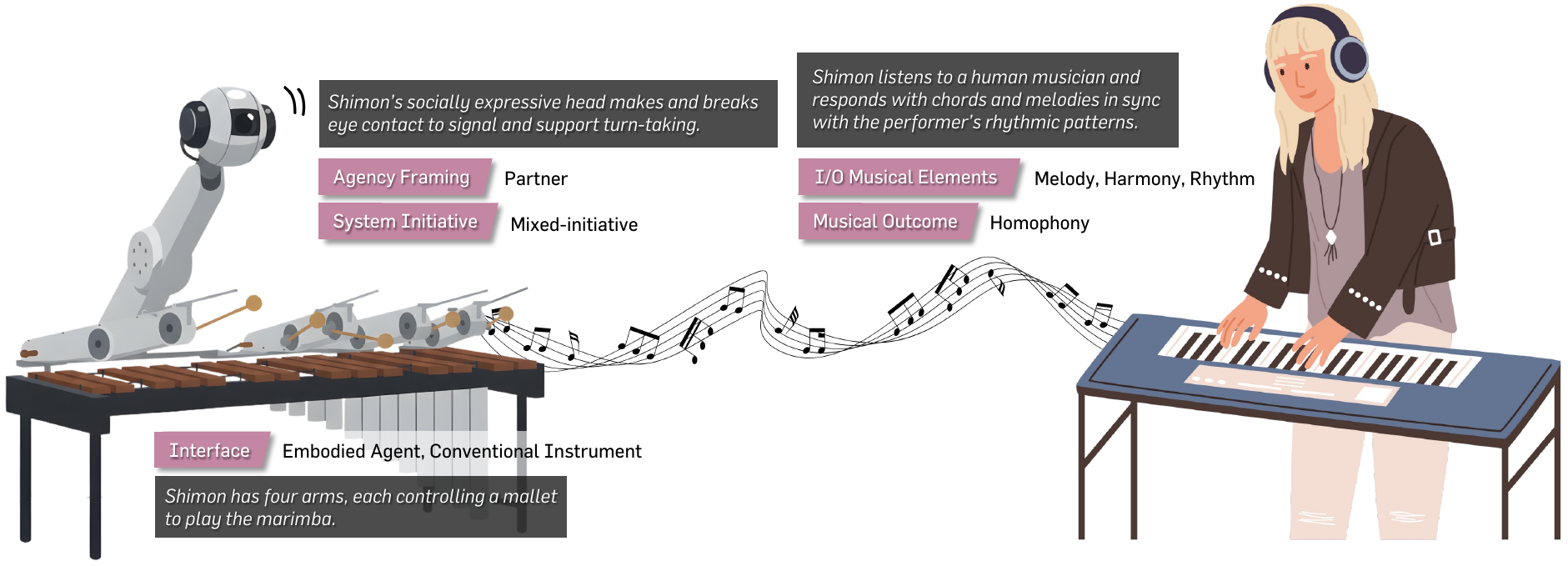}
  \vspace{.1em}
  \caption{
  \rev{
  Overview of Shimon~\citep{shimon} within the \interactionaspect{Interaction} aspect. 
  Shimon is an \interactioncode{embodied agent} performing on a \interactioncode{conventional instrument}---a marimba---using four robotic arms. 
  Shimon is framed as a musical \interactioncode{partner} in a \interactioncode{mixed-initiative} setting, using its socially expressive head to signal turn-taking behavior: looking toward the human signals them to lead, while looking toward the marimba indicates that Shimon will lead. 
  Together, Shimon and human performers create 
  \interactioncode{homophony} in jazz settings, producing tightly coupled melody–harmony textures. 
  Shimon listens to human \interactioncode{melody}, \interactioncode{harmony}, \interactioncode{rhythm}, producing harmonies or melodies shaped by the user’s rhythmic style.
  }
  }
  % \vspace{-.5em}
  \Description{
The figure illustrates the Shimon system interacting with a human musician. On the left, a robotic marimba player is shown: Shimon has four mechanical arms, each holding a mallet positioned above a marimba. Above the robot, a label explains that Shimon’s socially expressive head makes and breaks eye contact to support turn-taking. Two tags identify its \emph{Agency Framing} as “Partner” and its \emph{System Initiative} as “Mixed-initiative.” A label near the marimba indicates the \emph{Interface}: Shimon is an embodied agent that plays a conventional instrument, with a note stating that it has four arms to strike the marimba.
A staff of musical notes flows from the marimba toward a human musician on the right. The human performer is depicted wearing headphones and playing a keyboard. A caption near the musician states that Shimon listens to a human musician and responds with chords and melodies in sync with the performer’s rhythmic patterns. Two additional tags identify the \emph{I/O Musical Elements} as melody, harmony, and rhythm, and the \emph{Musical Outcome} as homophony.
Overall, the figure shows Shimon and a human musician engaging in coordinated musical interaction, exchanging melodic and harmonic material through real-time listening and expressive cues.
}
  \label{fig:designspace:interaction}
\end{figure*}

%% file: sections/05_03_designspace_technology.tex
The technology aspect of live music agents 
considers the technical specification of systems
that power their computational capabilities.
\rev{
The summary is provided in Tables~\ref{tab:technology-designspace-1}-\ref{tab:technology-designspace-2} in Appendix~\ref{appendix:tables}.
}

\subsubsection{Dimensions and Codes}

\paragraph{\technologydim{Model}}

The computational backbone of live music agents spans both traditional and modern paradigms.
The most common are \technologycode{stochastic process} methods (\codTechnologyModelStochasticProcessPct), including Markov chains and genetic algorithms~\citep{lucas2023a, genjam}. 
\technologycode{Task-specific DNNs} follows closely (\codTechnologyModelTaskSpecificDNNPct) using domain-trained CNNs, RNNs, or LSTMs~\citep{arai2023timtoshape, martin2019interactive, NIME22_25}. 
\technologycode{Classical ML} (\codTechnologyModelClassicalMLPct) includes modeling statistics with training data, such as support vector machines (SVMs) and linear regression~\citep{erdem2020raw, khallaghi2025squishysonics}. 
Some systems incorporate \technologycode{rule-based} components~\citep{voyager}, while others use \technologycode{shallow neural networks} such as MLPs with 1-3 layers ~\citep{franklin2024robocajon, shier2024real}. 
A smaller group relies on \technologycode{generative AI} transformers (\codTechnologyModelTransformerPct) trained on large datasets~\citep{musicfxdj, realchords, vampnet}.

\paragraph{\technologydim{Learning Algorithm}}
When models require training data, four learning paradigms are considered. 
Most systems use \technologycode{supervised learning} (\codTechnologyLearningAlgorithmSupervisedLearningPct), learning mappings from inputs to labeled outputs~\citep{shier2025designing, musicfxdj}. 
\technologycode{Unsupervised learning} discovers patterns in unlabeled data, such as latent spaces for performance~\citep{pianogenie}. 
\technologycode{Self-supervised learning} generates its own training signals, for example through masked prediction~\citep{vampnet}. 
The least common is \technologycode{reinforcement learning} (\codTechnologyLearningAlgorithmReinforcementLearningPct), framing music generation as actions optimized for long-term rewards~\citep{CullimoreHG14, realchords}.

% figure
\input{figures/design_space_technology}

\paragraph{\technologydim{Inference Objective}} 
Live music agents differ in functionality depending on the inference objectives their underlying models are designed to solve.
\technologycode{Unimodal generation} (\codTechnologyInferenceObjectiveUnimodalGenerationPct) creates new material from user inputs, in audio (e.g., timbre transfer~\citep{rave}) or symbolic domains (e.g., accompaniment or continuation~\citep{rlduet, continuator}). 
\technologycode{Classification} (\codTechnologyInferenceObjectiveClassificationPct) categorizes input data, such as object recognition in motion-based systems~\citep{kobayashi2020exsampling, juregui2019automatic}. 
\technologycode{Cross-modal generation} (\codTechnologyInferenceObjectiveCrossModalGenerationPct) translates between modalities, such as text-to-audio~\citep{nime2025_54} or control signals to symbolic contours~\citep{pianogenie}. 
\technologycode{Regression} (\codTechnologyInferenceObjectiveRegressionPct) predicts continuous values like tempo or sound parameters~\citep{raphael2001bayesian, hantrakul2018gesturernn}. 
A few systems use \technologycode{retrieval} (\codTechnologyInferenceObjectiveRetrievalPct), selecting close examples with respect to user input~\citep{spiremuse, reflexivelooper}.

\paragraph{\technologydim{Adaptation}}
While adaptation allows live music agents to evolve with users—through parameter updates in AI models or probability adjustments in stochastic systems—the majority of systems rely on \technologycode{no adaptation} (\codTechnologyAdaptationNoAdaptationPct), keeping parameters static after pre-training~\citep{kobayashi2023improvise+, hamano2013generating}. Some apply \technologycode{offline adaptation} (\codTechnologyAdaptationOfflineAdaptationPct), fine-tuning models on user data before performance~\citep{nime2025_54, jambot}. Others use \technologycode{online adaptation} (\codTechnologyAdaptationOnlineAdaptationPct), updating probabilities or weights during interaction~\citep{CemgilK01-0, jamfactory, evans2025repurposing, kitani2010improvgenerator, beyls2018motivated}.  A small number explore \technologycode{continual adaptation} (\codTechnologyAdaptationContinualAdaptationPct), enabling long-term incremental updates~\citep{derbinsky2011cognitive, shepardson2025evolving, assemblage}.

\paragraph{\technologydim{Technical Emphasis}}
Systems often embed technical considerations directly into design. The most common is \technologycode{latency} (\codTechnologyTechnologyDesiderataLatencyPct), addressed through smaller models~\citep{brochec2023toward}, latent representations~\citep{arai2023timtoshape, rave, shepardson2024tungnaa}, on-device inference~\citep{musicfxdj}, or anticipation mechanisms~\citep{realchords, jambot}. 
\technologycode{Efficiency} (\codTechnologyTechnologyDesiderataEfficiencyPct) is also emphasized, using parameter quantization or trading accuracy for resource feasibility~\citep{jambot, jiang2025improvised, wu2025gesturedriven}. 
Some systems support \technologycode{tempo adaptability}, aligning with natural or expressive tempo fluctuations~\citep{takase2020support, CemgilK01-0, jamfactory, jiang2025improvised, Collins08a}, while others assume fixed tempo~\citep{Collins08a, bachduet}. 
A few emphasize \technologycode{error handling} (\codTechnologyTechnologyDesiderataErrorHandlingPct), improving robustness or providing recovery mechanisms~\citep{realchords, ishida2004ism}.

\paragraph{\technologydim{Infrastructure}}
Technical infrastructure of a system supports the implementation and operation of the underlying technology.
Many systems rely on \technologycode{music programming environments} such as Max/MSP, Pure Data, or ChucK (\codTechnologyTechnologyInfrastructureMusicProgrammingEnvironmentPct)~\citep{NIME22_25, hamano2013generating, anders2018populous}. 
Other systems use \technologycode{general programming environments} like Python or JavaScript for development (\codTechnologyTechnologyInfrastructureGeneralProgrammingEnvironmentPct)~\citep{arai2023timtoshape, Ando14-2, BakhtB09}. 
To support AI components, developers use \technologycode{AI/ML frameworks} including PyTorch, TensorFlow, Keras, Magenta, or ChatGPT APIs~\citep{rave, jambot, hantrakul2018gesturernn, livecoding, harmonizing}. 
For communication, systems frequently employ \technologycode{music protocols} such as OSC and virtual MIDI~\citep{Gresham-Lancaster15, Kalonaris16, collins2010musical, turczan2019scale}. 
Applications may be implemented using \technologycode{software toolkits} like JUCE or Tone.js~\citep{jambot, davies2014improvasher, proctor2020a}, while hardware-oriented works use \technologycode{hardware toolkits} such as Arduino or Raspberry Pi for sensor and device integration~\citep{martin2019interactive, almeida2019amigo, keatch2014an}.

\paragraph{\technologydim{Runtime Requirements}}
The execution environment of a system shapes its accessibility, performance, and deployment context.
Most operate on \technologycode{commodity machines} such as laptops, desktops, or mobile devices (\codTechnologyRuntimeRequirementsCommodityMachinePct)~\citep{lucas2023a, fiebrink2010wekinator, aiterity}, sometimes needing consumer GPUs during training or inference~\citep{jambot, nime2025_54, vampnet}. 
Multimodal or mobile systems often depend on \technologycode{dedicated commodity hardware}, including LEDs for visualization~\citep{almeida2019amigo}, Kinect sensors~\citep{juregui2019automatic, Lepri16}, or EEG devices~\citep{filandrianos2020brainwaves}. 
Some projects use \technologycode{custom hardware} such as 3D-printed controllers or handcrafted instruments~\citep{pianogenie, aiterity, fiebrink2020reflections}. 
Others rely on \technologycode{cloud APIs} or hosted models for real-time inference~\citep{proctor2020a, harmonizing, jamai, uncannylove, musicfxdj}. 
Systems using large open-weight models may require \technologycode{high-performance compute}, including TPUs or GPUs like the A100~\citep{musicfxdj, nime2025_54}.

\paragraph{\technologydim{Integration}}
\rev{
The degree to which a system fits into musicians’ existing tools and workflows can influence its adoption.
The most seamless integrations are \technologycode{tool-integrated} systems that operate directly within digital audio workstations (DAWs) or music programming environments, often appearing as Max4Live devices, VST plugins, Max patches, or JUCE modules~\citep{sioros2011automatic, kobayashi2020exsampling, nime2025_54, waite2016church, wu2025gesturedriven, semilla, smith2012unsupervised, shier2024real}; yet these remain a minority (\codTechnologyWorkflowIntegrationToolIntegratedPct). 
In contrast, most systems appear as \technologycode{bespoke setups} (\codTechnologyWorkflowIntegrationBespokeSetupPct), including robots~\citep{shimon, gioti2019imitation, franklin2024robocajon}, custom instruments~\citep{davis2023etudb, privato2024stacco, marley2015gestroviser}, installations~\citep{Schedel2021RhumbLine, kobayashi2023improvise+}, or hardware relying on specialized sensors~\citep{NIME22_25, tahiroglu2016nonintrusive, benasher2013toward, filandrianos2020brainwaves}. 
While these support novel performance practices, they typically require dedicated equipment or spaces, limiting everyday use.
A smaller subset aims for broader accessibility through \technologycode{standalone} web, mobile, or desktop apps~\citep{knotts2021algorithmic, turczan2019scale, Kaliakatsos-Papakostas14, borovik2023realtime}.
Lastly, a few are offered only as \technologycode{source-only prototypes} (\eg GitHub repositories)~\citep{vampnet, songdriver} or \technologycode{developer toolkits} (\eg Python libraries or packages)~\citep{improtek, sioros2011automatic, shepardson2024tungnaa}, supporting experimentation and custom development rather than end-user deployment.
}

\subsubsection{Case Study: ReaLchords~\citep{realchords}}
\rev{
ReaLchords is an online accompaniment system that generates chords in real time in response to a user’s melody input (Figure~\ref{fig:designspace:technology}).
The system's \technologydim{Inference Objective} is \technologycode{unimodal generation}, where the model predicts the next chord token from preceding melody and chord tokens, both represented as symbolic sequences.

During training, the system learns to \textit{anticipate} future user melody inputs and to \textit{adapt} to unexpected user behavior.
This is achieved through finetuning a transformer-based \technologycode{generative AI} (\technologydim{Model}) prior to performance (\technologydim{Adaptation}—\technologycode{offline adaptation}). Technically, ReaLchords integrates multiple \technologydim{Learning Algorithms}: \technologycode{reinforcement learning} (RL), \technologycode{supervised learning}, and \technologycode{self-supervised learning}. The model is first pretrained via maximum likelihood estimation (MLE) (\technologycode{supervised learning}). RL finetuning then equips the model with anticipation and adaptation capabilities. Rewards are provided by \technologycode{self-supervised} models that evaluate musical coherence between melody and chords. Anticipation is further strengthened through a KL-divergence distillation from a teacher model with access to future melody tokens (\technologycode{supervised learning}) during finetuning.
This RL-based approach, which exposes the model to its own predictions and mistakes, enables ReaLchords to achieve low \technologycode{latency} and robust \technologycode{error handling}, allowing it to recover quickly from unexpected user notes, transpositions, or stylistic deviations (\technologydim{Technical Emphasis}).

In the user study, ReaLchords is delivered as a web application (\technologydim{Integration}---\technologycode{packaged standalone}) backed by an \technologycode{AI/ML framework} (\technologydim{Infrastructure}). 
Model inference runs on a remote server, keeping client-side computation lightweight and enabling deployment on \technologycode{commodity machines}, such as standard laptops, commonly used by musicians and creators (\technologydim{Runtime Requirements}).\footnote{Demo videos of ReaLchords: \greenurl{https://storage.googleapis.com/realchords/index.html}}
} % end of rev

%% file: figures/design_space_technology.tex
\begin{figure*}[t!]
  \includegraphics[width=\textwidth]{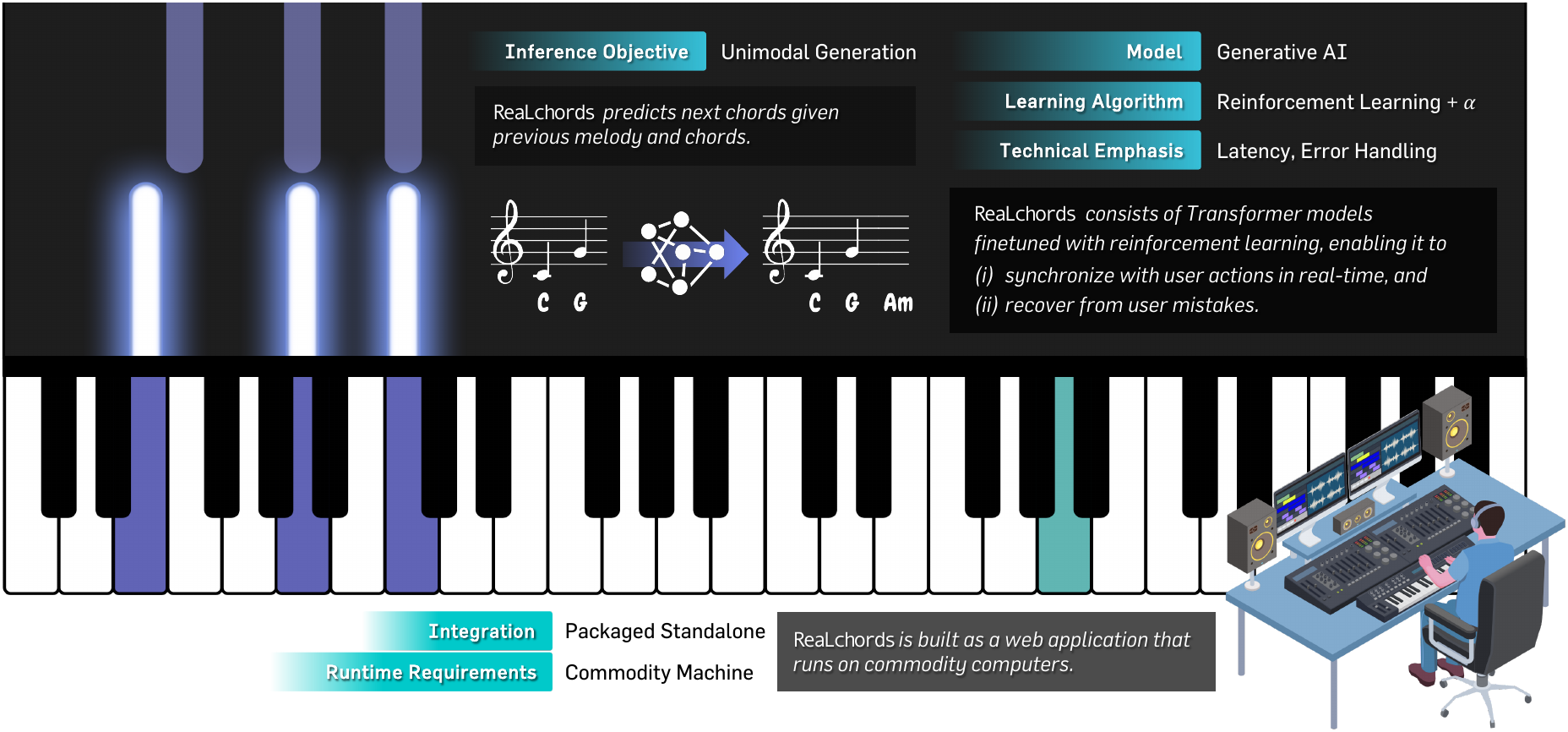}
  % \vspace{-.7em}
  \caption{
  \rev{
  Overview of ReaLchords~\citep{realchords} within the \technologyaspect{Technology} aspect. 
  ReaLchords performs \technologycode{unimodal generation}, predicting next-chord tokens from past melody and chord sequences. 
  The system uses transformer-based \technologycode{generative} models finetuned with a \technologycode{reinforcement learning} objective.
This improves \technologycode{latency} for real-time synchronization with user actions and enhances \technologycode{error handling}, enabling recovery from unexpected user inputs, mistakes, or stylistic deviations.
  In deployment, ReaLchords runs as a \technologycode{packaged standalone} web application, with model inference executed on a remote server to support real-time interaction on \technologycode{commodity machines}.
  }
  }
  % \vspace{-.5em}
  \Description{A diagram illustrating the ReaLchords system. The top panel shows the model’s inference objective---predicting the next chord from previous melody and chord tokens---along with labels for the underlying generative model, learning algorithms, and technical emphasis. A small musical example and a neural-network icon depict the melody-to-chord prediction process. Below, a piano keyboard is displayed with highlighted keys representing input notes and generated chords. At the bottom, the figure indicates that ReaLchords is deployed as a packaged standalone web application running on commodity machines. A user at a digital-audio workstation is shown interacting with the system.}
  \label{fig:designspace:technology}
\end{figure*}

% Overview of Shimon~\citep{shimon} within the \interactionaspect{Interaction} aspect. 
%   Shimon is an \interactioncode{embodied agent} performing on a \interactioncode{conventional instrument}---a marimba---using four robotic arms. 
%   Shimon is framed as a musical \interactioncode{partner} in a \interactioncode{mixed-initiative} setting, using its socially expressive head to signal turn-taking behavior: looking toward the human signals them to lead, while looking toward the marimba indicates that Shimon will lead. 
%   Together, Shimon and human performers create 
%   \interactioncode{homophony} in jazz settings, producing tightly coupled melody–harmony textures. 
%   Shimon listens to human \interactioncode{melody}, \interactioncode{harmony}, \interactioncode{rhythm}, producing harmonies or melodies shaped by the user’s rhythmic style.

%% file: sections/05_04_designspace_environment.tex
This aspect defines the broader environment in which human–AI musical collaboration is situated, encompassing the surrounding culture, institutional and economic conditions, and policy considerations.
\rev{
For a summary of this aspect, see Table~\ref{tab:ecosystem-designspace} in Appendix~\ref{appendix:tables}.}

% fig
\input{figures/design_space_ecosystem}

\subsubsection{Dimensions and Codes}

\paragraph{\ecosystemdim{Sociocultural Factors}}
This dimension describes the cultural values and social practices that shape how systems are designed and used. 
\ecosystemcode{Musical practice} covers the performance and production conventions systems draw from, such as  African-American aesthetics of multidominance~\citep{voyager} or ensemble formats like laptop orchestras~\citep{proctor2020a}. 
\ecosystemcode{Musical genre} reflects the stylistic traditions informing system behavior, for example pop music shaping verse-chorus forms~\citep{marchini2017rethinking} or EDM influencing the design of rhythm variation algorithms~\citep{vogl2017an}. \ecosystemcode{AI perception} also influences adoption: skepticism toward non-human performers can slow use~\citep{savery2024collaborationrobotsinterfaceshumans, tsiros2020towards}, while enthusiasm can encourage experimentation~\citep{dadabots}. Such skepticism often aligns with \ecosystemcode{cultural conservatism}, where users resist new tools that disrupt established workflows~\citep{tsiros2020towards, uncannylove}.

\paragraph{\ecosystemdim{Policy Considerations}}
This dimension describes legal, institutional, or organizational norms surrounding the system's design and use.
\ecosystemcode{Authorship} concerns include questions of whether creative credit belongs to musicians, system designers, or the sources of training data~\citep{jamfactory, assemblage}. 
\ecosystemcode{Data privacy} affects decisions such as relying on users’ own data rather than external assets, or running models on-device to protect information~\citep{wu2025gesturedriven, musicfxdj}. 
\ecosystemcode{Integrity} relates to plagiarism and ethical risks, including models reproducing fragments from training corpora~\citep{jiang2020when}, though this is less emphasized in entertainment contexts~\citep{yakura2021generate}. 
Finally, \ecosystemcode{personality rights} become relevant with voice synthesis and deepfake technologies, raising concerns about the appropriation of vocal identity~\citep{shepardson2024tungnaa}.

\paragraph{\ecosystemdim{Economic Consequences}}
\rev{
This dimension concerns how live music agents intersect with the economic realities of music-making, shaping labor, income, and broader valuation of musical work. Yet only a small portion of surveyed systems ($N=10$) explicitly engage with economic consequences. Within this subset, 70\% raise issues of \ecosystemcode{devaluing}, warning that AI-assisted performance may erode the perceived artistic value of human musicianship or diminish the role of human expertise in live contexts~\citep{trumpetai, savery2024collaborationrobotsinterfaceshumans}.
Additionally, 50\% highlight fears of \ecosystemcode{job replacement}, suggesting that live music agents could partially or wholly replace human performers~\citep{realchords, tsiros2020towards, jambot}, with some warning of deeper threats to musicians’ economic viability and continued participation in the industry~\citep{jiang2020when}.
}

\paragraph{\ecosystemdim{Musical-societal Consequences}}
This dimension captures how live music agents may reshape musical culture and practice. 
The most common theme is \ecosystemcode{reshaping idioms} (\codEcosystemMusicalSocietalConsequencesReshapingIdiomsPct), highlighting new creative values introduced by AI. These include how unpredictability alters the musician–instrument relationship~\citep{aiterity}, creates performative tension~\citep{aidj, intersections}, and enables new expressive possibilities such as performing with digital stylistic twins~\citep{jambot, assemblage}, repurposing band instruments into AI-augmented soloists~\citep{trumpetai, memachine}, or real-time integration of multimodal input like gesture, dance, or soundscapes~\citep{kobayashi2020exsampling, calmuswaves}. 

Other consequences include expanded accessibility through \ecosystemcode{democratization}~\citep{Kaliakatsos-Papakostas14, smith2012unsupervised, kitahara2017jamsketch}, new forms of \ecosystemcode{collaborations} across research communities and creative institutions~\citep{nash2020crowd, semilla}.
At a cultural level, systems may enable 
\ecosystemcode{cultural exchange} through automated style fusion \eg blending Chinese folk with counterpoint~\citep{jiang2020when}, 
or foster \ecosystemcode{revitalization} 
by drawing younger audiences to underrepresented or traditional music~\citep{10.1145/1254960.1254990}. 
Yet concerns persist around \ecosystemcode{misrepresentation}, particularly the marginalization of non-Western musical traditions in Western-trained models~\citep{jambot}, \ecosystemcode{skill dilution}, where reliance on technology risks diminishing musical competencies~\citep{tsiros2020towards}, and \ecosystemcode{over-reliance}, whereby musicians risk becoming dependent on such agents~\citep{tsiros2020towards}.

\subsubsection{Case Study: Channel-AI~\citep{tsiros2020towards}}
\rev{

Channel-AI is a live audio mixing agent that analyzes incoming audio and suggests sound parameter settings (Figure~\ref{fig:designspace:ecosystem}).
In developing the system, the authors foreground the perspectives and concerns of professional mixing engineers.
Many practitioners express negative \ecosystemcode{AI perception} and a broader \ecosystemcode{cultural conservatism} toward automation in creative audio work (\ecosystemdim{Sociocultural Factors}). 
Their skepticism arises from doubts 
that AI can match expert judgment on inherently subjective tasks, 
and
concerns that black-box recommendations are opaque or untrustworthy. 
Engineers also express anxiety about potential \ecosystemcode{job replacement} as AI features become more capable (\ecosystemdim{Economic Consequences}).
They additionally worry that automated tools may encourage \ecosystemcode{over-reliance} on presets and contribute to \ecosystemcode{skill dilution}, leading to a generation of less experienced practitioners with weakened listening skills (\ecosystemdim{Musical-societal Consequences}).
Lastly, 
practitioners emphasize the importance of 
preserving \ecosystemcode{authorship}, i.e., retaining 
creative control and ownership over the final sound (\ecosystemdim{Policy Considerations}).

In response to these concerns, the authors adopt several key decisions in the design of Channel-AI.
First, they minimize disruption to established workflows by ensuring that the system only operates when explicitly invoked 
and that all suggested parameters remain fully editable and reversible. 
The system also provides five adjustable levels of automation, 
allowing engineers to choose the degree of assistance they want rather than imposing a fixed mode of operation. 
Furthermore, the system communicates its capabilities and confidence levels clearly, 
enabling engineers to understand when and why a suggestion is being made. 
These choices result in a design that empowers music creation with AI assistance without displacing expert judgment.
} % end of dev

%% file: figures/design_space_ecosystem.tex
\begin{figure*}[t!]
  \includegraphics[width=\textwidth]{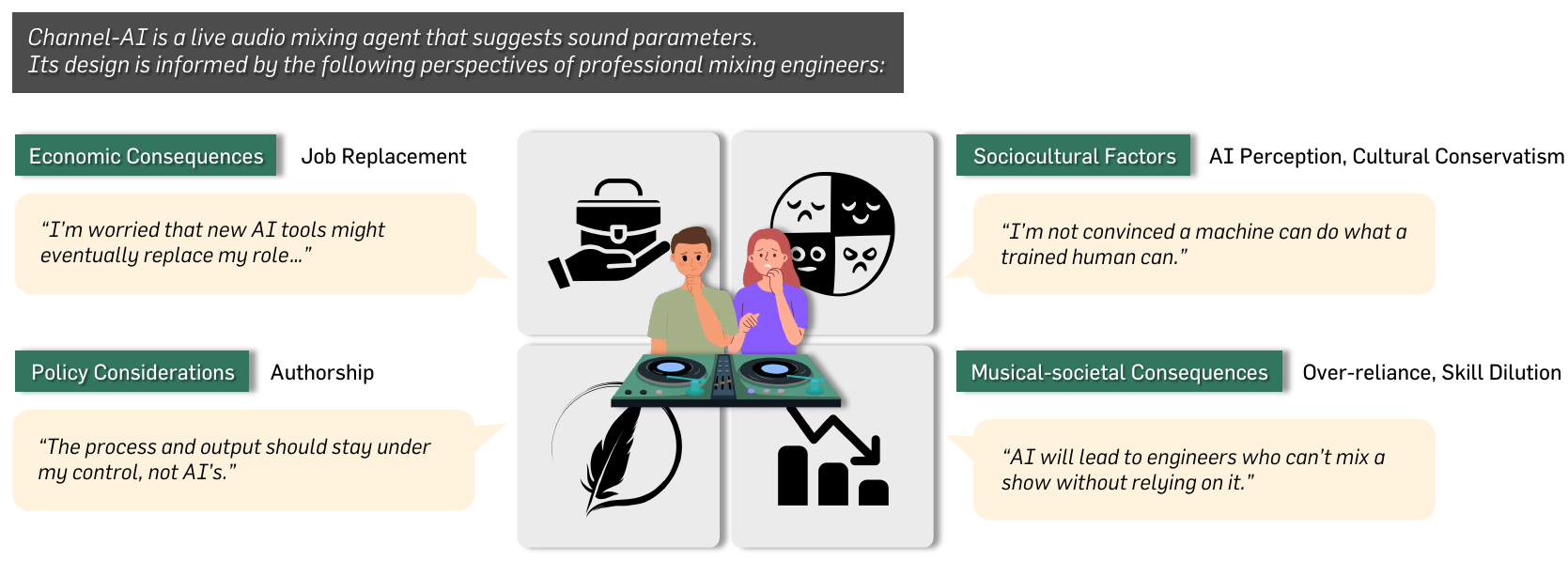}
  % \vspace{-.7em}
  \caption{\rev{Overview of Channel-AI~\citep{tsiros2020towards} within the  \ecosystemaspect{Ecosystem} aspect. 
  The design of Channel-AI is informed by professional mixing engineers’ perspectives across four dimensions. Engineers expressed \ecosystemdim{economic} concerns about AI tools replacing human roles; \ecosystemdim{policy}-related concerns that AI involvement might undermine human authorship; \ecosystemdim{sociocultural} concerns rooted in skepticism toward AI; and \ecosystemdim{musical-societal} concerns that automation may cause over-reliance and loss of expertise. 
  }
  }
  % \vspace{-.5em}
  \Description{
  This figure shows the design motivations behind Channel-AI, a live audio mixing agent, based on perspectives from professional mixing engineers. Four categories of concerns are illustrated: economic consequences (job replacement), policy considerations (authorship and control), sociocultural factors (AI perception and conservatism), and musical-societal consequences (over-reliance and skill dilution), each paired with a representative quote. At the center is an illustration of two engineers working at a mixing console, surrounded by icons representing these concerns.
  }
  \label{fig:designspace:ecosystem}
\end{figure*}

%% file: sections/06_casestudies.tex
\section{Design Space Use Cases}
\label{sec:scenario}

Design spaces commonly serve two purposes. 
First, they can be used to \textit{generate insights} through structured categorization, enabling researchers and practitioners to gain a systematic view of the field. 
Second, 
they can be used to \textit{generate new designs}: support researchers and designers to gain a deeper understanding of the actual design
choices and consequentially the generation
of new approaches~\citep{dsiiwa, langlots2024design}. 
In this section, 
we provide use cases of our design space to 
illustrate 
how our design space can support these purposes: generating insights 
(Section~\ref{sec:scenario:insights}) and generating new designs (Section~\ref{sec:scenario:newdesign}).
Many of these activities can be directly explored using our interactive visualizer, with additional guidance provided in Appendix~\ref{appendix:webtutorial}.

\input{figures/usecase_illustration}

\subsection{Generating Insights}
\label{sec:scenario:insights}

\subsubsection{\textbf{Understanding Trends and Gaps}}
\label{sec:scenario:insights:1}
The design space reveals both trends and gaps in live music agents: clustering across dimensions shows prevailing design approaches, while sparse or empty regions expose underexplored opportunities. These insights help stakeholders understand where
current efforts concentrate
and where 
future innovation is most needed (1-A in Figure~\ref{fig:usecase_illustration}).

For instance, policymakers involved in measuring risks for creative AI tools~\citep{rauh2024gaps, mocap, theft, labor_speech}, specifically live music agents, may turn to the \ecosystemaspect{Ecosystem} aspect to understand how frequently issues such as \ecosystemdim{Policy Considerations} [\ecosystemcode{authorship}, \ecosystemcode{integrity}] and \ecosystemdim{Economic Consequences} [\ecosystemcode{job replacement}] are discussed. The design space reveals that only a small fraction of systems explicitly discuss these topics.
Stakeholders can then move across aspects to investigate how these societal concerns intersect with other design choices. For instance, by exploring the \contextaspect{Usage Context} aspect, they can examine which \contextdim{Musical Contexts} tend to acknowledge policy-related challenges and compare the patterns between systems situated in \contextcode{popular music} and those in \contextcode{traditional music}. Similarly, they might analyze whether particular \contextdim{Value Emphasis}, such as \contextcode{control} or \contextcode{personalization}, co-occur with discussions of ethical considerations. 
% These views provide concrete evidence about where oversight is most urgently needed and where existing research has not yet addressed potential risks. 
By revealing which areas already incorporate ethical reflection and which remain largely unexamined, the design space helps policymakers prioritize topics for future regulation, identify communities that may need additional guidance or support, and craft policies that better reflect the realities of current practice.

\subsubsection{\textbf{Analyzing and Comparing Systems}}
\label{sec:scenario:insights:2}
Researchers can use the design space as an analytic lens to position and compare live music agents. 
By mapping systems onto relevant aspects and dimensions, 
they can construct a structured overview that highlights shared patterns, points of divergence, and distinctive characteristics (1-B in Figure~\ref{fig:usecase_illustration}).

Consider a researcher conducting a literature review on transformer-based accompaniment systems. 
They identified four relevant works: 
SongDriver~\citep{songdriver}, jam\_bot~\citep{jambot}, GrooveTransformer~\citep{evans2025repurposing}, and ReaLchords~\citep{realchords}.
To compare these systems more systematically, 
they map each onto the \technologyaspect{Technology} aspect of the design space.
From the analysis, they observe major differences in 
\technologydim{adaptation} method:
SongDriver~\citep{songdriver} does not employ adaptation (\technologycode{no adaptation}),
jam\_bot~\citep{jambot} and ReaLchords~\citep{realchords} adopt \technologycode{offline adaptation} strategies (by finetuning models before deployment),
and GrooveTransformer~\citep{evans2025repurposing} performs \technologycode{online adaptation} (by training Markov models in real time alongside its rhythm-generating transformer).
They also note contrasts in \technologydim{Technical Emphasis}.
While all four systems prioritize reducing \technologycode{latency} for responsive interaction,
ReaLchords places particular emphasis on \technologycode{error handling}, incorporating mechanisms to remain robust to user mistakes or unexpected musical deviations.
Conversely, jam\_bot emphasizes \technologycode{efficiency}, using model quantization to run on limited hardware during live performance.
From this comparison, the researcher gains insights into how distinct design priorities shape system behavior, which methodological combinations dominate current practice, and where gaps or opportunities---such as lightweight online adaptation of transformers---remain underexplored.

\subsection{Generating New Design}
\label{sec:scenario:newdesign}

\input{figures/trends_num_systems}

\subsubsection{\textbf{Designing a New Live Music Agent}}
\label{sec:scenario:newdesign:1}
The design space can support the creation of entirely new live music agents. 
When designers begin without a specific prototype, 
the dimensions function as a generative canvas: 
by surveying underexplored combinations, noting sparsely populated areas, or identifying emerging trends across studies, researchers can uncover opportunities for innovation (2-A in Figure~\ref{fig:usecase_illustration}).

For instance, suppose a designer is interested in \contextcode{live coding} practices and aims to develop a new agent that meaningfully complements this mode of performance.
By examining patterns in the \interactiondim{I/O modality} dimension, 
they observe that current systems typically support natural-language input that produces code, 
or code input that produces symbolic music. 
Yet the design space also reveals what is missing: there are no systems that generate new code in response to ongoing live coding activity, and no systems that listen to musical input and produce musical responses expressed in the form of code.
Recognizing these gaps opens space for novel live coding agents. A designer could develop an agent that listens to the performer's musical output and responds by generating live code that extends or contrasts with the performer's musical direction. Another possibility is a system that observes how the performer writes code in real time and proposes alternative patterns, transformations, or structural developments that can be incorporated directly into the live script. 

\subsubsection{\textbf{Iterating on an Existing System}}
\label{sec:scenario:newdesign:2}
The design space can also guide iterative refinement of existing systems. 
By mapping an existing system to the relevant dimensions, 
designers can more easily identify overlooked aspects, reconsider implicit assumptions, 
and explore alternatives that extend the system’s capabilities (2-B in Figure~\ref{fig:usecase_illustration}).

For instance,
suppose a research team has developed a web-based real-time `trading-fours'\footnote{``Trading fours'' is a method of soloing in jazz where band members exchange solos, each lasting four bars.} improvisation system for jazz pianists. 
After mapping their prototype to the design space, they observe that they have already addressed several important dimensions, including \contextdim{musical context}[\contextcode{jazz}], \interactiondim{temporal structure}[\interactioncode{turn-taking}], \technologydim{integration}[\technologycode{standalone}], and \ecosystemdim{sociocultural factors}[\ecosystemcode{musical practice}, \ecosystemcode{musical genre}].
However, the mapping also reveals dimensions that had been overlooked, such as \technologydim{technical emphasis}[\technologycode{tempo adaptability}, \technologycode{error handling}], \interactiondim{planning}[\interactioncode{tailoring}, \interactioncode{score}].
These gaps motivate the researchers to revisit and improve their design choices.
They may begin exploring how the underlying model might better adapt to performers' tempo changes and mistakes,
whether the system should support user-specific personalization through finetuning with user data,
and how incorporating a shared lead sheet---an essential artifact in many jazz settings---might improve user experience.
Through this iterative process, the design space serves as a reflective scaffold that 
helps the team avoid overlooking important design decisions and ultimately leads to a richer and more thoughtful design.

%% file: figures/usecase_illustration.tex
\begin{figure*}[t!]
  \includegraphics[width=\textwidth]{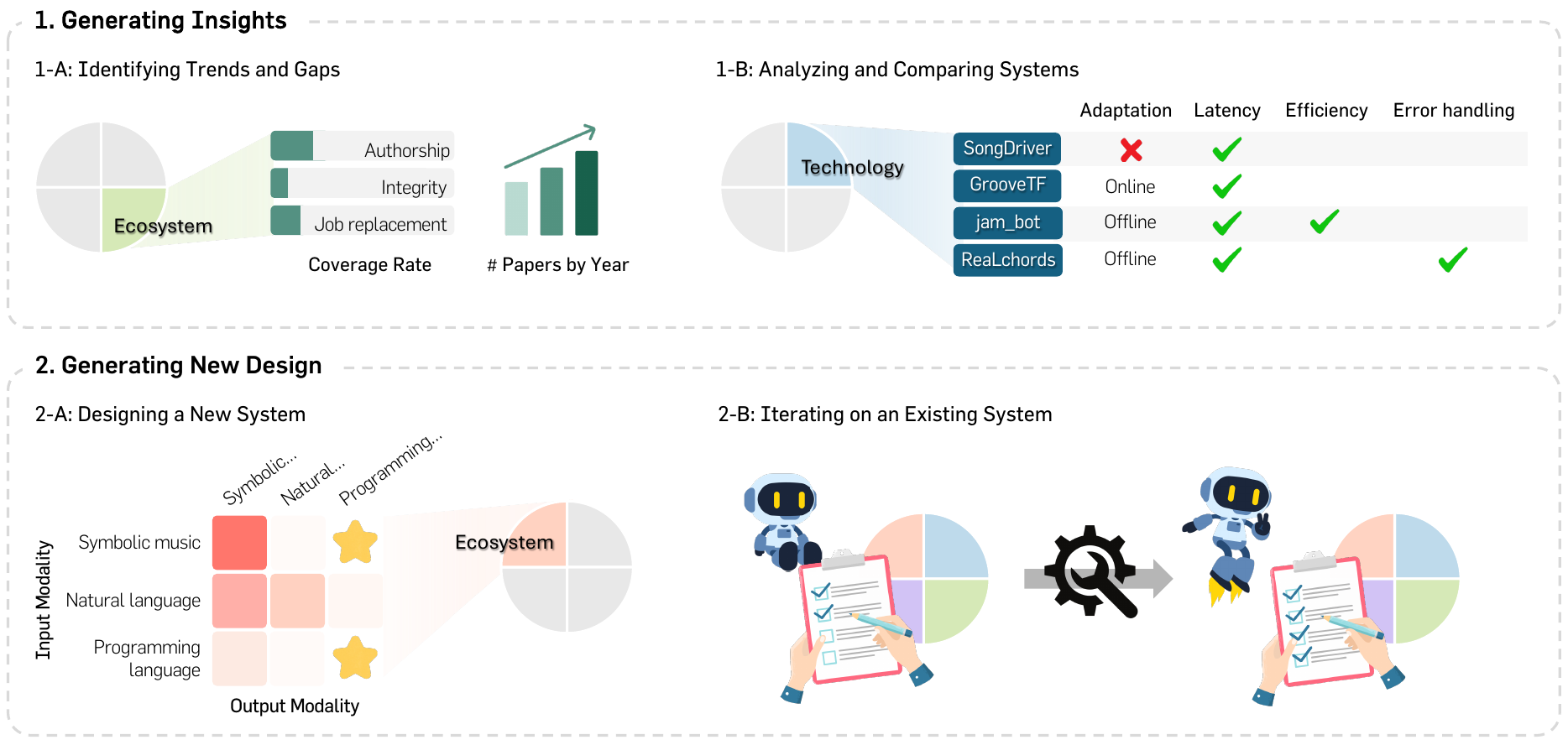}
  % \vspace{-.7em}
  \caption{
  \rev{
  Illustration of design space use cases. The design space framework supports two primary categories of activities: (1) generating insights and (2) generating new design. 
  In generating insights, 
  stakeholders can identify trends and gaps through various analytical approaches to identify patterns and evolution over time (1-A). They can also analyze and compare systems of interest by projecting them into specific dimensions and checking which codes they cover (1-B).
  In generating new design, researchers and designers can create new systems by exploring different combinations of dimensions and codes (2-A), and iterate on existing systems through systematic evaluation and refinement to improve functionality (2-B).
  }
  }
  % \vspace{-.5em}
  \Description{
  This figure illustrates two ways the design space can be applied: generating insights and generating new designs. The top row shows insight-generation tasks, including identifying trends and gaps through coverage and timeline charts, and comparing systems using a table of technical attributes such as adaptation type, latency, efficiency, and error handling. The bottom row shows design tasks, including designing a new system by selecting input and output modalities and ecosystem factors, and iterating on an existing system represented by a robot mascot, checklists, and adjustment icons. Overall, the figure highlights how the design space supports both analysis of existing work and creation or refinement of live music agents.
  }
  \label{fig:usecase_illustration}
\end{figure*}

%% file: figures/trends_num_systems.tex
\begin{figure*}[t!]
  \includegraphics[width=\textwidth]{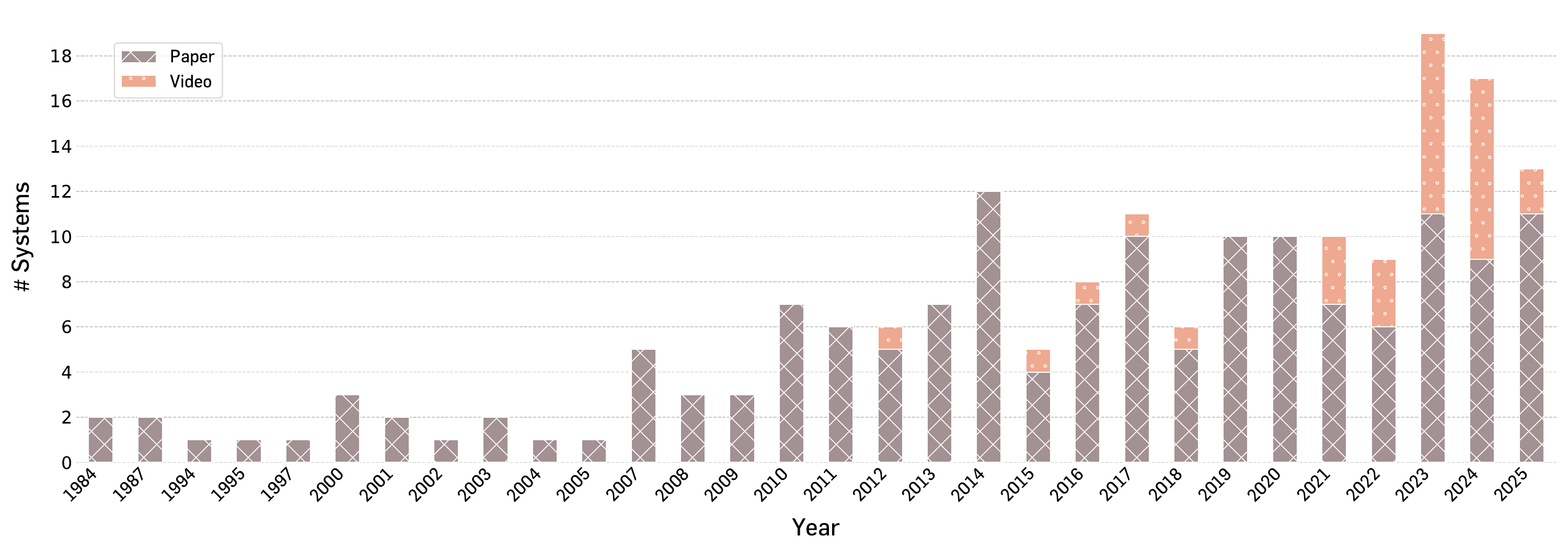}
  % \vspace{-1em}
  \caption{
  The number of publications and videos of live music agent systems by year (1984–2025). The figure shows steady growth since 2010, with video-based sharing rising rapidly after 2020.
  }
  % \vspace{-.5em}
  \Description{A vertical stacked bar chart shows the number of live music agent systems published per year from 1984 to 2025. The x-axis represents years, and the y-axis represents counts from 0 to 18. Bars are divided into two categories: papers (brown with crosshatch) and videos (orange with dots). From 1984 to 2005, the numbers remain low, rarely exceeding 2–3 systems per year. Beginning around 2010, the counts increase steadily, with most systems published as papers. After 2020, video-based dissemination becomes more common. The peak occurs in 2023 with 19 total systems, followed by 17 in 2024.}
  \label{fig:discussion:numpapers}
\end{figure*}

%% file: sections/07_discussion.tex
\section{Discussion}

\input{figures/trends_dimension_coverage}

\rev{
Building on the use cases presented in Section~\ref{sec:scenario}, we demonstrate the design space in action by applying it to generate concrete insights about the current state of live music agents and to explore new design opportunities. 
Concretely, we use the design space to identify trends and gaps in existing systems (Section~\ref{sec:discussion:insights}), and to propose novel design directions (Section~\ref{sec:discussion:design}).
}

\subsection{Generated Insights}
\label{sec:discussion:insights}

\subsubsection{\textbf{Societal Implications of Live Music Agents are Underexplored}}
\label{sec:discussion:insights:ethical}
%\cd{are Underexplored}}}
%Broadly, we identify
We observe
that 
the number of systems has grown steadily since 2010, with a sharp acceleration after 2020 (Figure~\ref{fig:discussion:numpapers}).
Importantly, we see a sharp rise in systems shared through video, reflecting a growing community of practitioners experimenting with and being exposed to live music agents.
Yet despite growing practitioner interest and further anticipated impact on the horizon, 
our analysis shows that \ecosystemaspect{Ecosystem} aspect of live music agents remain largely overlooked (Figure~\ref{fig:discussion:dimension}).
In particular, \ecosystemdim{policy considerations} and \ecosystemdim{economic consequences} receive far less attention than functional and technical concerns, even though these issues are central to real-world deployment and sustainability of such systems.
Ongoing debates in broader creative AI highlight what may lie ahead. 
Copyright disputes in art, writing, and music composition raise 
unresolved questions about \ecosystemcode{authorship} and \ecosystemcode{integrity}: 
Who should be credited when a machine contributes to the creative process? As its algorithmic pipeline blurs the roles of system designers, dataset contributors, and end-users, determining `who created what' becomes increasingly difficult~\citep{shroff2024ai, samuelson2023generative, morreale2021does, 10.1145/3706598.3713287}.
In parallel,
fears about \ecosystemcode{job replacement} point to possible shifts in creative economies, with impacts already visible in adjacent domains~\citep{arts8030115, doi:10.1177/01634437221077009, 10.1145/3600211.3604681}.

Beyond legal and economic considerations, 
\ecosystemdim{musical-societal consequences} warrant equal attention.
Just as programming has been reshaped by AI copilots~\citep{Akhilesh2025, Ray_2025}
and art by text-to-image generation models~\citep{10.1145/3475799},
advances in live music agents hold the potential to
\ecosystemcode{democratize} access to creative practice.
We are already witnessing this with offline text-to-music~\citep{10.1145/3706598.3713876, suno_Ai, udio_ai} and music deepfake models~\citep{feffer2023deepdrake, buko2023ghostwriter},
and live music agents are approaching a similar turning point as they become increasingly powerful, popular, and accessible~\citep{jambot, musicfxdj, strudel}.
This shift raises key questions: How will live music agents democratize access while avoiding \ecosystemcode{over-reliance}, \ecosystemcode{skill erosion} or homogenization in music? 
How might they foster \ecosystemcode{cultural exchange} without reinforcing inequities in whose traditions are represented (\ecosystemcode{misrepresentation})?
Crucially, how might they serve as unique augmentations of human creativity rather than replacements?
Anticipating these challenges requires proactive engagement from researchers, designers, policymakers, and practitioners.
Future work should integrate them into design and evaluation frameworks ensuring music agents evolve in ways that are not only technically robust but also socially responsible.

\subsubsection{\textbf{
% Improving Efficiency in 
% %Large Models
% Generative AI Models
% (Finding) 
Increasing Prevalence of Generative AI}}
\label{sec:discussion:insights:aitrend}

% \cd{Would maybe put this first? I think this is a key trend we want to highlight, and the societal implications dovetail nicely from it}
% \yewon{same idea but thought the fig 3 and 4 should come first :')}

Figure~\ref{fig:discussion:aimodel} shows the trend of AI models powering live music agents over time. 
\contextcode{Rule-based} methods, \contextcode{stochastic processes}, and \contextcode{classical ML} models initially dominated. 
%we see increase in DNN after 2012 (AlexNet~\citep{krizhevsky2012imagenet}) 
Unsurprisingly, we see an increase in the use of 
% \contextcode{shallow neural network models} and 
\contextcode{task-specific DNNs} following the dawn of deep learning in 2012~\citep{krizhevsky2012imagenet}. 
More recently, broadly capable \contextcode{generative AI} models have begun to power live music agents, though their proliferation is perhaps slower than that of generative AI in other areas of computing. 
One potential explanation is that live music interaction presents numerous technical challenges at odds with generative AI: 
\rev{computing resources in music are often constrained to commodity hardware~\citep{zhou2024local}, and
users demand very low control latency ($<100$ms)~\citep{bartlette2006effect, david2002problems}.}
In contrast, generative AI models usually involve dedicated hardware and high latency interaction loops. 
% \todo{Perhaps create a view for efficiency / low latency?}
An emerging line of work on generative AI for music places a keen emphasis on efficiency, e.g.,~through 
model compilation~\citep{zhou2024local}, 
anticipating user actions~\citep{jambot}, 
truncated context~\citep{musicfxdj}, distillation~\citep{novack2024presto}, and 
post training~\citep{novack2025fast}. 
Still, while these methods have enabled live interaction with symbolic models on commodity devices~\citep{zhou2024local} or audio models on dedicated hardware~\citep{musicfxdj},
%improved offline audio generation times, or enabled live audio interaction on dedicated hardware, 
we have yet to see a live music agent with broad audio generation capabilities that runs with low latency on commodity devices. 
We anticipate an increased presence of generative AI in live music agents once the trend lines of improvements to model efficiency and commodity hardware intersect.  %as further improvements are made to both model efficiency and commodity hardware. 
%confluence of further improvements to model efficiency and commodity hardware.

% and generative AI start to emerge since 2022 yet very few portion
% advance of generative AI for live music is challenging 
% live music: low latency, efficiency are key
% but large models high latency, memory usage, and large in size -- hard to be used as is 
% Recently proposed methods are 
%  - anticipating user actions~\citep{jambot}, simplifying / coarse? representations~\citep{musicfxdj}, or reinforcement learning~\citep{realchords}, latent diffusion models~\citep{huang2024musicstyletransferdiffusion} 
% yet still underexplored
% - cuda optimzation
% - speculative decoding

\input{figures/trends_ai_model}

\subsection{Opportunities for New Design}
\label{sec:discussion:design}

\subsubsection{\textbf{Customizing Live Music Agents}}
\label{sec:discussion:design:custom}

From our analysis, \contextcode{personalization} of live music agents is one of the values commonly emphasized in their design.  
Agents trained on a musician’s own data can act as a digital mirror of oneself, offering the unique experience of playing with a computational ``other'' that reflects, extends, and challenges one’s own musical practices~\citep{jambot, melbourne} (Figure~\ref{fig:discussion:design1}A).  
Yet surprisingly, most current models are deployed as-is (\technologycode{no adaptation}) or using \technologycode{offline adaptation} through fine-tuning on user data, and only a few exploring \technologycode{online} or \technologycode{continual} adaptation 
(Section~\ref{sec:designspace:technology}).  

We see several opportunities for advancing personalization in live music agents. For HCI, the challenge lies in designing interfaces that allow non-technical musicians to fine-tune models with their own data. Wekinator~\citep{fiebrink2010wekinator} exemplifies how musicians could engage directly with supervised learning models. 
Looking forward, the opportunity is to provide musicians with the ease and integration exemplified by no-code ML tools: environments where capturing musical inputs, retraining models, and versioning agents can be done fluidly, supporting personalization without technical expertise and creative flow~\citep{google2019teachablemachine, 10.1145/3411764.3445096, 10.1145/3196709.3196729}.
For AI practitioners, recent advances in test-time adaptation~\citep{liang2025comprehensive} and online reinforcement learning with human feedback (RLHF)~\citep{dong2024rlhfworkflowrewardmodeling} point to powerful strategies for enabling continual adaptation of modern AI. Applying these methods in a live music context raises important open questions, such as how feedback can be captured, represented, and acted upon in the highly time-sensitive environment of musical performance.

Broader challenges lie in supporting adaptation over extended timescales.  
Whereas offline and online learning adapt an agent to immediate events, continual or incremental learning techniques~\citep{HADSELL20201028, chen2018lifelong} may afford agents that gradually evolve with a musician’s changing style~\citep{10.1145/1279740.1279821}.  
This raises both design opportunities and open questions: how might long-term personalization affect a musician’s sense of authorship and agency? Could overly faithful digital ``copies'' collapse the productive tension of improvisation, whereas more interpretive evolutions open space for surprise and reflection?  
HCI research is well-positioned to examine these tensions, investigating how agents might balance the practical benefits of faithful mirroring with the creative opportunities of divergence~\citep{spiremuse, nottophd}, while AI advances in lifelong learning provide the technical basis for such adaptation. 
 
% \subsubsection{\textbf{Contingency as design material}}
\subsubsection{
%(Opportunity) 
\textbf{Contingency as Design Material}}
\label{sec:discussion:design:contingency}

While \technologycode{latency} has been widely emphasized as a core \technologydim{technical emphasis}, the human aspects of live music performance receive far less attention (Figure~\ref{fig:discussion:design1}B). In practice, musicians routinely introduce tempo fluctuations, expressive rhythmic choices (\eg rubato, swing, syncopation), and even mistakes. Within human ensembles, these are not failures but essential components of interaction. Performers adapt dynamically to varying tempo, treat rhythmic changes as non-verbal cues for shifts in style~\citep{doi:10.1177/0305735605056151}, and often value mistakes as sparks for improvisatory exploration~\citep{nime2023_9, crossan2003making}. 
% In jazz especially, as Miles Davis put it, ``\textit{It’s not the note you play that’s the wrong note---it’s the note you play afterwards that makes it right or wrong}.''
Current live music agents, by contrast, have not meaningfully engaged with these practices. 
Trained primarily on sanitized, error-free corpora and assuming fixed tempo, they tend to collapse when faced with expressive deviations or errors. 
Only a handful of efforts explore robustness or correction to contingencies~\citep{realchords, kondak2016active, bachduet}, but these remain isolated exceptions.

This gap raises design questions for future systems. 
When a human makes a mistake or changes in rhythm or tempo, should the agent act as a corrective partner absorbing these and maintaining coherence or as a creative partner that responds with new musical ideas? 
Should AI-generated ``mistakes'' be treated as opportunities for creativity, mirroring how human musicians reinterpret errors~\citep{caren2025melia}? 
More broadly, who holds agency in determining how contingencies are handled: the human, the AI, or both in negotiation? 
\rev{
HCI research on human-machine collaboration could offer frameworks for understanding such distributed agency by informing how systems surface and negotiate contingencies with performers~\citep{xambo2024agency}. 
Exploring such communication strategies that allow humans and AI to co-evolve from dynamic situations, rather than treating them as breakdowns, could open up new interaction paradigms and support richer forms of musical expressivity.
}
\input{figures/design_1}

% ======
% \subsubsection{\textbf{Integrating live music agents into musicians' workflow}}
% Observation: not much systems are tool-integrated
% Research opportunities: improving accessibility of the technology (e.g., DAW plugin) [4]

% \subsubsection{\textbf{Live coding agents}}
\subsubsection{
% (Opportunity) 
\textbf{Live Coding Agents}}
\label{sec:discussion:design:livecoding}
Despite \contextcode{live coding}'s prominence in computer music research~\citep{COLLINS_McLEAN_ROHRHUBER_WARD_2003, 04edcdf0816c4606bb03ff8203beba94, wangonthefly, mclean_2022_7219926, NIME22_36, slee2014}, 
we found surprisingly little attention given to it within the scope of live music agents (Figure~\ref{fig:discussion:design2}A).
This gap is notable when contrasted with the impact of AI programming assistants such as GitHub Copilot~\citep{10.1145/3597503.3608128}, 
which have transformed software practice. 
With only a few pioneering systems probing this intersection~\citep{strudel, Xambo2021Live, johnson2023musical}, 
there remains significant opportunity for future work to 
explore how AI might support live coders in generating musical patterns, transformations, or effects in real time. 
Beyond code assistance, other live music agents such as synthesizers~\citep{rave} or accompaniment models~\citep{tatar2018revive, realchords} could be integrated into live coding workflows, expanding improvisational possibilities and reshaping the ensemble dynamics of live coding practice.

Emerging practices of \textit{vibe coding}\footnote{Original post coining the term `vibe coding': \url{https://x.com/karpathy/status/1886192184808149383?lang=en}} further extend this horizon, 
where high-level natural language prompts orchestrate end-to-end software creation~\citep{Ray_2025}. 
Platforms like Emergentic.ai\footnote{\url{https://emergentic.ai/}} demonstrate how descriptive inputs can guide live coding agents~\citep{strudel}, 
suggesting workflows where performers cue intent semantically rather than programmatically.
Yet natural language is not the only modality for expressing musical goals; 
alternative prompts such as audio-based inputs~\citep{musicfxdj} 
or melodic contours~\citep{pianogenie, kitahara2017jamsketch} 
may align more closely with how musicians communicate intent. 
Taken together, these directions highlight live coding agents as fertile ground for advancing both accessibility and expressivity in human–AI musical improvisation.

% \subsubsection{\textbf{Live music agents in Extended Realities (XR)}}
\subsubsection{
% (Opportunity) 
\textbf{Live Music Agents in Extended Realities (XR)}}
\label{sec:discussion:design:xr}
% \interactioncode{XR interface} is a key area of interest in HCI due to its ability to deliver multimodal, context-aware, and spatially grounded experiences~\citep{10.1145/3173574.3173703}, while augmenting users with interactions that would be impossible in the physical world~\citep{10.1145/3173574.3173703}.
Although prior research has explored \interactioncode{XR interface} for music and audio~\citep{9328440, 10.1145/3654777.3676424}, its role in live music agents remains nascent (Figure~\ref{fig:discussion:design2}B), with a few recent works beginning to probe this intersection~\citep{wang2025ai, lucas2023a}. We see three key dimensions where XR could enhance live music agents. 
First, XR can \textit{augment the user} by enabling interactions and creative possibilities that would be impossible in the physical world~\citep{10765432, abtahi2019m, 10.1145/3173574.3173703}. XR has potential to foster new musical idioms in which musicians perform by engaging with virtual objects, agents, and environments~\citep{NIME20_23}, or enrich traditional instruments with intelligent overlays~\citep{9328440, wang2025ai, lucas2023a}. Furthermore, designing instruments in virtual reality (VR) may also democratize access to complex \interactioncode{custom instruments} without requiring specialized \interactioncode{bespoke setup}.

XR also enables \textit{multimodal and spatially grounded experiences}. By integrating more \interactioncode{I/O Modality} beyond sound, such as haptic~\citep{10.1145/2814895.2814918} and proprioceptive modalities~\citep{10.1145/3290605.3300905}, XR can make the music experience more immersive by both engaging more senses of the performer, and providing more context to the agent for more context-aware behaviors~\citep{10.1145/3586183.3606825, 10.1145/3708359.3712070}. Finally, XR allows agents to take on \textit{embodied or anthropomorphic forms}. Prior research has found that agent embodiment improved communication cues, social factors, and even alters how humans interact with them~\citep{568812, 10.1145/3491102.3517593}. While music agents have typically relied on physical robots~\citep{shimon} to provide such presence, XR enables virtual embodiments that can inhabit shared musical spaces, supporting more natural, expressive, and engaging co-performance.

\input{figures/design_2}

\subsection{Limitations and Future Work}
\label{sec:discussion:limitations}
Several limitations of the study warrant consideration.
First, although we followed existing practices in design space creation, 
only approximately 10\% of the corpus was double coded, 
while the remaining systems were coded by a single author.
As a result, some annotations may reflect individual interpretation errors or misinterpretations of authors' intentions. 
Second, the scope of our corpus is bounded by our definition of live music agents and our selection of venues and online platforms. 
Although we aimed for broader coverage across HCI, AI, and computer music fields, 
our dataset may underrepresent systems published in less-visible venues, non-English publications or commercial tools. 
Despite these limitations, we believe our design space provides a useful and robust framework for understanding the current landscape of live music agents and for innovating new designs.
As future work, we see opportunities to expand the corpus over time, incorporate community-driven contributions, and refine dimensions as new technologies and practices emerge. 
More broadly, we hope this design space serves as a living artifact that can be iteratively extended to support responsible, creative, and interdisciplinary development of live music agents.

%% file: figures/trends_dimension_coverage.tex
\begin{figure*}[t!]
  \includegraphics[width=\textwidth]{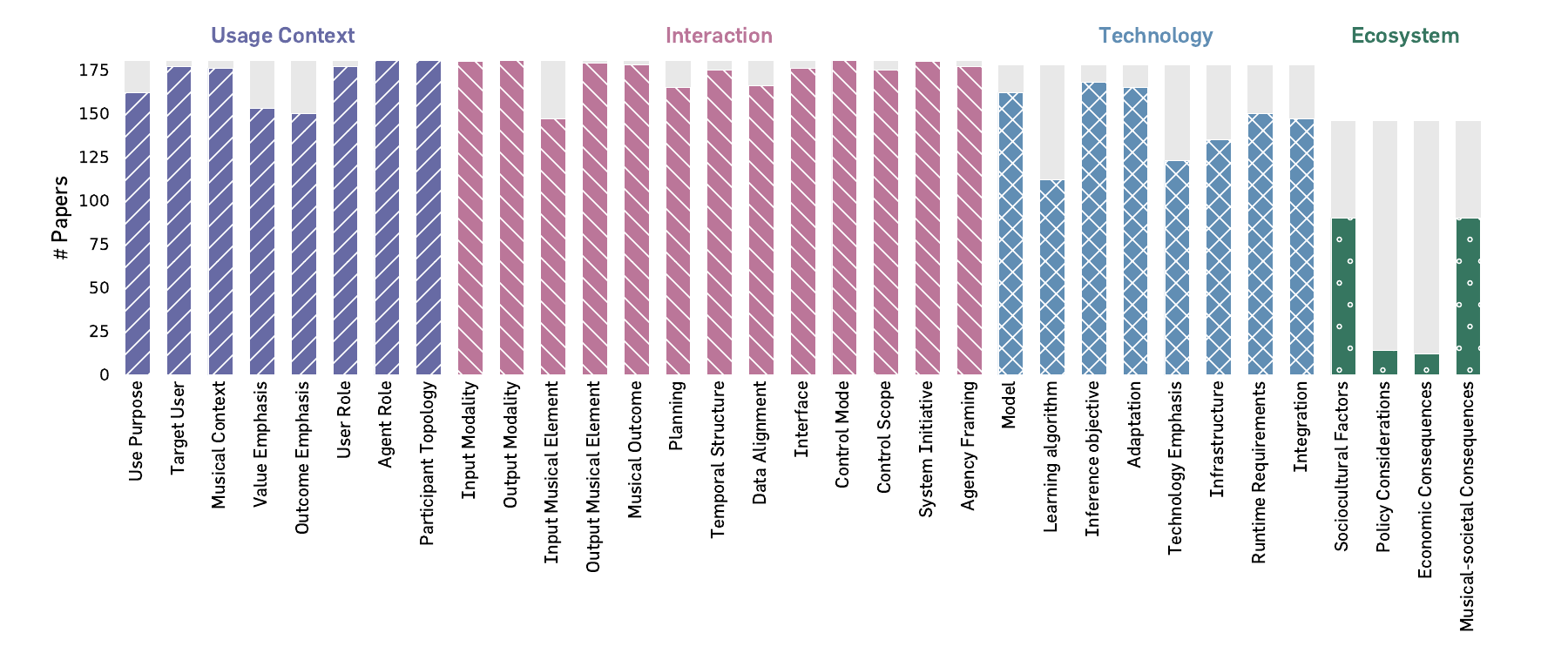}
  % \vspace{-.7em}
  \caption{
    Number of papers addressing each design space dimension. Gray bars mark the total papers associated with each aspect, and colored bars specify those addressing individual dimensions. 
    Dimensions in Ecosystem aspect receive comparatively little attention in existing live music agent systems.
  }
  % \vspace{-.5em}
  \Description{A vertical bar chart shows the number of papers coded under each dimension of the live music agents design space. Bars are grouped into four aspects: Usage Context (blue with diagonal stripes leaning right), Interaction (pink with diagonal stripes leaning left), Technology (blue crosshatch), and Ecosystem (green with dotted circles). Gray bars represent the total number of relevant papers for each aspect, while the colored patterned bars show counts for individual dimensions. Usage Context and Interaction dimensions are highly represented, Technology dimensions moderately, and Ecosystem dimensions are notably underrepresented, with very few papers coded for Policy Considerations and Economic Consequences.}
  \label{fig:discussion:dimension}
\end{figure*}

% \cd{one idea here: include sub bars (w/ a different visual texture) for each dimension reflecting the number of papers that were assigned multiple codes for this dimension. this is an important summary statistic that we can sneak in here with no additional space usage}

%% file: figures/trends_ai_model.tex
\begin{figure*}[t!]
  \includegraphics[width=\textwidth]{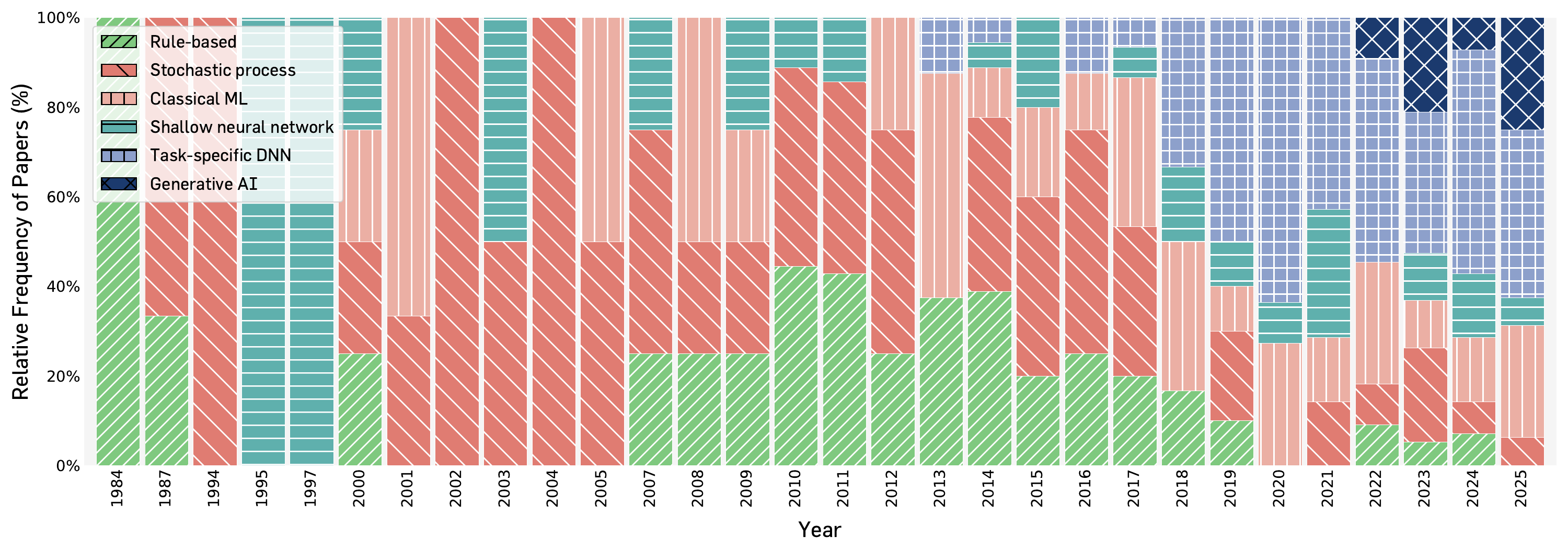}
  \vspace{-.5em}
  \caption{
  Relative frequency of different \technologycode{model} types used in live music agents over time (1984–2025). Rule-based, stochastic process, and shallow neural network approaches gradually decline, while task-specific deep neural networks (DNNs) gain attention after 2012. Since 2022, generative AI methods have emerged, though they represent only a small proportion.
  }
  \vspace{-.5em}
  \Description{A stacked bar chart shows the relative frequency of different computational models used in live music agent systems from 1984 to 2025. The x-axis represents years, and the y-axis represents percentage (0–100\%). Each bar is divided into categories: rule-based (green), stochastic process (red diagonal), classical ML (peach vertical), shallow neural networks (teal horizontal), task-specific DNNs (blue grid), and generative AI (dark blue crosshatch). Early systems (1980s–2000s) are dominated by rule-based and stochastic process methods. From 2012 onward, task-specific DNNs become increasingly common. Since 2022, generative AI appears as a new category, though still a small fraction compared to other models.}
  \label{fig:discussion:aimodel}
\end{figure*}

%% file: figures/design_1.tex
\begin{figure*}[t!]
  \includegraphics[width=.75\linewidth]{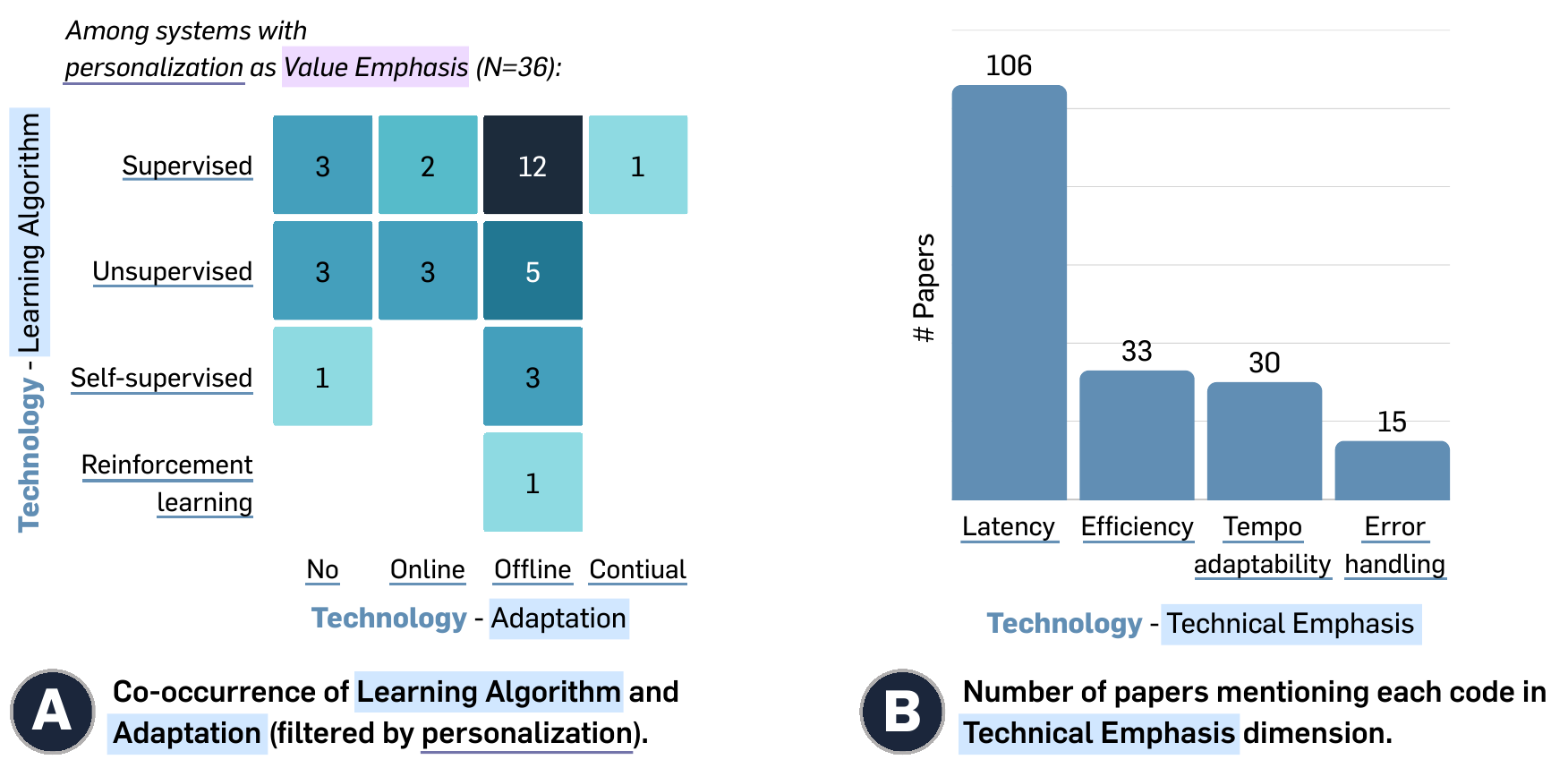}
  \vspace{-.7em}
  \caption{
  \rev{
Insights from the design space motivating new design directions for (A) customizing live music agents (B) and incorporating contingency as a design material. 
(A) Co-occurrence patterns between \technologydim{learning algorithm} and \technologydim{adaptation} show that \technologycode{online} and \technologycode{continual} adaptation remain rarely explored among systems emphasizing \contextcode{personalization}. These systems rely primarily on \technologycode{supervised} learning, with almost no use of \technologycode{reinforcement learning}.
(B) \technologycode{Error handling} is the least emphasized \technologydim{technical emphasis} in designing live music agents, especially when contrasted with the frequent attention to \technologycode{latency}.
}
  }
  \vspace{-.5em}
  \Description{
  This figure presents two analyses from the Technology aspect of the design space. The left heatmap shows how different learning algorithms co-occur with adaptation strategies among systems that emphasize personalization, with most using supervised learning and offline adaptation, and very few using online or continual adaptation. The right bar chart shows how frequently each technical emphasis code appears across papers, with latency mentioned far more often than efficiency, tempo adaptability, or error handling.
  }
  \label{fig:discussion:design1}
\end{figure*}

%% file: figures/design_2.tex
\begin{figure}[ht]
  \includegraphics[width=\linewidth]{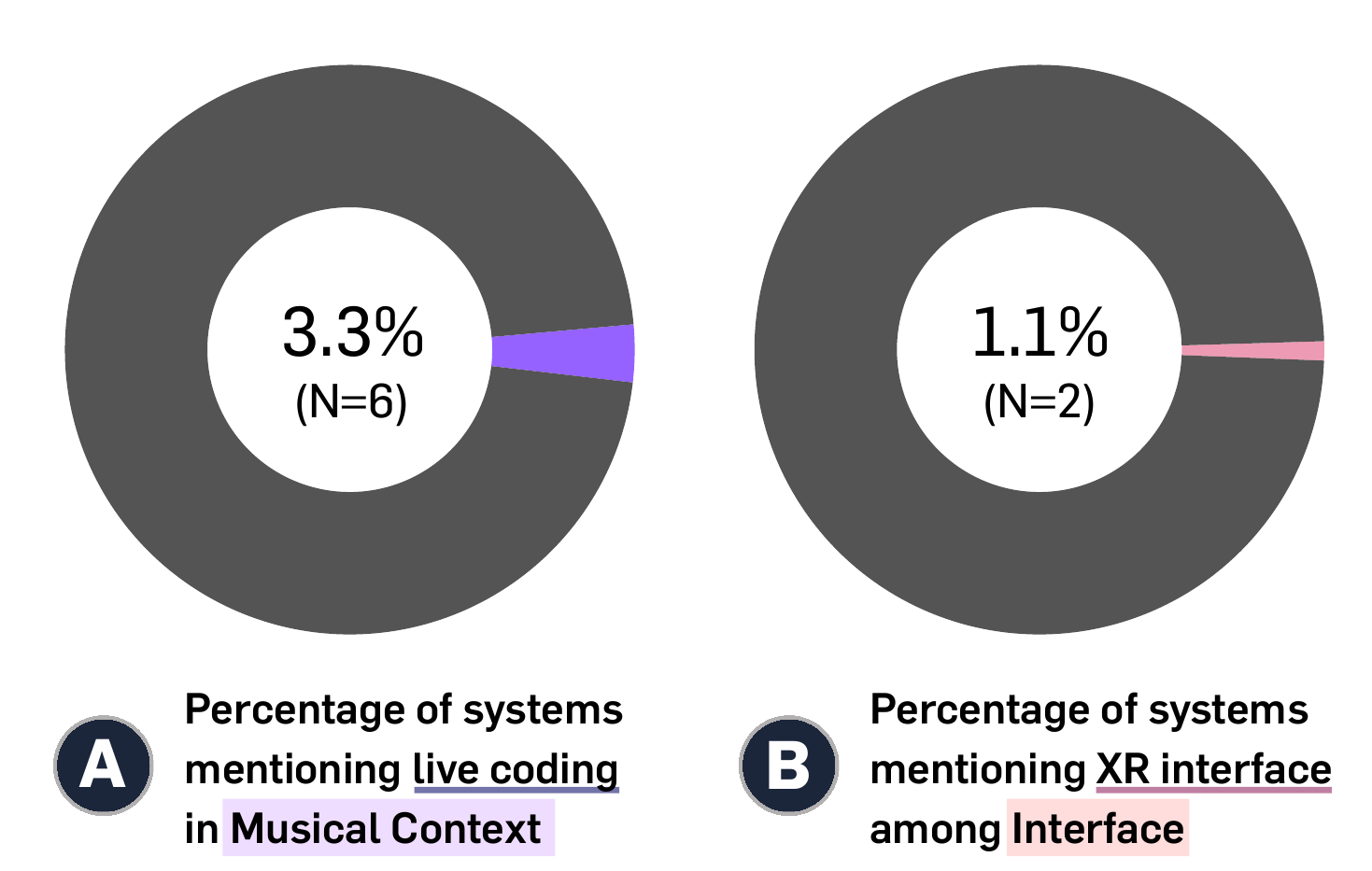}
  % \vspace{-.7em}
  \caption{
  \rev{
Insights from the design space motivating new design directions for (A) live coding agents and (B) extended-reality (XR)–equipped live music agents. 
(A) \contextcode{live coding} appears in only 3.3\% of systems, suggesting opportunities to explore agents that collaborate through code-based performance practices. 
(B) XR-based \interactiondim{interface} designs are even rarer (1.1\%), indicating a largely untapped space for spatial, embodied, or immersive live music interactions.
}
  }
  % \vspace{-.5em}
  \Description{
  This figure shows two donut charts summarizing how often certain design elements appear in surveyed systems. The left chart indicates that only 3.3\% of systems mention live coding in their musical context. The right chart shows that XR-based interfaces are even rarer, appearing in only 1.1\% of systems.
  }
  \label{fig:discussion:design2}
\end{figure}

%% file: sections/08_conclusion.tex
\section{Conclusion}
We introduce a design space for live musical agents that consolidates research across HCI, Computer Music, and AI. 
By analyzing \numtotalsystem systems from HCI, AI, and Computer Music literature and online videos, we identify \numdim dimensions and \numcode codes that characterize how these systems operate across usage context, interaction, technology, and ecosystem aspects. 
By highlighting recurring patterns as well as underexplored opportunities, our design space lays the groundwork for interdisciplinary collaboration and the creation of more compelling and responsible systems for live musical interaction. 

%% file: sections/_appendix.tex
\appendix 

% A
\section{Papers and Videos}
\label{appendix:data}

%% A-1
\subsection{Venues}
\label{appendix:data:venues}
To keep the number of papers reasonable, we decided to focus on
papers from the following venues in HCI, AI, and Computer Music. 
The papers were retrieved using ACM Digital Library, OpenReview, and official proceedings sites for Computer Music conferences.
We included all their paper tracks 
(e.g., CHI Late-Breaking Work and ACL Findings), but excluded
workshops and demos.

\begin{itemize}
    \item \textbf{HCI: } CHI, UIST, IUI, C\&C, DIS, CSCW  
    \item \textbf{AI: } NeurIPS, ICML, ICLR
    \item \textbf{Computer Music: } ISMIR, NIME, ICMC
\end{itemize}

%% A-2
\subsection{Search Queries}
\label{appendix:data:queries}

%%% A-2-1
\subsubsection{\textbf{Literature}}
\label{appendix:data:queries:papers}

The search query logic used to retrieve papers is listed below. We include morphological variants (e.g., \texttt{improvis*}, \texttt{jam*}) and both hyphenated/unhyphenated forms of \emph{real-time}. 
For Computer Music venues, we omit the \texttt{music} keyword since it is inherently implied.
Where fielded search is unavailable, we implemented a rate-limited crawler to collect proceedings and applied the same Boolean filter to parsed titles/abstracts.

\begin{framed}
\begingroup\small
\begin{verbatim}
( Title:(improvis* OR "real-time" OR live OR jam*)
  OR Abstract:(improvis* OR "real-time" OR live OR jam*) )
AND
( Title:(music) OR Abstract:(music) )
AND
( Title:(agent OR AI OR "artificial intelligence"
        OR intelligent OR generative OR autonomous)
  OR Abstract:(agent OR AI OR "artificial intelligence"
              OR intelligent OR generative OR autonomous) )
\end{verbatim}
\endgroup
\end{framed}

%%% A-2-2
\subsubsection{\textbf{Video}}
\label{appendix:data:queries:videos}

We list the search queries we used to collect videos on YouTube: 
`live music performance with AI,' `AI music improvisation,' `AI jam session,' `live music with neural networks,' `real-time music generation,' 'improvisation with ai,' `real-time AI synthesizer'

%% A-3
\subsection{Collected Resources}
\label{appendix:data:resources}

We list the collected systems according to their source: key papers, retrieved papers, and online videos.

% \cd{Can we ask an LLM to format these into a table for easier perusing? Columns like Type (Key, Crawled, Video), First Author (their last name), Year, Venue, Citation}
% \yewon{might submit as a supplementary resource}

\subsubsection{\textbf{Key Papers:}}
\label{appendix:data:resources:keypapers}
\citep{voyager,reflexivelooper,shimon,pianogenie,bachduet,realchords,spiremuse,genjam,somax2,rave,vampnet,jamfactory,vercoe1984synthetic,mimi4x,harmonix,fiebrink2020reflections,aiterity,assemblage,jambot,jiang2025improvised,shier2024real,reactiveaccompanist,bob,raphael2001bayesian,continuator,improtek,inasilentway,rlduet,songdriver,musicfxdj,dannenberg1984line,kondak2016active,fiebrink2010wekinator,privato2024stacco}

\subsubsection{\textbf{Retrieved Papers:}}
\label{appendix:data:resources:crawledpapers}
\citep{10.1145/1254960.1254990,marley2015gestroviser,turczan2019scale,sioros2011automatic,roberts2013enabling,naess2019physical,van2012mapping,tsiros2020towards,kobayashi2020exsampling,nash2020crowd,filandrianos2020brainwaves,shier2025designing,nime2025_54,shepardson2025evolving,evans2025repurposing,shepardson2024tungnaa,kobayashi2023improvise+,easthope2023snakesynth,marchini2017rethinking,mitchell2011soundgrasp,martin2019interactive,kapur2007integrating,xia2017improvised,godbehere2008wearable,donnarumma2012music,collins2010musical,byuksel2015braahms,jiang2020when,brochec2023toward,mcauliffe2023stochgran,polychronopoulos2022lem,dooley2021mytrombone,yakura2021generate,wang2019drumductor,gioti2019imitation,TerasakiTH17,Lepri16,LeffueK16,Gresham-Lancaster15,Fay15-1,CullimoreHG14,Ando14-2,BarateHL14,Kaliakatsos-Papakostas14,Hadjileontiadis14,VamvakousisR14,SarwateS14,EvansFGMO14,Spicer14-0,DealS13,KrekovicB12,Hoadley12,Yee-King11,AshS11,BakhtB09,Collins08a,WeinbergGRR07-0,Young07-2,JehanMF02,CemgilK01-0,HothkerH00,MoralesMW05,khallaghi2025squishysonics,arai2023timtoshape,wicaksono2022tapis,waite2016church,tahiroglu2016nonintrusive,smith2012unsupervised,delalez2017vokinesis,knotts2021algorithmic,takase2020support,stefani2024esteso,davies2014improvasher,eaton2014the,scurto2017shaping,hashida2007a,legroux2007vrroboser,gimenes2007musicianship,endo2012tweet,derbinsky2011cognitive,brown2018interacting,benasher2013toward,klooster2014in,tralie2024concatenatorbayesianapproachreal,franklin2024robocajon,tsuruoka2019soundwriter,anders2018populous,bing2017a,wang2025ai,huang2012melodicbrush,whalley2010generative,tahiroug2013pesi,vogl2017an,quintas2010glitch,NIME22_25,erdem2020raw,caren2025melia,lionetti2024muscleguided,davis2023etudb,lucas2023a,keatch2014an,ishida2004ism,hoskinson2003realtime,hantrakul2018gesturernn,juregui2019automatic,gaye2004in,ferguson2009an,eigenfeldt2008an,dahl2011tweetdreams,cherston2016musician,brown2010network,stark2014sound,savery2024collaborationrobotsinterfaceshumans,tatar2018revive,kitahara2017jamsketch,Schedel2021RhumbLine,proctor2020a,wu2025gesturedriven,borovik2023realtime,kitani2010improvgenerator,hamano2013generating,visi2017a,beyls2018motivated,almeida2019amigo,martelloni2023realtimepercussivetechniquerecognition,calmuswaves,oscar,roger1988new,interactive,dannerberg1987following, Xambo2021Live}

\begin{itemize}
    \item Duplicate paper pairs analyzed in combination: \citep{realchords, realjam}, \citep{spiremuse, thelle2023co}, \citep{bob, thom2001machine}
\end{itemize}

\subsubsection{\textbf{Videos:}}
\label{appendix:data:resources:videos}
\citep{dadabots,fruitgenie,semilla,magnetologues,sophtar,umwelt,bigyuki,trumpetai,iuai,jointai,pianoai,harmonizing,melbourne,improvai,aidj,magenta,strudel,jamai,livecoding,bionichaos,commentsai,improv,intersections,uncannylove,memachine,rave2,neurorack, Ircam}

\subsection{Positionality Statement}
The first through sixth authors primarily conducted the coding of papers and videos. The first three authors are PhD students with research and publication experience in HCI and music AI. 
The fourth author is a PhD student with publications in HCI and extensive experience in qualitative analysis and design space creation. The fifth author is a PhD student with research and publication experience in computer music. The sixth author is a senior undergraduate student with prior research experience in computer music; before coding, relevant HCI and qualitative research literature was shared, and the coding methods and design space were discussed in depth with the first and fifth authors. The overall coding process and design space development were further discussed and reviewed in weekly meetings with faculty researchers who are experts in computer music, AI, and HCI.

\rev{

\section{Utilization of the Living Artifact}
\label{appendix:webtutorial}

Here we demonstrate how our living artifact\footnote{\greenurl{https://live-music-agents.github.io/}} could be utilized to explore our design space.

\subsection{Explorer View}
The explorer view is the primary entry point for browsing annotated systems. 
It enables users to search for systems by title or interactively filter the dataset through a structured interface. 
Within this view, users can examine individual systems in depth and compare design choices across multiple systems (Section~\ref{sec:scenario:insights:2}). The case studies in Section~\ref{sec:designspace} were conducted using this view.

\begin{itemize}
\item Users select one of four high-level aspects (Usage Context, Interaction, Technology, Ecosystem) to reveal its dimensions.
\item Clicking on a dimension opens a dropdown list of codes; selecting codes acts as an inclusive filter (filtering for systems containing \textit{all} selected tags).
\item The main panel lists matching systems. Clicking a system card opens a detail page with metadata, video demonstrations (when available), and the complete set of coded tags.
\end{itemize}

\subsection{Frequency View}
The frequency view 
provides a quantitative overview of 
how systems are distributed across the design space, highlighting trending and underrepresented areas (Section~\ref{sec:scenario:insights:1}). 
The analyses shown in Figure~\ref{fig:discussion:dimension} (Section~\ref{sec:discussion:insights:ethical}), which summarize the number of papers addressing each dimension, were generated using this view.
The discussions in Sections~\ref{sec:discussion:design:contingency}–\ref{sec:discussion:design:xr} also draw on insights derived from this view.

\begin{itemize}
\item At the top level, users can view the global coverage of the four main aspects.
\item Selecting an aspect displays a bar chart of its constituent dimensions. 
\item Clicking on a specific dimension bar drills down to the code level (e.g., clicking \interactiondim{Input Modality} reveals the split between \interactioncode{audio waveform}, \interactioncode{symbolic music}, and others).
\item Clicking a specific code bar retrieves the list of systems tagged with that code.
\end{itemize}

\subsection{Trends View}
The trends view supports 
analysis of historical trajectories and emerging patterns in the field. 
It allows users to examine how dimensions and codes evolve over time, revealing shifts in research focus, the emergence of new interaction paradigms, and the adoption of specific technologies (Section~\ref{sec:scenario:insights:1}).
The insights shown in Figure~\ref{fig:discussion:numpapers} and Figure~\ref{fig:discussion:aimodel}, 
as discussed in Sections~\ref{sec:discussion:insights:ethical} and~\ref{sec:discussion:insights:aitrend}, were generated using this view.

\begin{itemize}
\item Users can toggle the `Total Systems' overlay to view the overall growth of the field against specific trends.
\item The sidebar allows users to plot multiple codes simultaneously to compare their prevalence over time (e.g., comparing the trajectory of \technologycode{Task-specific DNN} vs. \technologycode{Generative AI}.
\item The `Year Bin' adjusts the temporal resolution (e.g., 3-year bins to smooth out volatility in data with lower sample sizes).
\item We provide `Presets' (e.g., \textit{"Emerging Musical Contexts"}) that automatically configure the chart to help users understand how to use this view effectively.
\end{itemize}

\subsection{Co-occurrence View}
The co-occurrence view visualizes relationships between 
design dimensions using a matrix heatmap, 
revealing clusters of design choices that frequently appear together.
By exploring these pairwise relationships, users can identify common patterns as well as underexplored combinations of codes (Section~\ref{sec:scenario:newdesign:1}).
This view informed the analysis in Section~\ref{sec:discussion:design:custom}, where we examined the co-occurrence between \technologydim{Learning Algorithm} and \technologydim{Adaptation}. 

\begin{itemize}
\item Users assign dimensions to the vertical and horizontal axes (e.g., mapping \contextdim{User Role} against \contextdim{Agent Role}).
\item The intensity of each cell indicates the number of systems sharing that specific pair of codes. 
\item A `Filter Data' section in the sidebar enables conditional analysis, e.g., a user can filter the dataset by the code \contextcode{jazz} before generating a heatmap, effectively asking: `\textit{In Jazz systems, how does Input Modality correlate with Agent Role?}'
\item Similar to the Trends view, `Selection Presets' are provided to help users understand how to use this view effectively.
\end{itemize}

\section{Design Space}
\label{appendix:tables}
Below, we provide summary tables for the design spaces presented in Section~\ref{sec:designspace}.
Following prior work~\citep{dsiiwa, dsida, MacLean01091991}, 
each dimension is framed as a question, with codes serving as potential answers. 
Tables~\ref{tab:task-designspace-1}–\ref{tab:task-designspace-2} present the \contextaspect{Usage Context} aspect;
Tables~\ref{tab:interaction-designspace-1}–\ref{tab:interaction-designspace-3} present the \interactionaspect{Interaction} aspect;
Tables~\ref{tab:technology-designspace-1}–\ref{tab:technology-designspace-2} summarize the \technologyaspect{Technology} aspect;
and Table~\ref{tab:ecosystem-designspace} summarizes the \ecosystemaspect{Ecosystem} aspect.

\input{figures/appendix_visualizer}
\FloatBarrier

\input{tables/context-design-space}
\input{tables/interaction-design-space}
\input{tables/technology-design-space}
\input{tables/environment-design-space}

}

%% file: figures/appendix_visualizer.tex
\begin{figure*}[ht!]
  \includegraphics[width=0.95\textwidth]{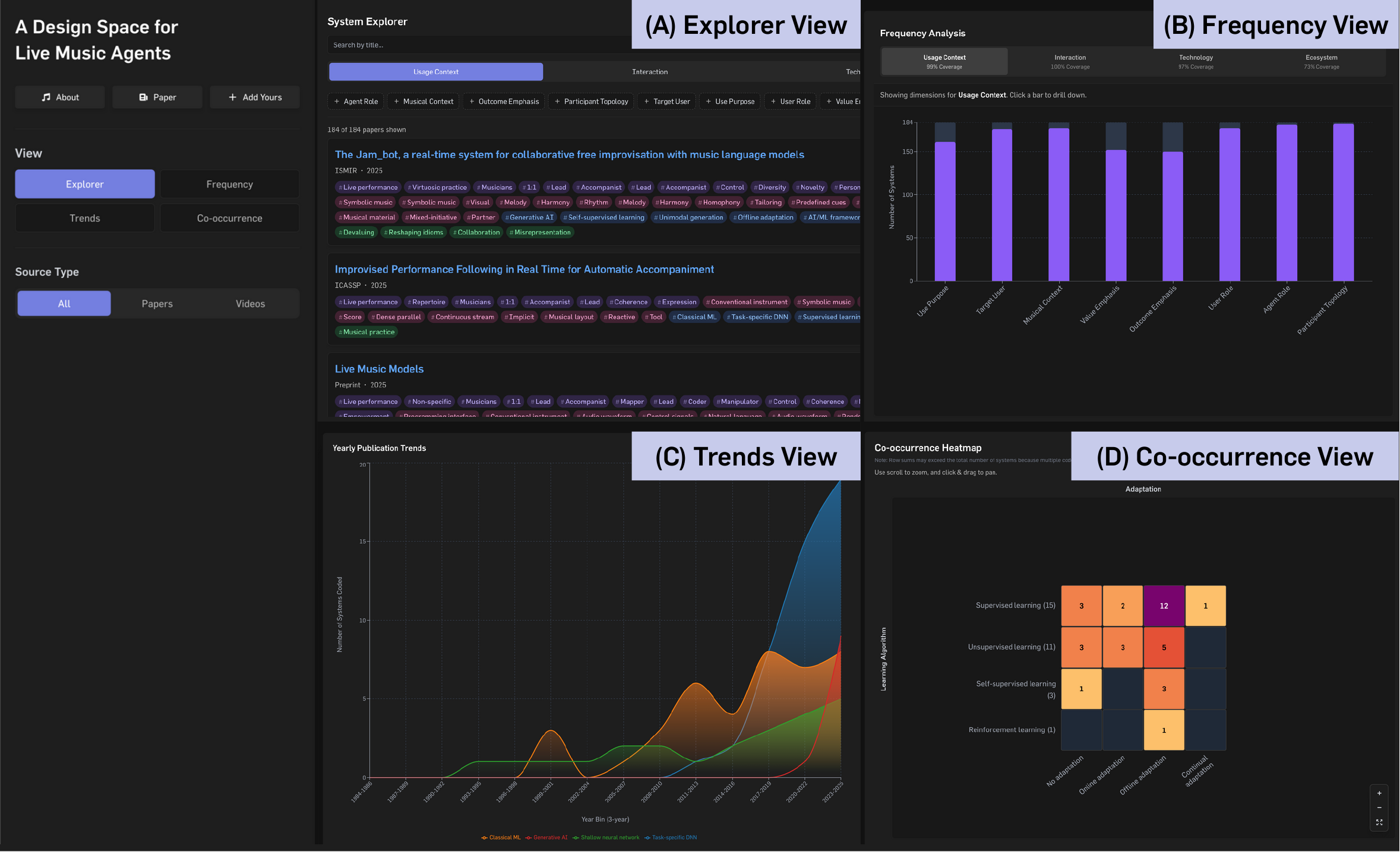}
  % \vspace{-.7em}
  \caption{
  \rev{
 Overview of the four main views in our interactive design-space visualizer.
(A) Explorer View: browse, filter, and inspect individual systems and their coded attributes.
(B) Frequency View: examine how frequently each dimension and code appears across the 184 systems.
(C) Trends View: analyze temporal trajectories and emerging patterns across dimensions and codes.
(D) Co-occurrence View: visualize pairwise relationships between dimensions to reveal common and underexplored design combinations.
  }
  }
  % \vspace{5em}
  \Description{
  Overview of our interactive web visualizer, which supports four complementary modes of analysis: (A) the Explorer View for browsing, searching, and filtering annotated systems; (B) the Frequency View for showing how often each design code appears across the dataset; (C) the Trends View for visualizing temporal patterns in system design choices; and (D) the Co-occurrence View for examining how codes from different dimensions appear together.
  }
  \label{fig:visualizer}
\end{figure*}

%% file: tables/context-design-space.tex
\aspect{\contextaspect{Usage Context}}{tab:task-designspace-1}{tab:task-designspace-2}{\aspUseContextNum}{\aspUseContextPct}{
    \dimension{Use Purpose}{Use Purpose}{mutedviolet}{Which musical activity is the system designed for?}{\dimUseContextUsePurposeNum}{\dimUseContextUsePurposePct}{
    
        \code{Live performance}{Live performance}{
        %Time-sensitive
        Real-time
        music performance intended for an audience
        }{\citep{shimon}}{\textbf{\codUseContextUsePurposeLivePerformanceNum}}{\textbf{\codUseContextUsePurposeLivePerformancePct}}
        
        \code{Composition}{Composition}{Open-ended musical exploration for brainstorming and composition}{\citep{spiremuse}}{\codUseContextUsePurposeCompositionNum}{\codUseContextUsePurposeCompositionPct}

        \code{Entertainment}{
        Recreation
        % CHRIS: slight preference for ``Recreation'' or ``Play``? Entertainment feels overloaded w/ the Entertainment *Industry*
        }{Recreational musical activity serving entertainment purposes}{\citep{harmonix}}{\codUseContextUsePurposeEntertainmentNum}{\codUseContextUsePurposeEntertainmentPct}
        
        \code{Skill acquisition}{Skill acquisition}{Interactive practice or learning activities for developing musical skills}{\citep{almeida2019amigo}}{\codUseContextUsePurposeSkillAcquisitionNum}{\codUseContextUsePurposeSkillAcquisitionPct}
        
        \code{Art installation}{Art installation}{A site-specific, interactive 
        %sound 
        music 
        environment engaging visitors}{\citep{kobayashi2023improvise+}}{\codUseContextUsePurposeArtInstallationNum}{\codUseContextUsePurposeArtInstallationPct}
    }

    \dimension{Target User}{Target User}{mutedviolet}{Who is the intended user of the system?}{\dimUseContextTargetUserNum}{\dimUseContextTargetUserPct}{

    \code{Musicians}{Musicians}{Skilled users who perform, compose, or produce music}{\citep{inasilentway}}{\textbf{\codUseContextTargetUserMusiciansNum}}{\textbf{\codUseContextTargetUserMusiciansPct}}
    
    \code{Novice users}{Novice users}{Early-stage users with minimal musical background}{\citep{pianogenie}}{\codUseContextTargetUserNoviceUsersNum}{\codUseContextTargetUserNoviceUsersPct}
    
    \code{Audience}{Audience}{Participants who experience or observe the performance}{\citep{nash2020crowd}}{\codUseContextTargetUserAudienceNum}{\codUseContextTargetUserAudiencePct}
    
    }
    
    \dimension{Musical Context}{Musical Context}{mutedviolet}{Which musical style 
    %and norms 
    or practice 
    shape the context in which the system operates?}{\dimUseContextMusicalContextNum}{\dimUseContextMusicalContextPct}{

        \code{Non-specific}{Non-specific}{Broad, unspecified musical context without reference to a particular style}{\citep{musicfxdj}}{\textbf{\codUseContextMusicalContextNonSpecificNum}}{\textbf{\codUseContextMusicalContextNonSpecificPct}}

        \code{Electronic music}{Electronic music}{Group of music genres employing electronic musical instruments}{\citep{vogl2017an}}{\codUseContextMusicalContextElectronicMusicNum}{\codUseContextMusicalContextElectronicMusicPct} % in its creation

        \code{New music}{New music}{Contemporary experimental practices emphasizing innovation}{\citep{fiebrink2020reflections}}{\codUseContextMusicalContextNewMusicNum}{\codUseContextMusicalContextNewMusicPct}
    
        \code{Jazz}{Jazz}{
        %African-American rooted music 
        Music tradition 
        marked by improvisation, % and 
        %complex composition, 
        % complexity, 
        African-American roots
        }{\citep{voyager}}{\codUseContextMusicalContextJazzImprovisationNum}{\codUseContextMusicalContextJazzImprovisationPct}

        % \code{Electroacoustic music}{Electroacoustic music}{Music that integrates electronically produced sound with acoustic sound sources}{\citep{Lepri16}}{\codUseContextMusicalContextElectroacousticMusicNum}{\codUseContextMusicalContextElectroacousticMusicPct}
        
        \code{Western classical}{Western classical}{
        %Western 
        European 
        concert music between Baroque and Romantic periods}{\citep{rlduet}}{\codUseContextMusicalContextWesternClassicalNum}{\codUseContextMusicalContextWesternClassicalPct}
        
        \code{Popular music}{Popular music}{
        %Mass-mediated genres 
        Widely consumed modern Western music genres 
        such as pop, rock, and hip-hop}{\citep{marchini2017rethinking}}{\codUseContextMusicalContextPopularMusicNum}{\codUseContextMusicalContextPopularMusicPct}

        \code{Score-based}{
        Repertoire
        %Score-based
        % CHRIS: Maybe ``Repertoire''?
        }{Notation-driven performance of a fixed composition}{\citep{TerasakiTH17}}{\codUseContextMusicalContextScoreBasedNum}{\codUseContextMusicalContextScoreBasedPct}
        
        \code{Traditional music}{Traditional music}{Musical practices reflecting the musical heritage of particular culture
        % CHRIS: Feels more culture-centric than community-centric. E.g. punk music is community-centric, but it is not traditional music
        %or community
        }{\citep{kapur2007integrating}}{\codUseContextMusicalContextTraditionalMusicNum}{\codUseContextMusicalContextTraditionalMusicPct}
        
        \code{Live coding}{Live coding}{Real-time coding as a performative act of music making}{\citep{strudel}}{\codUseContextMusicalContextLiveCodingNum}{\codUseContextMusicalContextLiveCodingPct}
        
        \code{Virtuosic practice}{Virtuosic practice}{
        % CHRIS: is this where Jordan Rudess ended up? feels like an arbitrary category relative to the others, but not opposed
        A technique-centric practice emphasizing speed, precision, and control
        }{\citep{jambot}}{\codUseContextMusicalContextVirtuosicPracticeNum}{\codUseContextMusicalContextVirtuosicPracticePct}

    }
    
\dimension{User Preference}{
    %User Preference
    % Desiderata
    Value Emphasis}{mutedviolet}{What human values or musical desiderata are emphasized in the design of the agent or system?}{\dimUseContextUserPreferenceNum}{\dimUseContextUserPreferencePct}{
    
        \code{Control}{Control}{%The extent to which 
        User can directly and predictably steer agent output}{\citep{dooley2021mytrombone}}{\textbf{\codUseContextUserPreferenceControlNum}}{\textbf{\codUseContextUserPreferenceControlPct}}

        \code{Coherence}{Coherence}{
        %The extent to which 
        System's output is musically consistent
        %, structured, 
        and aligned with user expectations}{\citep{realchords}}{\codUseContextUserPreferenceCoherenceNum}{\codUseContextUserPreferenceCoherencePct}
        
        \code{Novelty}{Novelty}{
        %The extent to which 
        System produces novel or unexpected output}{\citep{privato2024stacco}}{\codUseContextUserPreferenceNoveltyNum}{\codUseContextUserPreferenceNoveltyPct}
        
        \code{Diversity}{Diversity}{
        %The extent to which 
        System produces varied outputs}{\citep{shier2025designing}}{\codUseContextUserPreferenceDiversityNum}{\codUseContextUserPreferenceDiversityPct}
        
        \code{Personalization}{Personalization}{
        %The extent to which 
        System adapts to the user's preferences, musical style, or interaction history}{\autocitep{jambot}}{\codUseContextUserPreferencePersonalizationNum}{\codUseContextUserPreferencePersonalizationPct}
        
    }
    
    \dimension{Expected impact}{
    %Expected impact
    Outcome Emphasis}{mutedviolet}{
    %What user impacts does the system enable?
    What musical outcomes are emphasized in the design of the agent or system?
    }{\dimUseContextExpectedImpactNum}{\dimUseContextExpectedImpactPct}{

        \code{Exploration}{Exploration}{Trying new musical directions or breaking habits}{\citep{shepardson2024tungnaa}}{\textbf{\codUseContextExpectedImpactExplorationNum}}{\textbf{\codUseContextExpectedImpactExplorationPct}}

        \code{Empowerment}{Empowerment}{
        %\parbox[t][2\baselineskip][t]{\linewidth}{Supports users with the ability to achieve beyond their limits \\ Enable performing in a way that they wouldn't have been able to do by themselves}
        Achieving outcomes beyond their limits
        % the ability to achieve beyond their limits, or enables performing in a way that they would not have been able to do alone
        }{\citep{pianogenie}}{\codUseContextExpectedImpactEmpowermentNum}{\codUseContextExpectedImpactEmpowermentPct}

        \code{Engagement}{Engagement}{Feeling involved in the musical experience}{\citep{fruitgenie}}{\codUseContextExpectedImpactEngagementNum}{\codUseContextExpectedImpactEngagementPct}
    
        \code{Expression}{Expression}{Conveying emotions, ideas, or identities through music}{\citep{caren2025melia}}{\codUseContextExpectedImpactExpressionNum}{\codUseContextExpectedImpactExpressionPct}

        \code{Immersion}{Immersion}{Uninterrupted musical engagement where users lose track of time}{\citep{lucas2023a}}{\codUseContextExpectedImpactImmersionNum}{\codUseContextExpectedImpactImmersionPct}

        \code{Reflection}{Reflection}{Reflecting on their musical practices or artistic vision}{\citep{jambot}}{\codUseContextExpectedImpactReflectionNum}{\codUseContextExpectedImpactReflectionPct}
        
        \code{Delegation}{Delegation}{Reducing cognitive effort by delegating tasks to the system}{\citep{byuksel2015braahms}}{\codUseContextExpectedImpactDelegationNum}{\codUseContextExpectedImpactDelegationPct}

    }
}

\aspect{\contextaspect{Usage Context}}{tab:task-designspace-2}{tab:task-designspace-1}{\aspUseContextNum}{\aspUseContextPct}{
    
    \dimension{User Role}{User Role}{mutedviolet}{What musical function does the human user perform in real time?}{\dimUseContextUserRoleNum}{\dimUseContextUserRolePct}{
    
        \code{Lead}{Lead}{Generates the primary musical material such as melody}{\citep{reactiveaccompanist}}{\textbf{\codUseContextUserRoleLeadNum}}{\textbf{\codUseContextUserRoleLeadPct}} %  or full track

        \code{Manipulator}{Manipulator}{Shapes musical output by adjusting parameters}{\citep{mcauliffe2023stochgran}}{\codUseContextUserRoleManipulatorNum}{\codUseContextUserRoleManipulatorPct} %  or values
        
        \code{Non-musical performer}{Non-musical performer}{Contribute to music through non-musical actions such as motion}{\citep{legroux2007vrroboser}}{\codUseContextUserRoleNonMusicalPerformerNum}{\codUseContextUserRoleNonMusicalPerformerPct}

        \code{Conductor}{Conductor}{Guides or coordinates 
        %agent's 
        % CHRIS: when possible, let's make the agent / user roles symmetric. human participants may conduct other human participants
        participant's 
        musical actions}{\citep{proctor2020a}}{\codUseContextUserRoleConductorNum}{\codUseContextUserRoleConductorPct}

        \code{Accompanist}{Accompanist}{Generates harmonic, rhythmic, or textural support to the lead}{\citep{reflexivelooper}}{\codUseContextUserRoleAccompanistNum}{\codUseContextUserRoleAccompanistPct}
    
        \code{Coder}{Coder}{Generates or manipulates music by writing code or prompt}{\citep{dadabots}}{\codUseContextUserRoleCoderNum}{\codUseContextUserRoleCoderPct}

        \code{Mixer}{Mixer}{Selects and/or blends musical materials (\eg samples) into a continuous flow}{\citep{nime2025_54}}{\codUseContextUserRoleMixerNum}{\codUseContextUserRoleMixerPct}

    }

    \dimension{Agent Role}{Agent Role}{mutedviolet}{What musical function does the agent 
    %play 
    perform  
    in real time?}{\dimUseContextAgentRoleNum}{\dimUseContextAgentRolePct}{

        \code{Accompanist}{Accompanist}{Generates harmonic, rhythmic, or textural support to the lead}{\citep{realchords}}{\textbf{\codUseContextAgentRoleAccompanistNum}}{\textbf{\codUseContextAgentRoleAccompanistPct}}

        \code{Mapper}{Mapper}{
        Translates non-musical or abstract input (e.g., text or gestures) into musical output
        }{\citep{calmuswaves}}{\codUseContextAgentRoleMapperNum}{\codUseContextAgentRoleMapperPct}
    
        \code{Lead}{Lead}{Generates the primary musical material such as a melody}{\citep{bob}}{\codUseContextAgentRoleLeadNum}{\codUseContextAgentRoleLeadPct}
        
        \code{Remixer}{Remixer}{Modifies 
        % existing 
        musical materials (e.g., through sound synthesis parameter control)}{\citep{davies2014improvasher}}{\codUseContextAgentRoleRemixerNum}{\codUseContextAgentRoleRemixerPct}

        \code{Controller}{Controller}{
        Translates 
        musically-intentional 
        % musically-grounded 
        % control signals 
        control signals 
        % (\eg button presses, knob turns)
        (\eg knob turns) 
        %(e.g., from a digital musical instrument) 
        % (\eg sound parameters, contours) 
        into musical output
        }{\citep{naess2019physical}}{\codUseContextAgentRoleControllerNum}{\codUseContextAgentRoleControllerPct}

        \code{Interpreter}{Interpreter}{Reflects 
        % or represents 
        musical content through other modalities (e.g., visualization)}{\citep{klooster2014in}}{\codUseContextAgentRoleInterpreterNum}{\codUseContextAgentRoleInterpreterPct}
        
        \code{Conductor}{Conductor}{Guides or coordinates participant's musical actions}{\citep{uncannylove}}{\codUseContextAgentRoleConductorNum}{\codUseContextAgentRoleConductorPct}
        
        \code{Evaluator}{Evaluator}{Provides 
        % evaluative 
        feedback to guide or refine the musical content}{\citep{tsiros2020towards}}{\codUseContextAgentRoleEvaluatorNum}{\codUseContextAgentRoleEvaluatorPct}
        
    }
    
    \dimension{Collaboration Structure}{
    %Collaboration Structure
    %Collaborative Topology
    Participant Topology}{mutedviolet}{
    %Who is involved and how are they grouped in the creative process?
    %How is creative control distributed among human and AI participants?
    What distribution of participating entities (humans : agents) is involved in live musical activities?
    }{\dimUseContextCollaborationStructureNum}{\dimUseContextCollaborationStructurePct}{
    
        % CHRIS: I prefer to drop ``collaboration'', not a strong pref though
        % CHRIS: using texttt for monospace alignment
        \code{1:1 collaboration}{\texttt{1:1}}{A 
        % single 
        human user interacts with an 
        % single 
        AI agent}{\citep{kondak2016active}}{\textbf{\codUseContextCollaborationStructureOneOneCollaborationNum}}{\textbf{\codUseContextCollaborationStructureOneOneCollaborationPct}}
        
        \code{N:1 collaboration}{\texttt{N:1}}{Multiple human users interact with 
        a 
        % single 
        shared AI agent}{\citep{WeinbergGRR07-0}}{\codUseContextCollaborationStructureNOneCollaborationNum}{\codUseContextCollaborationStructureNOneCollaborationPct}
        
        \code{N:N collaboration}{\texttt{N:N}}{Multiple human users interact with multiple AI agents
        % CHRIS: Not sure I understand what ``N pairs'' means? Also, doesn't feel like it needs clarification
        %(e.g., N pairs or mixed ensemble)
        }{\citep{Schedel2021RhumbLine}}{\codUseContextCollaborationStructureNNCollaborationNum}{\codUseContextCollaborationStructureNNCollaborationPct}

        \code{1:N collaboration}{\texttt{1:N}}{A 
        % single 
        human user interacts with multiple AI agents}{\citep{mimi4x}}{\codUseContextCollaborationStructureOneNCollaborationNum}{\codUseContextCollaborationStructureOneNCollaborationPct}

        \code{Agent only}{\texttt{0:N} (Agent only)}{
        %Only
        One or more AI agents 
        %generate 
        create
        music without direct human involvement}{\citep{genjam}}{\codUseContextCollaborationStructureAgentOnlyNum}{\codUseContextCollaborationStructureAgentOnlyPct}
        
    }
    
}

%% file: tables/interaction-design-space.tex
\aspectwithcaption{\interactionaspect{Interaction}}{tab:interaction-designspace-1}{tab:interaction-designspace-2,tab:interaction-designspace-3}{\aspInteractionNum}{\aspInteractionPct}{    
    \dimension{I/O modality}{I/O Modality}{mutedpink}{What types of music or non-music representations are used for system input or output?
    %in real time?
    % CHRIS: redundant IMO... we could append ``in real time'' to the defining question for all dimensions but we shouldn't
    }{\dimInteractionInputModalityNum}{\dimInteractionInputModalityPct}{

        \codeio{Raw audio waveform}{
        Audio waveform
        }{Continuous audio signals, 
        % without symbolic abstraction, 
        e.g., real-time microphone input
        % or generated waveforms
        }{\autocitep{rave}}{\textbf{\codInteractionInputModalityRawAudioWaveformNum}}{\textbf{\codInteractionInputModalityRawAudioWaveformPct}}{\autocitep{somax2}}{\textbf{\codInteractionOutputModalityRawAudioWaveformNum}}{\textbf{\codInteractionOutputModalityRawAudioWaveformPct}}

        \codeio{Symbolic music}{Symbolic music}{Structured music data such as notes, chords, rhythms (e.g., MIDI, MusicXML)}{\citep{jambot}}{\codInteractionInputModalitySymbolicMusicNum}{\codInteractionInputModalitySymbolicMusicPct}{\autocitep{bachduet}}{\codInteractionOutputModalitySymbolicMusicNum}{\codInteractionOutputModalitySymbolicMusicPct}
        
        \codeio{Control signals}{Control signals}{
        % Control signals 
        Signals 
        that modulate system generation (\eg sound parameters, trigger events)}{\citep{pianogenie}}{\codInteractionInputModalityControlSignalsNum}{\codInteractionInputModalityControlSignalsPct}{\autocitep{shier2024real}}{\codInteractionOutputModalityControlSignalsNum}{\codInteractionOutputModalityControlSignalsPct}

        \codeio{Gesture}{Gesture}{Physical 
        % or motion-based 
        % CHRIS: if we're grouping input and output together here, probably makes sense for the codes to not refer to inptu or output sepcifically (allow motion output as a potenital option)
        %input captured 
        information 
        via body movement, pose, or controller interaction (\eg 
        % Leap Motion, 
        Kinect)}{\autocitep{fiebrink2010wekinator}}{\codInteractionInputModalityGestureNum}{\codInteractionInputModalityGesturePct}{\autocitep{kapur2007integrating}}{\phantom{0}\codInteractionOutputModalityGestureNum\phantom{0}}{\codInteractionOutputModalityGesturePct}

        \codeio{Visual}{Visual}{Graphical 
        % or visual 
        representation of music (e.g., 
        % real-time 
        spectrograms, note displays, waveform plots)}{\autocitep{kitahara2017jamsketch}}{\codInteractionInputModalityVisualNum}{\codInteractionInputModalityVisualPct}{\autocitep{mimi4x}}{\phantom{0}\codInteractionOutputModalityVisualNum\phantom{0}}{\codInteractionOutputModalityVisualPct}

        \codeio{Natural language}{Natural language}{Linguistic (text or spoken) data}{\citep{musicfxdj}}{\codInteractionInputModalityNaturalLanguageNum}{\codInteractionInputModalityNaturalLanguagePct}{\autocitep{harmonizing}}{\phantom{0}\codInteractionOutputModalityNaturalLanguageNum\phantom{0}}{\codInteractionOutputModalityNaturalLanguagePct}

        \codeio{Physiological data}{Physiological data}{Physiological (e.g., EEG) readings from on-body devices}{\autocitep{hamano2013generating}}{\codInteractionInputModalityOnBodySensorDataNum}{\codInteractionInputModalityOnBodySensorDataPct}{-}{\phantom{0}\codInteractionOutputModalityOnBodySensorDataNum\phantom{0}}{\codInteractionOutputModalityOnBodySensorDataPct}
                
        \codeio{Exogenous sensor data}{Exogenous sensor data}{External, offstage sensor readings
        % CHRIS: this one feels oddly specific, though I'm not sure what it might be referring to. can we try to redefine it in a way that could be interpreted by someone just reading the design space, rather than a corresponding paper?
        }{\autocitep{nash2020crowd}}{\phantom{0}\codInteractionInputModalityExogenousSensorDataNum\phantom{0}}{\codInteractionInputModalityExogenousSensorDataPct}{-}{\codInteractionOutputModalityExogenousSensorDataNum}{\codInteractionOutputModalityExogenousSensorDataPct}

        \codeiolast{Programming language}{Programming language}{
        Programming languages for music making (\eg Tidal Cycles)
        }{\autocitep{livecoding}}{\phantom{0}\codInteractionInputModalityProgrammingLanguageNum\phantom{0}}{\codInteractionInputModalityProgrammingLanguagePct}{\autocitep{strudel}}{\codInteractionOutputModalityProgrammingLanguageNum}{\codInteractionOutputModalityProgrammingLanguagePct}

    }

    \dimension{I/O musical element}{I/O Musical Element}{mutedpink}{What musical elements does the system receive as input or generate as output?}{\dimInteractionInputMusicalElementNum}{\dimInteractionInputMusicalElementPct}{

        \codeio{Melody}{Melody}{A sequence of musical notes forming a tune}{\autocitep{realchords}}{\textbf{\codInteractionInputMusicalElementMelodyNum}}{\textbf{\codInteractionInputMusicalElementMelodyPct}}{\autocitep{BarateHL14}}{\textbf{\codInteractionOutputMusicalElementMelodyNum}}{\textbf{\codInteractionOutputMusicalElementMelodyPct}}
        
        \codeio{Rhythm}{Rhythm}{Pattern of note durations, timing, and beat emphasis in the music / percussion}{\autocitep{improtek}}{\codInteractionInputMusicalElementRhythmNum}{\codInteractionInputMusicalElementRhythmPct}{\autocitep{harmonix}}{\codInteractionOutputMusicalElementRhythmNum}{\codInteractionOutputMusicalElementRhythmPct}

        \codeio{Sound texture}{Sound texture}{Non-melodic sonic materials (\eg vocalizations, soundscapes, sonic gestures)}{\autocitep{caren2025melia}}{\codInteractionInputMusicalElementSoundTextureNum}{\codInteractionInputMusicalElementSoundTexturePct}{\autocitep{godbehere2008wearable}}{\codInteractionOutputMusicalElementSoundTextureNum}{\codInteractionOutputMusicalElementSoundTexturePct}

        \codeio{Harmony}{Harmony}{A set of simultaneous notes that form chords or harmonic progressions}{\autocitep{marchini2017rethinking}}{\codInteractionInputMusicalElementHarmonyNum}{\codInteractionInputMusicalElementHarmonyPct}{\autocitep{bing2017a}}{\codInteractionOutputMusicalElementHarmonyNum}{\codInteractionOutputMusicalElementHarmonyPct}

        \codeio{Timbre}{Timbre}{Spectral/timbral qualities distinguishing sources}{\autocitep{shepardson2025evolving}}{\codInteractionInputMusicalElementTimbreNum}{\codInteractionInputMusicalElementTimbrePct}{\autocitep{hoskinson2003realtime}}{\codInteractionOutputMusicalElementTimbreNum}{\codInteractionOutputMusicalElementTimbrePct}
        
        \codeio{Control parameters}{Control parameters}{Non-note controls for tempo, 
        % meter, 
        dynamics, 
        spatialization, 
        and FX (\eg gain level, gating)}{\autocitep{evans2025repurposing}}{\codInteractionInputMusicalElementControlParametersNum}{\codInteractionInputMusicalElementControlParametersPct}{\autocitep{hantrakul2018gesturernn}}{\codInteractionOutputMusicalElementControlParametersNum}{\codInteractionOutputMusicalElementControlParametersPct}
        
        \codeio{Rendered track}{
        %Rendered track
        Multitrack music
        }{An audio rendering 
        that integrates 
        multiple musical tracks into a unified recording
        }{\autocitep{musicfxdj}}{\codInteractionInputMusicalElementRenderedTrackNum}{\codInteractionInputMusicalElementRenderedTrackPct}{\autocitep{whalley2010generative}}{\phantom{0}\codInteractionOutputMusicalElementRenderedTrackNum\phantom{0}}{\codInteractionOutputMusicalElementRenderedTrackPct}

        \codeiolast{Lyrics}{Lyrics}{Linguistic content guiding music}{\autocitep{shepardson2024tungnaa}}{\phantom{0}\codInteractionInputMusicalElementLyricsNum\phantom{0}}{\codInteractionInputMusicalElementLyricsPct}{\autocitep{harmonizing}}{\codInteractionOutputMusicalElementLyricsNum}{\codInteractionOutputMusicalElementLyricsPct}
        
    }
}

\aspect{\interactionaspect{Interaction}}{tab:interaction-designspace-2}{tab:interaction-designspace-1,tab:interaction-designspace-3}{\aspInteractionNum}{\aspInteractionPct}{
    \dimension{Musical outcome}{Musical Outcome}{mutedpink}{What is the musical texture resulting from the interaction between participants?}{\dimInteractionMusicalOutcomeNum}{\dimInteractionMusicalOutcomePct}{

        \code{Textural}{Textural}{Stack of 
        % distinct 
        sound layers emphasizing timbral and rhythmic density
        % rather than melodic hierarchy
        }{\autocitep{marley2015gestroviser}}{\textbf{\codInteractionMusicalOutcomeTexturalNum}}{\textbf{\codInteractionMusicalOutcomeTexturalPct}}

        \code{Homophony}{Homophony}{One melodic line clearly supported by accompaniment}{\autocitep{sioros2011automatic}}{\codInteractionMusicalOutcomeHomophonyNum}{\codInteractionMusicalOutcomeHomophonyPct}

        \code{Polyphony}{Polyphony}{Multiple independent melodic lines}{\autocitep{turczan2019scale}}{\codInteractionMusicalOutcomePolyphonyNum}{\codInteractionMusicalOutcomePolyphonyPct}

        \code{Non-specific}{Non-specific}{
        Mentions music being generated without enough detail to identify the outcome texture}{\autocitep{martin2019interactive}}{\codInteractionMusicalOutcomeNonSpecificNum}{\codInteractionMusicalOutcomeNonSpecificPct}  

        \code{Monophony}{Monophony}{A single melodic line}{\autocitep{knotts2021algorithmic}}{\codInteractionMusicalOutcomeMonophonyNum}{\codInteractionMusicalOutcomeMonophonyPct}

        \code{Heterophony}{Heterophony}{Simultaneous variations of one melodic line}{\autocitep{brown2018interacting}}{\codInteractionMusicalOutcomeHeterophonyNum}{\codInteractionMusicalOutcomeHeterophonyPct} 
    }
    
    \dimension{Planning}{Planning}{mutedpink}{What is prepared or shared among participants before performance?}{\dimInteractionPlanningNum}{\dimInteractionPlanningPct}{

        \code{No planning}{No planning}{All musical decisions are made live}{\citep{dahl2011tweetdreams}}{\textbf{\codInteractionPlanningNoPlanningNum}}{\textbf{\codInteractionPlanningNoPlanningPct}}

        \code{User configuration}{User configuration}{Set system presets (e.g., model hyperparameter)}{\citep{collins2010musical}}{\codInteractionPlanningUserConfigurationNum}{\codInteractionPlanningUserConfigurationPct}

        \code{Tailoring}{Tailoring}{
        %Feed user data to customize the system
        Specific user data is fed to customize the system
        }{\citep{inasilentway}}{\codInteractionPlanningTailoringNum}{\codInteractionPlanningTailoringPct}            
        
        \code{Score planning}{
        %Score planning
        Score
        }{Pre-planned score (e.g. a lead sheet) or score elements (\eg chords, tempo)}{\citep{reflexivelooper}}{\codInteractionPlanningScorePlanningNum}{\codInteractionPlanningScorePlanningPct}

        \code{Material prep}{Material prep}{Pre-select and load musical materials (e.g., sound samples)}{\citep{mimi4x}}{\codInteractionPlanningMaterialPrepNum}{\codInteractionPlanningMaterialPrepPct}

        \code{Predefined cues}{Predefined cues}{Predefine mappings between cues and behaviors}{\citep{mitchell2011soundgrasp}}{\codInteractionPlanningPredefinedCuesNum}{\codInteractionPlanningPredefinedCuesPct}
    
        \code{Timeline}{
        %Timeline
        Initiative timeline
        }{Schedule who plays when (e.g., alternation or lead changes format)}{\citep{bob}}{\codInteractionPlanningTimelineNum}{\codInteractionPlanningTimelinePct}
        
    }
    
    \dimension{Temporal structure}{Temporal Structure}{mutedpink}{How are musical contributions temporally structured between participants?}{\dimInteractionTemporalStructureNum}{\dimInteractionTemporalStructurePct}{
        \code{Dense parallel}{Dense parallel}{Musical contributions occur simultaneously in dense overlapping layers}{\citep{realchords}}{\textbf{\codInteractionTemporalStructureDenseParallelNum}}{\textbf{\codInteractionTemporalStructureDenseParallelPct}}
        
        \code{Sparse parallel}{Sparse parallel}{Musical contributions occur simultaneously at low density}{\citep{naess2019physical}}{\codInteractionTemporalStructureSparseParallelNum}{\codInteractionTemporalStructureSparseParallelPct}

        \code{Hybrid}{Hybrid}{Temporal structure shifts dynamically during performance}{\citep{shimon}}{\codInteractionTemporalStructureHybridNum}{\codInteractionTemporalStructureHybridPct}

        \code{Unstructured}{Unstructured}{Temporal structure is free-form;
        % and unconstrained, 
        %; follow not consistent temporal rule
        no consistent temporal rule
        }{\citep{Gresham-Lancaster15}}{\codInteractionTemporalStructureUnstructuredNum}{\codInteractionTemporalStructureUnstructuredPct}
        
        \code{Turn-taking}{Turn-taking}{Musical contributions alternate in back-and-forth pattern between 
        participants
        % CHRIS: when possible, let's always try to future proof by removing distinctions between users and agents
        }{\citep{continuator}}{\codInteractionTemporalStructureTurnTakingNum}{\codInteractionTemporalStructureTurnTakingPct}

    }

    \dimension{Data alignment}{Data Alignment}{mutedpink}{How is data among participants synchronized during performance?}{\dimInteractionDataAlignmentNum}{\dimInteractionDataAlignmentPct}{

        \code{Continuous stream}{Continuous stream}{Data is synced in real time as a constant data stream}{\citep{takase2020support}}{\textbf{\codInteractionDataAlignmentContinuousStreamNum}}{\textbf{\codInteractionDataAlignmentContinuousStreamPct}}
    
        \code{Background trigger}{Background trigger}{
        Data is shared when specific triggers 
        or event occur
        % System listens passively and reacts to specific triggers or events (data is shared when trigger happen)
        }{\citep{almeida2019amigo}}{\codInteractionDataAlignmentBackgroundTriggerNum}{\codInteractionDataAlignmentBackgroundTriggerPct}
        
        \code{Periodic}{Periodic}{Data is shared at regular intervals or checkpoints}{\citep{knotts2021algorithmic}}{\codInteractionDataAlignmentPeriodicNum}{\codInteractionDataAlignmentPeriodicPct}
            
    }
    
    \dimension{Interface}{Interface}{mutedpink}{What hardware or software interface facilitates the interaction during the performance?}{\dimInteractionInterfaceNum}{\dimInteractionInterfacePct}{
    
        \code{Graphical user interface}{Graphical user interface}{System offers a standard software interface with keyboard, mouse, touch control, etc.}{\citep{BakhtB09}}{\textbf{\codInteractionInterfaceGraphicalUserInterfaceNum}}{\textbf{\codInteractionInterfaceGraphicalUserInterfacePct}}

        \code{Conventional instrument}{Conventional instrument}{System connects to commonly available acoustic or digital instrument}{\citep{stefani2024esteso}}{\codInteractionInterfaceConventionalInstrumentNum}{\codInteractionInterfaceConventionalInstrumentPct}

        \code{Sensor device}{Sensor device}{System connects to specialized hardware sensors to collect user data}{\citep{filandrianos2020brainwaves}}{\codInteractionInterfaceSensorDeviceNum}{\codInteractionInterfaceSensorDevicePct}      
        
        \code{Custom instrument}{Custom instrument}{System connects to or is embedded in a custom-built instrument}{\citep{privato2024stacco}}{\codInteractionInterfaceCustomInstrumentNum}{\codInteractionInterfaceCustomInstrumentPct}
       
        \code{Stage visual}{Stage visual}{System connects to an audience-facing output projection LED walls
        %or projectors
        }{\citep{shepardson2025evolving}}{\codInteractionInterfaceStageVisualNum}{\codInteractionInterfaceStageVisualPct}
 
        \code{Visual programming interface}{Programming interface}{
        %Visual programming interface / live coding interface
        System includes a programming environment (\eg visual programming interface)
        }{\citep{livecoding}}{\codInteractionInterfaceProgrammingInterfaceNum}{\codInteractionInterfaceProgrammingInterfacePct}

        \code{Embodied agent}{Embodied agent}{System includes a physical, embodied output (\eg robot arm)}{\citep{shimon}}{\codInteractionInterfaceEmbodiedAgentNum}{\codInteractionInterfaceEmbodiedAgentPct}

        \code{XR interface}{XR interface}{System offers an interface through VR, AR, or XR environments}{\citep{wang2025ai}}{\codInteractionInterfaceXRInterfaceNum}{\codInteractionInterfaceXRInterfacePct}

        \code{DJ Gear}{DJ gear}{System connects to DJ-specific gears (\eg professional controller, mixer, turntable)}{\citep{tsiros2020towards}}{\codInteractionInterfaceDJGearNum}{\codInteractionInterfaceDJGearPct}
    }
    
}

\aspect{\interactionaspect{Interaction}}{tab:interaction-designspace-3}{tab:interaction-designspace-1,tab:interaction-designspace-2}{\aspInteractionNum}{\aspInteractionPct}{

    \dimension{Control Mode}{Control Mode}{mutedpink}{How does the user steer the system in real time?}{\dimInteractionControlModeNum}{\dimInteractionControlModePct}{

        \code{Implicit}{Implicit}{User's 
        %unconscious 
        implicit 
        actions are 
        %taken as system's input 
        interpreted by the system 
        to generate output}{\citep{realchords}}{\textbf{\codInteractionControlModeImplicitNum}}{\textbf{\codInteractionControlModeImplicitPct}}

        \code{Explicit}{Explicit}{User can %consciously 
        explicitly
        control system generation via buttons, sliders, etc.}{\citep{musicfxdj}}{\codInteractionControlModeExplicitNum}{\codInteractionControlModeExplicitPct}
        
        \code{No control}{No control}{User cannot control the system}{\citep{genjam}}{\codInteractionControlModeNoControlNum}{\codInteractionControlModeNoControlPct}
        
    }

    \dimension{Control Scope}{Control Scope}{mutedpink}{What aspects of the agent's output can be steered by user in real time?}{\dimInteractionControlScopeNum}{\dimInteractionControlScopePct}{

        \code{Musical material}{Musical material}{Immediate musical contents 
        such as 
        (\eg harmony, rhythm, 
        % or %melody
        phrasing)}{\citep{continuator}}{\textbf{\codInteractionControlScopeMusicalMaterialNum}}{\textbf{\codInteractionControlScopeMusicalMaterialPct}}

        \code{Style direction}{Global style}{High-level stylistic attributes of music 
        %such as 
        (\eg timbre, genre, instrumentation, mood)}{\citep{musicfxdj}}{\codInteractionControlScopeStyleDirectionNum}{\codInteractionControlScopeStyleDirectionPct}
        
        \code{Behavioral direction}{High-level behavior}{Agent's high-level role, autonomy, or interactive behavior}{\citep{spiremuse}}{\codInteractionControlScopeBehavioralDirectionNum}{\codInteractionControlScopeBehavioralDirectionPct}

        \code{Musical structure}{Musical layout}{
        %High-level
        Layout of music (\eg tempo, meter, or key)
        %/ which layer to play
        }{\citep{vogl2017an}}{\codInteractionControlScopeMusicalStructureNum}{\codInteractionControlScopeMusicalStructurePct}
    }
    
    \dimension{System initiative}{System Initiative}{mutedpink}{How is system initiative distributed across %interaction?
    participants?
    }{\dimInteractionSystemInitiativeNum}{\dimInteractionSystemInitiativePct}{
    
        \code{Reactive}{Reactive}{System only reacts to user actions}{\autocitep{visi2017a}}{\textbf{\codInteractionSystemInitiativeReactiveNum}}{\textbf{\codInteractionSystemInitiativeReactivePct}}
        
        \code{Mixed-initiative}{Mixed-initiative}{System alternates between reacting and initiating}{\autocitep{BarateHL14}}{\codInteractionSystemInitiativeMixedInitiativeNum}{\codInteractionSystemInitiativeMixedInitiativePct}

        \code{Proactive}{Proactive}{System initiates material or leads the direction}{\autocitep{Hoadley12}}{\codInteractionSystemInitiativeProactiveNum}{\codInteractionSystemInitiativeProactivePct}

    }
    
    \dimension{Agency framing}{Agency Framing}{mutedpink}{How is the system's agency defined and used?}{\dimInteractionAgencyFramingNum}{\dimInteractionAgencyFramingPct}{        
        \code{Tool}{Tool}{
        % Frames system as 
        A controllable tool/instrument to realize user intent}{\autocitep{umwelt}}{\textbf{\codInteractionAgencyFramingToolNum}}{\textbf{\codInteractionAgencyFramingToolPct}}
        
        \code{Hybrid}{Hybrid}{
        %Frames system as 
        A mix of partner and tool capable of independent contribution yet directed by user}{\autocitep{donnarumma2012music}}{\codInteractionAgencyFramingHybridNum}{\codInteractionAgencyFramingHybridPct}

        \code{Partner}{Partner}{
        %Frames system as 
        An autonomous partner with its own intentions}{\autocitep{inasilentway}}{\codInteractionAgencyFramingPartnerNum}{\codInteractionAgencyFramingPartnerPct}

    }
    % {\todo{Explain \% and number aggregation for I/O dimension}}
}

%% file: tables/technology-design-space.tex
\aspect{\technologyaspect{Technology}}{tab:technology-designspace-1}{tab:technology-designspace-2}{\aspTechnologyNum}{\aspTechnologyPct}{

    \dimension{Model}{Model}{mutedblue}{What is the type of the underlying computational or AI model(s)?}{\dimTechnologyModelNum}{\dimTechnologyModelPct}{

        \code{Stochastic process}{Stochastic process}{Nondeterministic probabilistic processes that sample outcomes 
        %without learned weights
        based on simple statistics
        }{\citep{continuator}}{\textbf{\codTechnologyModelStochasticProcessNum}}{\textbf{\codTechnologyModelStochasticProcessPct}}

        \code{Task-specific DNN}{Task-specific DNN}{Deep neural network (DNN) trained for a narrow task or domain}{\citep{lionetti2024muscleguided}}{\codTechnologyModelTaskSpecificDNNNum}{\codTechnologyModelTaskSpecificDNNPct}

        \code{Classical ML}{Classical ML}{Statistical machine learning (ML) models without neural layers}{\citep{scurto2017shaping}}{\codTechnologyModelClassicalMLNum}{\codTechnologyModelClassicalMLPct}
        
        \code{Rule-based}{Rule-based}{Deterministic rules or algorithms without training data}{\citep{godbehere2008wearable}}{\codTechnologyModelRuleBasedNum}{\codTechnologyModelRuleBasedPct}

        \code{Shallow neural network}{Shallow neural network}{One to three hidden layers used for simple tasks}{\citep{HothkerH00}}{\codTechnologyModelShallowNeuralNetworkNum}{\codTechnologyModelShallowNeuralNetworkPct}
        
        \code{Transformer}{Generative AI}{
        %Broadly trained attention model adaptable across tasks and domains
        Large model trained on large datasets and adaptable to many tasks and domains
        % CHRIS: this really should be broader than ``Transformer''. the architecture is not the salient detail here, but rather the training paradigm and task specificity. we should do a consistency pass after the deadline to make sure all large gen AI models ended up here and not in task-specific DNN
        }{\citep{vampnet}}{\codTechnologyModelTransformerNum}{\codTechnologyModelTransformerPct}
        
    }
    
    \dimension{Learning algorithm}{Learning Algorithm}{mutedblue}{How is the underlying model trained?}{\dimTechnologyLearningAlgorithmNum}{\dimTechnologyLearningAlgorithmPct}{
    
        \code{Supervised learning}{Supervised learning}{Model learns from labeled data with known input-output pairs}{\citep{fiebrink2010wekinator}}{\textbf{\codTechnologyLearningAlgorithmSupervisedLearningNum}}{\textbf{\codTechnologyLearningAlgorithmSupervisedLearningPct}}
        
        \code{Unsupervised learning}{Unsupervised learning}{Model learns from unlabeled data by finding patterns and structure}{\citep{smith2012unsupervised}}{\codTechnologyLearningAlgorithmUnsupervisedLearningNum}{\codTechnologyLearningAlgorithmUnsupervisedLearningPct}
        
        \code{Self-supervised learning}{Self-supervised learning}{Model learns from unlabeled data by generating supervisory signals from the data itself}{\citep{rave}}{\codTechnologyLearningAlgorithmSelfSupervisedLearningNum}{\codTechnologyLearningAlgorithmSelfSupervisedLearningPct}
        
        \code{Reinforcement learning}{Reinforcement learning}{Model learns by interacting with environment and reward signals}{\citep{rlduet}}{\codTechnologyLearningAlgorithmReinforcementLearningNum}{\codTechnologyLearningAlgorithmReinforcementLearningPct}
        
    }
    
    \dimension{Inference objective}{Inference Objective}{mutedblue}{How does the underlying model transforms inputs into outputs based on the task objective?}{\dimTechnologyInferenceObjectiveNum}{\dimTechnologyInferenceObjectivePct}{

        \code{Unimodal generation}{Unimodal generation}{Model generates outputs in the same modality as the input}{\citep{bachduet}}{\textbf{\codTechnologyInferenceObjectiveUnimodalGenerationNum}}{\textbf{\codTechnologyInferenceObjectiveUnimodalGenerationPct}}
        
        \code{Classification}{Classification}{Model assigns input to predefined categories or labels}{\citep{martelloni2023realtimepercussivetechniquerecognition}}{\codTechnologyInferenceObjectiveClassificationNum}{\codTechnologyInferenceObjectiveClassificationPct}

        \code{Cross-modal generation}{Cross-modal generation}{Model generates outputs in a different modality from the input}{\citep{arai2023timtoshape}}{\codTechnologyInferenceObjectiveCrossModalGenerationNum}{\codTechnologyInferenceObjectiveCrossModalGenerationPct}
    
        \code{Regression}{Regression}{Model predicts a continuous value based on input data}{\citep{CemgilK01-0}}{\codTechnologyInferenceObjectiveRegressionNum}{\codTechnologyInferenceObjectiveRegressionPct}

        \code{Retrieval}{Retrieval}{Model selects or ranks existing content based on user input}{\citep{spiremuse}}{\codTechnologyInferenceObjectiveRetrievalNum}{\codTechnologyInferenceObjectiveRetrievalPct}
        
    }
    
    \dimension{Adaptation}{Adaptation}{mutedblue}{How does the underlying model adapt to the user (e.g., by updating their parameters)?}{\dimTechnologyAdaptationNum}{\dimTechnologyAdaptationPct}{
    
        \code{No adaptation}{No adaptation}{Model parameters are used as-is and not updated}{\citep{hamano2013generating}}{\textbf{\codTechnologyAdaptationNoAdaptationNum}}{\textbf{\codTechnologyAdaptationNoAdaptationPct}}

        \code{Offline adaptation}{Offline adaptation}{Model is fine-tuned on user data prior to a session}{\citep{nime2025_54}}{\codTechnologyAdaptationOfflineAdaptationNum}{\codTechnologyAdaptationOfflineAdaptationPct}
        
        \code{Online adaptation}{Online adaptation}{Model adapts on-the-fly during a session based on user behavior}{\citep{CemgilK01-0}}{\codTechnologyAdaptationOnlineAdaptationNum}{\codTechnologyAdaptationOnlineAdaptationPct}

        \code{Continual adaptation}{Continual adaptation}{Model adapts incrementally across sessions}{\citep{assemblage}}{\codTechnologyAdaptationContinualAdaptationNum}{\codTechnologyAdaptationContinualAdaptationPct}
        
    }

    \dimension{Technology desiderata}{Technical Emphasis}{mutedblue}{Which practical needs 
    %shape 
    are emphasized in the 
    %model 
    technical 
    design and deployment?}{\dimTechnologyTechnologyDesiderataNum}{\dimTechnologyTechnologyDesiderataPct}{
    
        \code{Latency}{Latency}{
        %Needs to 
        Minimize inference delay and remain responsive to user inputs}{\citep{brochec2023toward}}{\textbf{\codTechnologyTechnologyDesiderataLatencyNum}}{\textbf{\codTechnologyTechnologyDesiderataLatencyPct}}
        
        \code{Efficiency}{Efficiency}{
        %Needs to 
        Run within the available computing and memory resources}{\citep{jambot}}{\codTechnologyTechnologyDesiderataEfficiencyNum}{\codTechnologyTechnologyDesiderataEfficiencyPct}

        \code{Tempo adaptability}{Tempo adaptability}{
        %Needs to 
        Follow natural tempo fluctuations and expressive timing}{\citep{jamfactory}}{\codTechnologyTechnologyDesiderataTempoAdaptabilityNum}{\codTechnologyTechnologyDesiderataTempoAdaptabilityPct}
                
        \code{Error handling}{Error handling}{
        %Needs to 
        Handle mistakes or unexpected user actions}{\citep{realchords}}{\codTechnologyTechnologyDesiderataErrorHandlingNum}{\codTechnologyTechnologyDesiderataErrorHandlingPct}
        
    }

}

\aspect{\technologyaspect{Technology}}{tab:technology-designspace-2}{tab:technology-designspace-1}{\aspTechnologyNum}{\aspTechnologyPct}{

    \dimension{Technology infrastructure}{Infrastructure}{mutedblue}{What technical infrastructure supports the implementation and operation of the underlying technology?}{\dimTechnologyTechnologyInfrastructureNum}{\dimTechnologyTechnologyInfrastructurePct}{
    
        \code{Music programming environment}{Music programming\phantom{00000} environment}{Platforms for real-time audio synthesis and processing (\eg Max/MSP)}{\citep{NIME22_25}}{\textbf{\codTechnologyTechnologyInfrastructureMusicProgrammingEnvironmentNum}}{\textbf{\codTechnologyTechnologyInfrastructureMusicProgrammingEnvironmentPct}}
        
        \code{General programming environment}{General programming\phantom{000} environment}{Common-purpose programming languages (\eg Python)}{\citep{dahl2011tweetdreams}}{\codTechnologyTechnologyInfrastructureGeneralProgrammingEnvironmentNum}{\codTechnologyTechnologyInfrastructureGeneralProgrammingEnvironmentPct}

        \code{AI/ML framework}{AI/ML framework}{Libraries and APIs for AI (\eg PyTorch, ChatGPT API)
        % CHRIS: After deadline, I might suggest we try to disentangle APIs from AI/ML frameworks
        }{\citep{vampnet}}{\codTechnologyTechnologyInfrastructureAIMLFrameworkNum}{\codTechnologyTechnologyInfrastructureAIMLFrameworkPct}   
        
        \code{Protocol}{Music protocol}{Standards for communication between music systems (\eg OSC)}{\citep{Gresham-Lancaster15}}{\codTechnologyTechnologyInfrastructureProtocolNum}{\codTechnologyTechnologyInfrastructureProtocolPct}
        
        \code{Software toolkit}{Software toolkit}{Software packages for audio, interaction, and media development (\eg JUCE)}{\citep{van2012mapping}}{\codTechnologyTechnologyInfrastructureSoftwareToolkitNum}{\codTechnologyTechnologyInfrastructureSoftwareToolkitPct}
        
        \code{Hardware toolkit}{Hardware toolkit}{Prototype kits for physical sensing and control (\eg Arduino)}{\citep{kapur2007integrating}}{\codTechnologyTechnologyInfrastructureHardwareToolkitNum}{\codTechnologyTechnologyInfrastructureHardwareToolkitPct}

    }
    
    \dimension{Runtime requirements}{Runtime Requirements}{mutedblue}{What execution environment does the system depend on?}{\dimTechnologyRuntimeRequirementsNum}{\dimTechnologyRuntimeRequirementsPct}{
    
        \code{Commodity machine}{Commodity machine}{Runs on a standard laptop/desktop or mobile device}{\citep{fiebrink2010wekinator}}{\textbf{\codTechnologyRuntimeRequirementsCommodityMachineNum}}{\textbf{\codTechnologyRuntimeRequirementsCommodityMachinePct}}

        \code{Dedicated commodity hardware}{Dedicated commodity\phantom{000} hardware}{Requires a particular off-the-shelf device beyond the computer (\eg depth camera, brain electronic sensors)}{\citep{wang2025ai}}{\codTechnologyRuntimeRequirementsDedicatedCommodityHardwareNum}{\codTechnologyRuntimeRequirementsDedicatedCommodityHardwarePct}
                
        \code{Custom hardware}{Custom hardware}{Requires a non-retail or custom-built device (e.g., 3D-printed controller)}{\citep{fiebrink2020reflections}}{\codTechnologyRuntimeRequirementsCustomHardwareNum}{\codTechnologyRuntimeRequirementsCustomHardwarePct}
        \code{Cloud API}{Cloud API}{Requires access to remote servers 
        %with API license
        via APIs
        }{\citep{musicfxdj}}{\codTechnologyRuntimeRequirementsCloudAPINum}{\codTechnologyRuntimeRequirementsCloudAPIPct}

        \code{High-performance compute}{High-performance\phantom{0000} compute}{Requires pro/multi-GPU or accelerators (e.g., A100, H100, TPU) to meet desired latency or quality}{\citep{nime2025_54}}{\codTechnologyRuntimeRequirementsHighPerformanceComputeNum}{\codTechnologyRuntimeRequirementsHighPerformanceComputePct}

    }
    
    \dimension{Workflow Integration}{Integration}{mutedblue}{How closely does the system match musicians' existing tools and setup?}{\dimTechnologyWorkflowIntegrationNum}{\dimTechnologyWorkflowIntegrationPct}{
    
        \code{Bespoke setup}{Bespoke setup}{Requires non-standard devices outside common musician setups}{\citep{aiterity}}{\textbf{\codTechnologyWorkflowIntegrationBespokeSetupNum}}{\textbf{\codTechnologyWorkflowIntegrationBespokeSetupPct}}

        \code{Packaged standalone}{Standalone}{Provides ready-to-run desktop/web/mobile apps requiring new workflow adoption}{\citep{vogl2017an}}{\codTechnologyWorkflowIntegrationPackagedStandaloneNum}{\codTechnologyWorkflowIntegrationPackagedStandalonePct}
    
        \code{Source-only prototype}{Source-only prototype}{Runs from code repository or Jupyter notebook requiring developer steps to run}{\citep{johnson2023musical}}{\codTechnologyWorkflowIntegrationSourceOnlyPrototypeNum}{\codTechnologyWorkflowIntegrationSourceOnlyPrototypePct}

        \code{Tool-integrated}{Tool-integrated}{Runs within or alongside common music tools (e.g., VST Plugins)}{\citep{semilla}}{\codTechnologyWorkflowIntegrationToolIntegratedNum}{\codTechnologyWorkflowIntegrationToolIntegratedPct}
        
        \code{Developer toolkit}{Developer toolkit}{Provides libraries, APIs, or subpatches could be used with programming environment}{\citep{improtek}}{\codTechnologyWorkflowIntegrationDeveloperToolkitNum}{\codTechnologyWorkflowIntegrationDeveloperToolkitPct}
        
    }
    
}

%% file: tables/environment-design-space.tex
\aspect{\ecosystemaspect{Ecosystem}}{tab:ecosystem-designspace}{}{\aspEcosystemNum}{\aspEcosystemPct}{

    \dimension{Sociocultural factors}{Sociocultural Factors}{cozygreen!50}{What cultural values and social practices surround the system's design and use?}{\dimEcosystemSocioculturalFactorsNum}{\dimEcosystemSocioculturalFactorsPct}{
        \code{Musical practice}{Musical practice}{Established conventions guiding performance and composition}{\citep{proctor2020a}}{\textbf{\codEcosystemSocioculturalFactorsMusicalPracticeNum}}{\textbf{\codEcosystemSocioculturalFactorsMusicalPracticePct}}
        
        \code{Musical genre}{Musical genre}{Distinct genres, styles, or traditions shaping music-making}{\citep{marchini2017rethinking}}{\codEcosystemSocioculturalFactorsMusicalGenreNum}{\codEcosystemSocioculturalFactorsMusicalGenrePct}

        \code{AI perception}{AI perception}{Prevailing views of AI's role and legitimacy in music-making}{\citep{dadabots}}{\codEcosystemSocioculturalFactorsAIPerceptionNum}{\codEcosystemSocioculturalFactorsAIPerceptionPct}

        \code{Cultural conservatism}{Cultural conservatism}{Preference for preserving established practices over innovation}{\citep{tsiros2020towards}}{\codEcosystemSocioculturalFactorsCulturalConservatismNum}{\codEcosystemSocioculturalFactorsCulturalConservatismPct}
    }

    \dimension{Policy factors}{Policy Considerations}{cozygreen!50}{What legal, institutional, or organizational norms surround the system's design and use?}{\dimEcosystemPolicyFactorsNum}{\dimEcosystemPolicyFactorsPct}{
        \code{Copyright concerns}{Authorship}{Potential infringement or challenges to users’ authorship over musical outcomes}{\citep{jamfactory}}{\textbf{\codEcosystemPolicyFactorsCopyrightConcernsNum}}{\textbf{\codEcosystemPolicyFactorsCopyrightConcernsPct}}

        \code{Data privacy}{Data privacy}{Rules and expectations around collection, storage, and use of personal data}{\citep{wu2025gesturedriven}}{\codEcosystemPolicyFactorsDataPrivacyNum}{\codEcosystemPolicyFactorsDataPrivacyPct}

        \code{Plagiarism}{Integrity}{Ethical concerns about uncredited borrowing or non-infringing imitation}{\citep{jiang2020when}}{\codEcosystemPolicyFactorsPlagiarismNum}{\codEcosystemPolicyFactorsPlagiarismPct}
        
        \code{Personality rights}{Personality rights}{Legal and ethical limits on using real people in media or AI outputs}{\citep{shepardson2024tungnaa}}{\codEcosystemPolicyFactorsPersonalityRightsNum}{\codEcosystemPolicyFactorsPersonalityRightsPct}

    }

    \dimension{Economic Consequences}{Economic Consequences}{cozygreen!50}{How will the system influence the economic landscape of music industry?}{\dimEcosystemEconomicConsequencesNum}{\dimEcosystemEconomicConsequencesPct}{

        \code{Devaluing}{Devaluing}{Lower the financial or cultural value of human musicianship}{\citep{savery2024collaborationrobotsinterfaceshumans}}{\textbf{\codEcosystemEconomicConsequencesDevaluingNum}}{\textbf{\codEcosystemEconomicConsequencesDevaluingPct}}

        % \code{Deskilling}{Deskilling}{Reduce demand for specialized expertise or long-term training}{\citep{}}{\codEcosystemEconomicConsequencesDeskillingNum}{\codEcosystemEconomicConsequencesDeskillingPct}

        \code{Job replacement}{Job replacement}{Replace human creative labor with AI systems}{\citep{realchords}}{\codEcosystemEconomicConsequencesJobReplacementNum}{\codEcosystemEconomicConsequencesJobReplacementPct}

    }

    \dimension{Musical-societal consequences}{Musical-societal Consequences}{cozygreen!50}{How will the systems change music scene and more broadly society?}{\dimEcosystemMusicalSocietalConsequencesNum}{\dimEcosystemMusicalSocietalConsequencesPct}{
        \code{Reshaping idioms}{Reshaping idioms}{Invent new musical forms, aesthetics, or performance practices}{\citep{assemblage}}{\textbf{\codEcosystemMusicalSocietalConsequencesReshapingIdiomsNum}}{\textbf{\codEcosystemMusicalSocietalConsequencesReshapingIdiomsPct}}

        \code{Democratization}{Democratization}{Promote wider participation in music creation}{\citep{Kaliakatsos-Papakostas14}}{\codEcosystemMusicalSocietalConsequencesDemocratizationNum}{\codEcosystemMusicalSocietalConsequencesDemocratizationPct}
    
        \code{Cross-domain collaboration}{Collaboration}{Foster artist–researcher partnerships bridging artistic and scientific/tehnical expertise}{\citep{nash2020crowd}}{\codEcosystemMusicalSocietalConsequencesCrossDomainCollaborationNum}{\codEcosystemMusicalSocietalConsequencesCrossDomainCollaborationPct}

        \code{Skill dilution}{Skill dilution}{Erode traditional musical competencies by reducing their depth}{\citep{tsiros2020towards}}{\codEcosystemMusicalSocietalConsequencesSkillDilutionNum}{\codEcosystemMusicalSocietalConsequencesSkillDilutionPct}

        \code{Cultural exchange}{Cultural exchange}{Foster style fusion and dialogue between different musical traditions or regions}{\citep{jiang2020when}}{\codEcosystemMusicalSocietalConsequencesCulturalExchangeNum}{\codEcosystemMusicalSocietalConsequencesCulturalExchangePct}

        \code{Revitalization}{Revitalization}{Stimulate interest in endangered instruments or marginalized genres}{\citep{10.1145/1254960.1254990}}{\codEcosystemMusicalSocietalConsequencesRevitalizationNum}{\codEcosystemMusicalSocietalConsequencesRevitalizationPct}

        \code{Misrepresentation}{Misrepresentation}{Distort or misrepresent musical communities, practices, or traditions}{\citep{jambot}}{\codEcosystemMusicalSocietalConsequencesMisrepresentationNum}{\codEcosystemMusicalSocietalConsequencesMisrepresentationPct}

        \code{Over-reliance}{Over-reliance}{Induce over-reliance on systems or AI outputs}{\citep{tsiros2020towards}}{\codEcosystemMusicalSocietalConsequencesOverRelianceNum}{\codEcosystemMusicalSocietalConsequencesOverReliancePct}
    }
}